\documentclass{aa}  

\usepackage{graphicx}
\usepackage{natbib}
\usepackage{amssymb}
\usepackage{txfonts}
\usepackage{makecell}
\bibpunct{(}{)}{;}{a}{}{,}
\usepackage{xcolor}
\usepackage{xspace}
\usepackage{listings}
\usepackage[colorlinks = true,
           linkcolor = blue,
           urlcolor  = blue,
           citecolor = blue,
           anchorcolor = blue]{hyperref}
\defcitealias{hunt+2023}{HR23}
\defcitealias{castroginard+2022}{CG22}
\defcitealias{casamiquela+2021}{C21}

\def\Gaia{{\it Gaia }}

\def\gaiaeso{\emph{Gaia}-ESO }
\DeclareRobustCommand{\ion}[2]{\textup{#1\,\textsc{\lowercase{#2}}}}
\DeclareRobustCommand{\teff}{T_{\mathrm{eff}}}
\DeclareRobustCommand{\logg}{\log g}
\DeclareRobustCommand{\mh}{\mathrm{[M/H]}}
\DeclareRobustCommand{\vmic}{\varv_\mathrm{mic}}

\DeclareRobustCommand{\ispec}{\texttt{iSpec} }
\DeclareRobustCommand{\kms}{\mathrm{km \ s}^{-1}}
\DeclareRobustCommand{\vr}{\varv_\mathrm{r}}
\DeclareRobustCommand{\vz}{\varv_\mathrm{z}}
\DeclareRobustCommand{\AV}{A_\mathrm{V}}
\DeclareRobustCommand{\kmskpc}{\mathrm{km} \ {\mathrm{s}}^{-1} \ {\mathrm{kpc}}^{-1}}

\DeclareRobustCommand{\rgc}{R_{\mathrm{GC}}}
\DeclareRobustCommand{\rguide}{R_{\mathrm{guide}}}
\DeclareRobustCommand{\rbirth}{R_{\mathrm{birth}}}
\DeclareRobustCommand{\rperi}{R_{\mathrm{peri}}}
\DeclareRobustCommand{\rapo}{R_{\mathrm{apo}}}
\DeclareRobustCommand{\zmax}{Z_{\mathrm{max}}}
\DeclareRobustCommand{\omegabar}{{\Omega}_{\mathrm{bar}}}
\DeclareRobustCommand{\rbar}{R_{\mathrm{bar}}}
\DeclareRobustCommand{\Af}{A_{\mathrm{f}}}
\DeclareRobustCommand{\phibar}{{\phi}_{\mathrm{bar}}}

\makeatletter
\newcommand\makebig[2]{%
  \@xp\newcommand\@xp*\csname#1\endcsname{\bBigg@{#2}}%
  \@xp\newcommand\@xp*\csname#1l\endcsname{\@xp\mathopen\csname#1\endcsname}%
  \@xp\newcommand\@xp*\csname#1r\endcsname{\@xp\mathclose\csname#1\endcsname}%
}
\makeatother

\makebig{bigggg}{4.0}

\begin{document}

   \title{The high-altitude, inner-disc, and chemically peculiar 
   open cluster\\ UBC~1052\thanks{Based on data acquired with the Very Large Telescope (VLT) at the European Southern Observatory (ESO) in Paranal, Chile (ESO Programme ID: 111.255H.001).}}

   \author{Judit Donada\inst{\ref{inst:iccub},\ref{inst:fqa},\ref{inst:ieec}} 
          \and Laia Casamiquela\inst{\ref{inst:lira_paris}}
          \and Friedrich Anders\inst{\ref{inst:iccub},\ref{inst:fqa},\ref{inst:ieec}}
          \and Lola Balaguer-N\'{u}\~{n}ez\inst{\ref{inst:iccub},\ref{inst:fqa},\ref{inst:ieec}}
          \and Sergi Blanco-Cuaresma\inst{\ref{inst:harvard_usa},\ref{inst:unidistance_suisse}}
          \and Xavier Luri\inst{\ref{inst:ieec},\ref{inst:iccub},\ref{inst:fqa}}
          \and Ditte Slumstrup\inst{\ref{inst:grantecan},\ref{inst:iac}}
          \and Carme Jordi\inst{\ref{inst:ieec},\ref{inst:iec_bcn}}
          \and Alfred Castro-Ginard\inst{\ref{inst:iccub},\ref{inst:fqa},\ref{inst:ieec}}
          \and Ricardo Carrera\inst{\ref{inst:inaf_bologna}}
          \and Josep Manel Carrasco\inst{\ref{inst:iccub},\ref{inst:fqa},\ref{inst:ieec}}}

    \institute{{Institut de Ci\`encies del Cosmos (ICCUB), Universitat de Barcelona (UB), Mart\'i i Franqu\`es 1, 08028 Barcelona, Spain\\
    \email{jdonada@fqa.ub.edu} \label{inst:iccub}}
    \and{Departament de Física Qu\`antica i Astrof\'isica (FQA), Universitat de Barcelona (UB), Mart\'i i Franqu\`es 1, 08028 Barcelona, Spain \label{inst:fqa}}
    \and{Institut d'Estudis Espacials de Catalunya (IEEC), Esteve Terradas, 1, Edifici RDIT, Campus PMT-UPC, 08860 Castelldefels, Spain\label{inst:ieec}}
    \and{LIRA, Observatoire de Paris, Université PSL, Sorbonne Université, Université Paris Cité, CY Cergy Paris Université, CNRS, 92190 Meudon, France\label{inst:lira_paris}}
    \and{Harvard-Smithsonian Center for Astrophysics, 60 Garden Street, Cambridge, MA 02138, USA \label{inst:harvard_usa}}
    \and{Faculty of Psychology, UniDistance Suisse, Brig, Switzerland \label{inst:unidistance_suisse}}
    \and{GRANTECAN, Cuesta de San José s/n, E-38712 Breña Baja, La Palma, Spain \label{inst:grantecan}}
    \and{Instituto de Astrofísica de Canarias, E-38205 La Laguna, Tenerife, Spain \label{inst:iac}}
    \and{Institut d'Estudis Catalans (IEC), Carrer del Carme, 47, 08001 Barcelona, Spain \label{inst:iec_bcn}}
    \and{INAF - Osservatorio di Astrofisica e Fisica dello Spazio, via P. Gobetti 93/3, 40129 Bologna, Italy \label{inst:inaf_bologna}\\}
    }

   \date{Received 8 January 2026; accepted 27 March 2026}


  \abstract
   {Of the thousands of newly discovered open clusters (OCs) thanks to the exquisite precision of the {\it Gaia} mission data, only a small fraction has been observed with high-resolution spectroscopy. Particularly, the population of OCs in the inner disc at relatively high altitudes ($Z$) from the Galactic plane remains poorly studied. Few such high-|$Z$| inner-disc OCs have been detected, and most are sparse groupings of stars that still await confirmation as real OCs.}
   {We performed a detailed spectroscopic analysis of the high-|$Z$| inner-disc OC UBC~1052, an old cluster located at a cylindrical galactocentric radius $\rgc = 6.1$ kpc, where it is one of a few OCs situated at a considerable altitude ($Z = 340$ pc).}
   {We used FLAMES/VLT to acquire high signal-to-noise ratio UVES spectra of four red clump (RC) members ($G\sim14$ mag), from which we derived high-precision radial velocities and local thermodynamic equilibrium chemical abundances for 23 elements. A strict line-by-line differential analysis was carried out using a reference RC star and a solar analogue in the OC M~67, allowing us to derive very precise abundances for each star (a median precision in [X/H] of $\simeq0.06$ dex). We also acquired GIRAFFE spectra for other candidate member stars and derived their radial velocities.}
   {We determine that UBC~1052 has an age of $2.25\pm0.25$ Gyr, a distance of $3.11 \pm 0.07$ kpc, an extinction of $\AV =1.23$ mag, and a mean radial velocity of $\overline{\vr} =34.0 \pm 0.6$ $\kms$.
   We find that the four RC stars have fully compatible chemical abundances, thus confirming UBC~1052 as a real OC. It has [Fe/H] $= +0.05 \pm 0.01$ dex, and with [X/H] dispersions among the four stars $<0.03$ dex for 20 elements, we give conservative limits for chemical inhomogeneities at $\simeq0.05$ dex for these species.}
   {UBC~1052 stands out as the oldest and highest-|$Z$| inner-disc OC studied at high resolution to date, being located in the poorly sampled inner Galactic region where old OCs and OCs with large maximum excursions from the plane are scarce. Its relatively low [Fe/H] at its $\rgc$ suggests it is a rare candidate for an inward-migrated OC in the inner disc. Its detailed abundance pattern (e.g. [Ba/Zr] and [Nd/Y]) shows some interesting features that appear to be unique in the current census of OCs studied at high resolution, making it an interesting object for potential strong chemical-tagging searches for already dispersed member stars.}


   \maketitle
   \nolinenumbers
%

\section{Introduction}
\label{sec:introduction}

\begin{figure*}[h!]
{\includegraphics[width = 0.9905\textwidth]{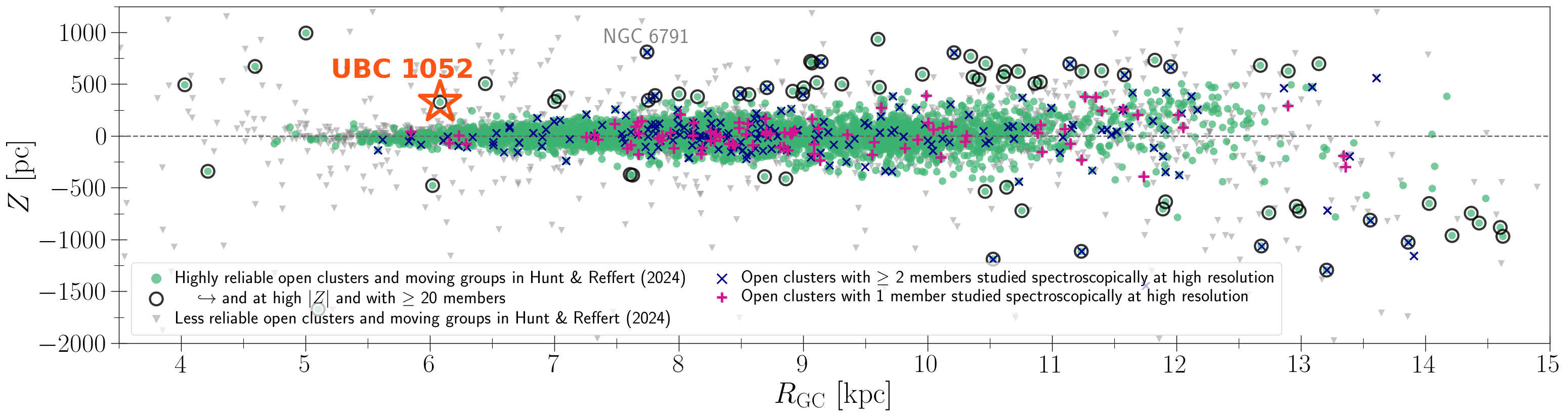}}
\caption{Height above the Galactic plane ($Z$) as a function of the galactocentric radius ($\rgc$) for the OCs and moving groups in \citet{hunt+2024}. The most reliable sample identified in \citetalias{hunt+2023} is represented as green circles, and circled in black are those with at least 20 members and located at an angle with respect to the plane measured from the Galactic centre larger than 2.5 degrees. UBC~1052 is marked by the red star. OCs studied spectroscopically at high resolution are shown as magenta pluses (only one studied member) and blue crosses (if at least one of the studies observed two members or more). We represent the OCs studied in the large spectroscopic surveys \gaiaeso \citep{Randich+2022}, GALAH \citep{Spina+2021}, and APOGEE DR19 OCCAM \citep{Otto+2025}; and in the dedicated studies at higher resolution OCCASO \citep{Carbajo+2024}, SPA \citep{Frasca+2019, Casali+2020, Zhang+2021SPA, DalPonte2025}, and OSTTA \citep{Carrera+2022}.}
\label{fig:rz_plane}
\end{figure*}

Open clusters (OCs) have long served as valuable indicators of the structure of the Milky Way disc \citep[e.g.][]{Trumpler1930, Janes1982, Cantat-Gaudin2020, Castro-Ginard2021} as well as tracers of its star formation and nucleosynthesis history \citep[e.g.][]{Janes1988, Friel1995, Hu2025}. The data releases delivered by ESA's {\it Gaia} mission \citep{GaiaCollaboration2016Mission} in recent years have significantly enhanced both the detection \citep[e.g.][]{Castro-Ginard2018, Castro-Ginard2019, Castro-Ginard2020, castroginard+2022, Ferreira2020, Ferreira2021, Casado2021, Qin2021, Qin2023} and characterisation of OCs \citep[e.g.][]{GaiaCollaboration2018Babusiaux, Bossini2019, Cantat-Gaudin2020, Dias+2021, Cavallo+2024} and revolutionised our understanding of young stellar aggregations (\citealt{Brown2021, Cantat-Gaudin2022, Cantat-Gaudin2024}).

One of the conundrums still under debate is related to the occurrence rate of high-altitude (high-|$Z$|) OCs, usually referred to as high-latitude OCs: clusters that are located far from the Galactic plane. Since most OCs are younger than 1 Gyr (\citealt{Wielen1971, Anders2021, hunt+2024}), it is hard to conceive that many of these objects may have moved far from the disc plane on short timescales (\citealt{Martinez-Medina2016} and references therein). However, until ten years ago, it was believed that many such high-|$Z$| OCs existed, and some of them even bear NGC designations, meaning that their discoveries date back to the 19th century or before. For example, more than a hundred new such candidates were reported in pre-{\it Gaia} data by the Milky Way Star Cluster survey \citep{Kharchenko2013, Schmeja2014, Scholz2015}. Thanks to the precision of the \Gaia DR2 data \citep{GaiaCollaboration2018Babusiaux}, \citet{Cantat-Gaudin2018} were able to show that all of these objects were spurious discoveries, some of them even coinciding with nearby dwarf galaxies. In a follow-up paper, \citet{Cantat-GaudinAnders2020} also debunked many of the old high-|$Z$| objects with NGC names as spurious detections that could be ruled out as OC candidates thanks to {\it Gaia}. Nevertheless, the enhanced astrometric precision of the following {\it Gaia} EDR3 data \citep{GaiaCollaboration2021} allowed for the detection of hundreds of new OC candidates (e.g. \citealt{castroginard+2022}, hereafter CG22), including some particularly intriguing ones at high altitudes (highlighted in their Fig.~8). Later, among the new OC candidates detected in \citet[][hereafter HR23]{hunt+2023} using {\it Gaia} DR3 data \citep{GaiaCollaboration2023Vallenari} there were also high-|$Z$| objects, although most of them do not belong to the highly reliable sample (see Fig.~\ref{fig:rz_plane}). So several intriguing questions about the high-altitude OCs remain: How many are there, why are they there, what is their dynamical origin, and why are they not disrupted yet?

The high-|$Z$| objects discovered in {\it Gaia} (E)DR3 data by \citetalias{castroginard+2022} and \citetalias{hunt+2023} are mostly older than 1 Gyr. Their survival is linked to processes that redistribute stellar populations in the Galaxy, such as radial migration and disc heating, making them key benchmarks for studying the mechanisms that drive the evolution of the Galactic disc \citep[e.g.][]{Lamers2005, Jilkova2012, Gustafsson+2016, Martinez-Medina+2017}. And given that most star formation occurs in the thin disc, studying these distant high-|$Z$| OCs through spectroscopic follow-up is crucial for understanding their nature and what their chemical abundances reveal about their origins. In this context, we conducted high-resolution spectroscopic observations of one such newly discovered cluster: UBC~1052, an old OC with a prominent red clump (RC) located at a cylindrical galactocentric radius of $\rgc \simeq 6$ kpc and approximately 340~pc above the Galactic plane (see Figs.~8 and 9 in \citetalias{castroginard+2022}).

High-resolution spectroscopic observations of OCs are necessary to precisely characterise the chemical abundance profile of the Galactic disc and its evolution with age \citep[e.g.][]{Spina+2022, Magrini+2023, Carbajo+2024, DalPonte2025, Otto+2025}.
The main advantage of OCs over field stars in this respect is that the mean chemical abundances can be determined much more precisely $(\sigma_{\textrm{[X/H]}}\propto1/\sqrt{N-1}$, with $N$ being the number of observed stars), assuming that the molecular cloud from which they were born was homogeneous and that no evolutionary processes affect the photospheric abundances observed now for stars in different evolutionary stages. Also, their ages and distances are much better constrained than for field stars. The scarcity of OCs towards the central Galaxy, however, makes it difficult to use OCs to constrain the chemical evolution of the inner Galactic disc. Also, only a few of the innermost OCs in the current census have been observed spectroscopically at high resolution to date, and they are located at low |$Z$|. In this inner-disc region, there are very few reliable detections of high-|$Z$| OCs, and most of the candidates are sparse groupings of stars that still await confirmation as real OCs. The study of UBC~1052 is a very valuable addition in this sense: As of this paper, it is the innermost high-altitude OC characterised spectroscopically at high resolution (see Fig.~\ref{fig:rz_plane}). In this paper, we carry out a strict line-by-line differential chemical abundance analysis. It produces the most precise results among the different traditional spectroscopic analysis methods (that is, non-data-driven approaches), but for now it has not been applied systematically at a large scale to analyse the high-resolution spectra collected in large spectroscopic surveys devoted to OCs. 

This paper is structured as follows: In Sect.~\ref{sec:data} we describe the new and archival spectroscopic data gathered to study the cluster, and the spectroscopic analysis is detailed in Sect.~\ref{sec:method}. We present the results in Sect.~\ref{sec:results} and discuss them in Sect.~\ref{sec:discussion}, focusing on UBC~1052's chemical composition and its potential origin in the context of the Galactic radial metallicity gradient. Our conclusions are presented in Sect.~\ref{sec:conclusions}.

\begin{figure*}
\includegraphics[width=.33\textwidth]{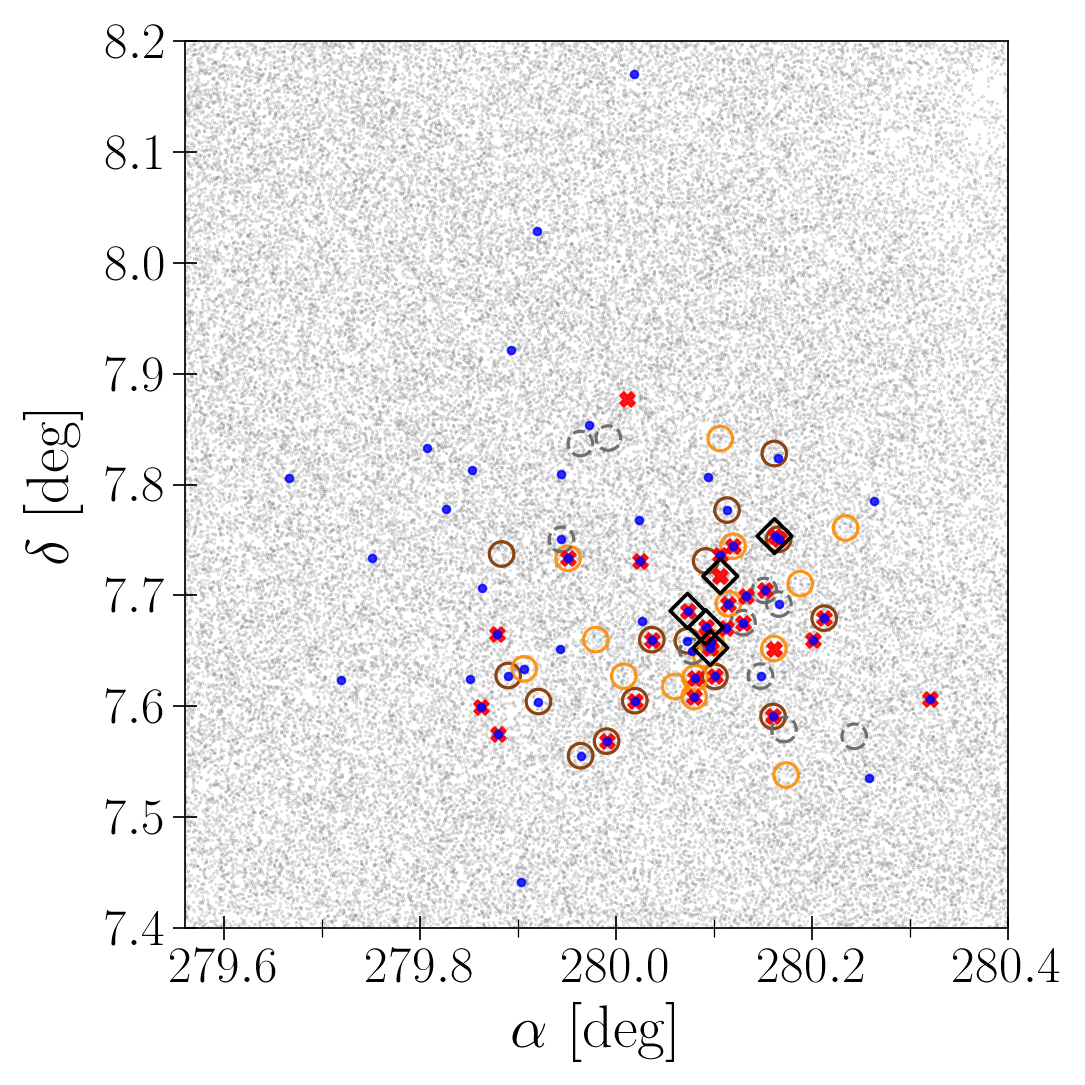} 
\includegraphics[width=.33\textwidth]{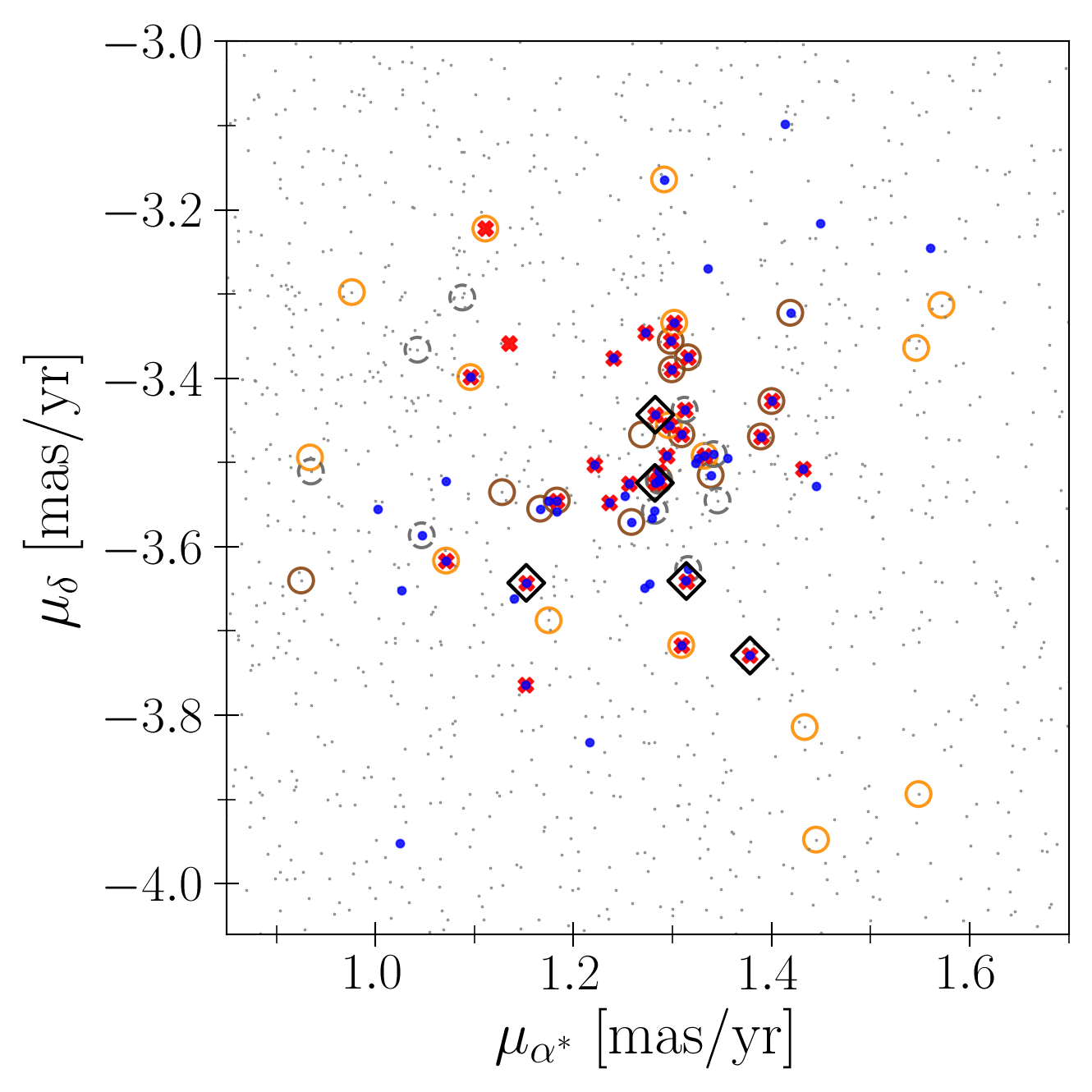} 
\includegraphics[width=.33\textwidth]{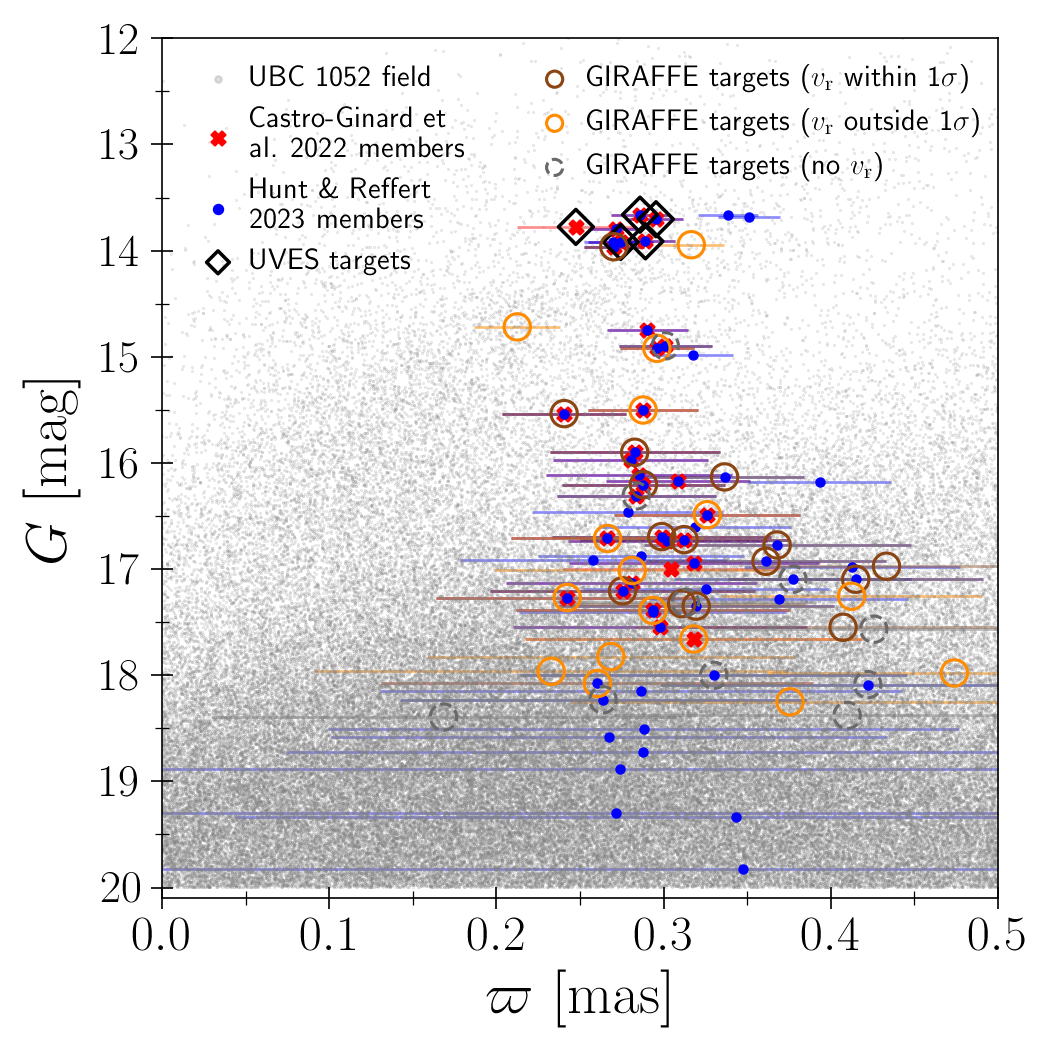} 
\caption{{\it Gaia} DR3 view of UBC~1052. \textit{Left}: On-sky projection of UBC~1052. \textit{Centre}: Proper motion distribution. \textit{Right}: $G$-band magnitude as a function of the parallax. UBC~1052 members identified in \citetalias{castroginard+2022} are represented as red crosses, and the ones identified in \citetalias{hunt+2023} are represented as blue circles. The targets observed using UVES are indicated as black open squares. Those observed using GIRAFFE are indicated as open circles, differentiating those for which their derived radial velocity (see Appendix~\ref{app_sec:giraffe}) is compatible with our estimated mean cluster $\vr$ within $1\sigma$ (brown circles) and those for which it is not (orange circles), and those for which the radial velocity cannot be derived (dashed grey circles). The cluster field stars in {\it Gaia} DR3 catalogue brighter than $G = 20$ mag are represented as grey dots.}
\label{fig:gaiadr3}
\end{figure*}

\section{Data} \label{sec:data}

\subsection{UBC~1052 membership list}

UBC~1052 was first identified by \citetalias{castroginard+2022} using the DBSCAN clustering algorithm \citep{Ester1996} on {\it Gaia} EDR3 data, which revealed 31 probable members down to a magnitude of $G = 18$ mag. Later, \citetalias{hunt+2023} confirmed the existence of the cluster with {\it Gaia} DR3 data using the HDBSCAN algorithm \citep{McInnes2017} and reaching fainter magnitudes. They classified UBC~1052 as a highly reliable cluster and found 59 members brighter than $G = 20$ mag, 28 of which overlap with the \citetalias{castroginard+2022} sample. However, the \citetalias{hunt+2023} membership list likely contains more contaminants, as suggested by the larger parallax dispersion (see the right panel of Fig.~\ref{fig:gaiadr3} and Table~\ref{table_params_ubc1052}). Our target selection relied on the initial catalogue of \citetalias{castroginard+2022}, which was the only one available when our observations were carried out.

\subsection{FLAMES/VLT observations and archival spectra} \label{subsec:data_VLT}

We obtained spectra for 46 potential members (Fig.~\ref{fig:gaiadr3}) using FLAMES \citep{Pasquini+2002}, the VLT’s multi-object spectrograph that simultaneously feeds two instruments: GIRAFFE and UVES. With UVES \citep{Dekker+2000}, we gathered high-resolution ($R \sim 47\,000$), high signal-to-noise ratio (S/N $\simeq 70 - 80$) spectra of five RC high-probability members (the largest number possible), which enabled a precise chemical abundance analysis (see Sect.~\ref{sec:method}). We used the Red Arm cross-disperser unit CD3 (central wavelength: 580 nm, wavelength range: 480-680 nm). For the remaining 41 targets, we used GIRAFFE fibres with three medium-resolution gratings: HR11 (559-584 nm, $R \sim 27\,000$), HR13 (611-640 nm, $R \sim 24\,000$), and HR14A (630-669 nm, $R \sim 17\,000$). The analysis of the GIRAFFE spectra is limited to deriving radial velocities ($\vr$) to assess their compatibility with UBC~1052’s mean $\vr$ estimated using UVES spectra (see Appendix~\ref{app_sec:giraffe}).

Although the vast majority of UBC~1052’s members are within FLAMES’s field of view (25 arcmin diameter), due to fibre positioning constraints some initially selected cluster members could not be observed. Ultimately, of the 46 observed stars, 31 were confirmed as UBC~1052 members (21 in \citetalias{castroginard+2022}, 29 in \citetalias{hunt+2023}, and 19 in both catalogues), while 15 served as filler targets with very low membership probabilities. The five RC stars observed with UVES are UBC~1052 members according to \citetalias{castroginard+2022}, and four of them are members according to \citetalias{hunt+2023}.

Observations were carried out in service mode in June and August 2023 during Period 111 (Programme ID: 111.255H.001), with seeing < $0.8''$. For each UVES target we acquired nine exposures of 950 s each. However, since a UVES fibre broke during the observations, for one of the RC targets we only obtained three exposures (and it had to be subsequently discarded in our analysis due to the low S/N; see Sect.~\ref{subsec:method_final_cluster_abunds}). This star is precisely the only UVES target not included in the UBC~1052 membership list of \citetalias{hunt+2023} (but it does appear in \citetalias{castroginard+2022}; see Fig.~\ref{fig:cmd}). For each GIRAFFE target three exposures were acquired, each of 2850 s and with a different grating.
The UVES spectra were reduced with the \texttt{EsoReflex} software \citep{Freudling+2013}, optimising the parameters of each pipeline workflow for the observed settings and S/N. The reduced GIRAFFE spectra were downloaded from the ESO science archive.

To derive solar-scaled differential abundances for the RC stars observed with UVES, we required three additional UVES spectra (see Sect.~\ref{sec:method}), which we retrieved from public archives:
\begin{enumerate}
\item {\it Gaia} DR3 604918144751101440, RC star in M~67: We used a S/N $\sim$ 130 UVES spectrum available in the ESO Archive\footnote{\url{http://archive.eso.org/wdb/wdb/adp/phase3_spectral/form}; Arcfile: ADP.2020-07-06T21:57:21.694.fits, Programme ID: 079.C-0131(A)}.
\item {\it Gaia} DR3 604914949295282816, solar-type star in M~67: We used the UVES spectra acquired by \citet{onehag+2011}\footnote{\url{https://archive.eso.org/wdb/wdb/eso/eso_archive_main/query?prog_id=082.D-0726}, Programme ID: 082.D-0726(A)}, obtaining a combined spectrum with S/N $\sim$ 110.
\item Sun: We used a S/N $\sim$ 380 UVES solar spectrum comprised in the {\it Gaia} FGK benchmark stars public spectral library\footnote{\url{https://www.blancocuaresma.com/s/benchmarkstars}} \citep{BlancocuaresmaLibrary+2014}.
\end{enumerate}

\section{Analysis of UVES spectra}
\label{sec:method}
In order to achieve high precisions in the determinations of the chemical abundances of the RC stars in UBC~1052 observed with UVES, we performed a strict line-by-line differential analysis. For each studied absorption line, it entails determining absolute abundances using the same methodology for a star under investigation and a reference star. This analysis cancels out to a certain degree the systematic uncertainties due to unaccounted blends and poor atomic line parameters, and it also helps to mitigate the possible departures from the assumptions of plane-parallel atmospheres (1D) and local thermodynamic equilibrium (LTE), yielding significantly reduced uncertainties (see e.g. \citealt{nissen+2018}, for a review). This technique relies on choosing a reference star that has atmospheric parameters (APs) as close as possible to those of the studied star. Thus, it is especially useful for the analysis of solar analogues with respect to the Sun since it directly yields the conventional solar-scaled abundances. Other works have applied this strategy to other stellar types, analysing them differentially with respect to a reference star with similar APs \citep[e.g.][]{casamiquela+2020}. This is what we did by analysing the RC stars in UBC~1052 differentially with respect to a RC member of the OC M~67. We then translated these differential abundances to the solar scale by performing an additional strict line-by-line differential analysis of a solar analogue in M~67 with respect to the Sun. Assuming that the RC star and the solar analogue in M~67 have the same abundance pattern for the studied elements, the solar-scaled abundances of UBC~1052 RC stars were obtained summing up their differential abundance with respect to M~67 to the differential abundance obtained for the solar analogue in M~67 with respect to the Sun (see Sect.~\ref{subsec:method_diff_abunds}). The same strategy to compute solar-scaled abundances was used in \citet{casamiquela+2021} to analyse RC stars in 47 OCs and in \citet{casamiquela+2022} for the OC UBC 274, using the Hyades as the auxiliary OC in both cases.

The assumption that the element abundances are the same for a RC star and a solar-type star in the same OC implies intra-cluster chemical homogeneity (at the level of the measurement uncertainties), and it only holds for elements whose abundances do not significantly change during stellar evolution. Physical processes that can cause differences between abundances in dwarf and giant stars are, for instance, atomic diffusion and mixing (e.g. \citealt{Semenova+2020}). However, differences in the measured abundances in dwarf and giant stars can also be due to spectral analysis effects, such as correlations between APs and abundances, non-LTE effects, or because unidentified line blends are stronger in a certain type of star (e.g. \citealt{Blanco-Cuaresma+2015}).
In the case of M~67, several studies have found evidence of atomic diffusion \citep{Onehag2014, BertelliMotta+2018,souto+2019}, observing a general decrease of [X/H] with increasing mass for main-sequence (MS) stars that reaches a minimum at the turn-off. Nevertheless, regarding the comparison of solar-type stars with RC stars, according to available atomic diffusion models there should not be any difference in their surface abundances for most of our studied elements \citep{souto+2019}. Still, on the experimental side, some studies have found variations between the measured abundances of solar-type and RC stars in M~67 for a few elements (\citealt{Gao+2018} and \citealt{souto+2019}). However, the uncertainties in the abundances are large, and there is also a large abundance scatter among G-type MS stars and, separately, among RC stars; additionally, some of the observed abundance differences are erased when performing non-LTE corrections. Thus, the results are not conclusive. In our analysis, we only found significant differences in the estimated absolute abundances of the RC star in M~67 and the solar analogue in M~67 for Na. Therefore, for all the possible chemical species except for Na we computed solar-scaled abundances through the aforementioned methodology, which involves two different differential analyses and the use of stars in an auxiliary OC.

We chose M~67 as the auxiliary OC because (i) it has publicly available high-S/N UVES spectra for a large number of members, including RC and solar-type stars, (ii) the membership lists for M~67 are very reliable (enabling us to ascertain the membership of the two stars used with very high confidence), and (iii) its age is closer to that of UBC~1052 (4.5 Gyr vs. 2.25 Gyr) than any other well-studied OC. Since the RC position slightly moves in the Kiel diagram as age and metallicity change, the fact that UBC~1052 and M~67 have similar ages reduces the differences in the APs between the two clusters' RC stars (see Sect.~\ref{subsec:atmospheric_params}). The closer the studied and reference stars are in terms of APs, the more the differential abundance analysis mitigates systematic offsets, which are highly dependent on these parameters.

The RC star in M~67 used as reference ({\it Gaia} DR3 604918144751101440, $G=10.24$ mag) was chosen among the RC members in M~67 with high-S/N public UVES spectra that had the most similar APs to those of the studied UBC~1052 RC stars. It is a high-probability member of M~67 according to both \cite{Cantat-Gaudin2020} and \citetalias{hunt+2023}. As the solar-type star in M~67 studied differentially with respect to the Sun we chose the well-known solar twin M67-1194 ({\it Gaia} DR3 604914949295282816, $G= 14.45$ mag), a high-confidence member extensively studied in the literature \citep[e.g.][]{onehag+2011, Liu+2016}.

The spectroscopic analysis of the UVES spectra (both from our own observations and from the archive) was carried out using {\tt iSpec} \citep{blancocuaresma+2014,blancocuaresma2019}. We made use of a similar version of the pipeline we have been using in our latest works \citep[e.g. see Sect.~3 in][for further details]{casamiquela+2020}, which is an adaptation of the pipeline used in \citet{blancocuaresma+2018}. In the following subsections we summarise the steps of the analysis.

\subsection{Determination of radial velocities}
\label{subsec:method_RVs}
We first performed the barycentric correction for each UVES spectrum. 
Then, if more than one spectrum were available for a star, they were combined to increase the S/N. The radial velocity in the solar barycentric reference frame was determined through cross-correlation with the high-S/N NARVAL solar spectrum from the {\it Gaia} FGK benchmark stars library\footnote{See footnote 3.} and then the combined spectrum was shifted to the rest frame. Strong telluric absorption lines were masked out in the subsequent analysis. 

\subsection{Determination of atmospheric parameters}
\label{subsec:method_APs}
We used spectral synthesis to compute the effective temperature ($\teff$), surface gravity ($\logg$), overall metallicity ($\mh$), [$\alpha$/M], and microturbulence ($\vmic$).
To generate synthetic spectra we used the radiative transfer code SPECTRUM \citep{gray+1994} and the one-dimensional stratification MARCS atmospheric models\footnote{\url{http://marcs.astro.uu.se/}} \citep{gustafsson+2008}, both of which assume LTE. The data used for the atomic absorption lines were version six of the \gaiaeso line list \citep{heiter+2021}.

The continuum normalisation and determination of the APs were performed in an iterative way. First, the continuum of the target spectrum was fit with quadratic splines, applying a median and maximum filter in order to exclude absorption lines. The target spectrum was normalised by dividing it over this fitted continuum, and then the APs were determined by comparing the uncertainty-weighted fluxes of a set of observed features in the spectrum with a synthetic spectrum, varying the APs of the latter until a least-squares algorithm converged. The regions for which this fitting was performed are absorption lines for which a solar abundance within $\pm 0.05$ dex with respect to the solar abundance of reference was obtained when analysing a solar spectrum with {\tt iSpec} (as explained in \citealt{blancocuaresma2019}, Sect.~3.5). The exact region around each absorption line where the fit was performed was adjusted automatically to cover well the line profile. In addition, the wings of H$\alpha$, H$\beta$, and \ion{Mg}{I} b triplet lines were included to better constrain $\teff$ and $\logg$. In a second step, a synthetic template with the previously determined APs was used to re-normalise the target spectrum, for which the final APs were determined using the same least-squares algorithm as before.
The projected equatorial rotational velocity $\varv~{\rm sin}i$, the macroturbulence parameter, and the spectral resolution are degenerate and difficult to disentangle for giant stars. We applied the strategy described in \citet{blancocuaresma2019}: we used a fixed value for $\varv~{\rm sin}i = 1.6$ km ${\rm s}^{-1}$, computed the macroturbulence using the empirical relation for giants used in the \gaiaeso Survey (built-in function implemented in {\tt iSpec}), and only left the spectral resolution free, thus accounting for all broadening effects.

\subsection{Differential chemical abundances}
\label{subsec:method_diff_abunds}
\begin{figure}
{\includegraphics[width = 0.495\textwidth]{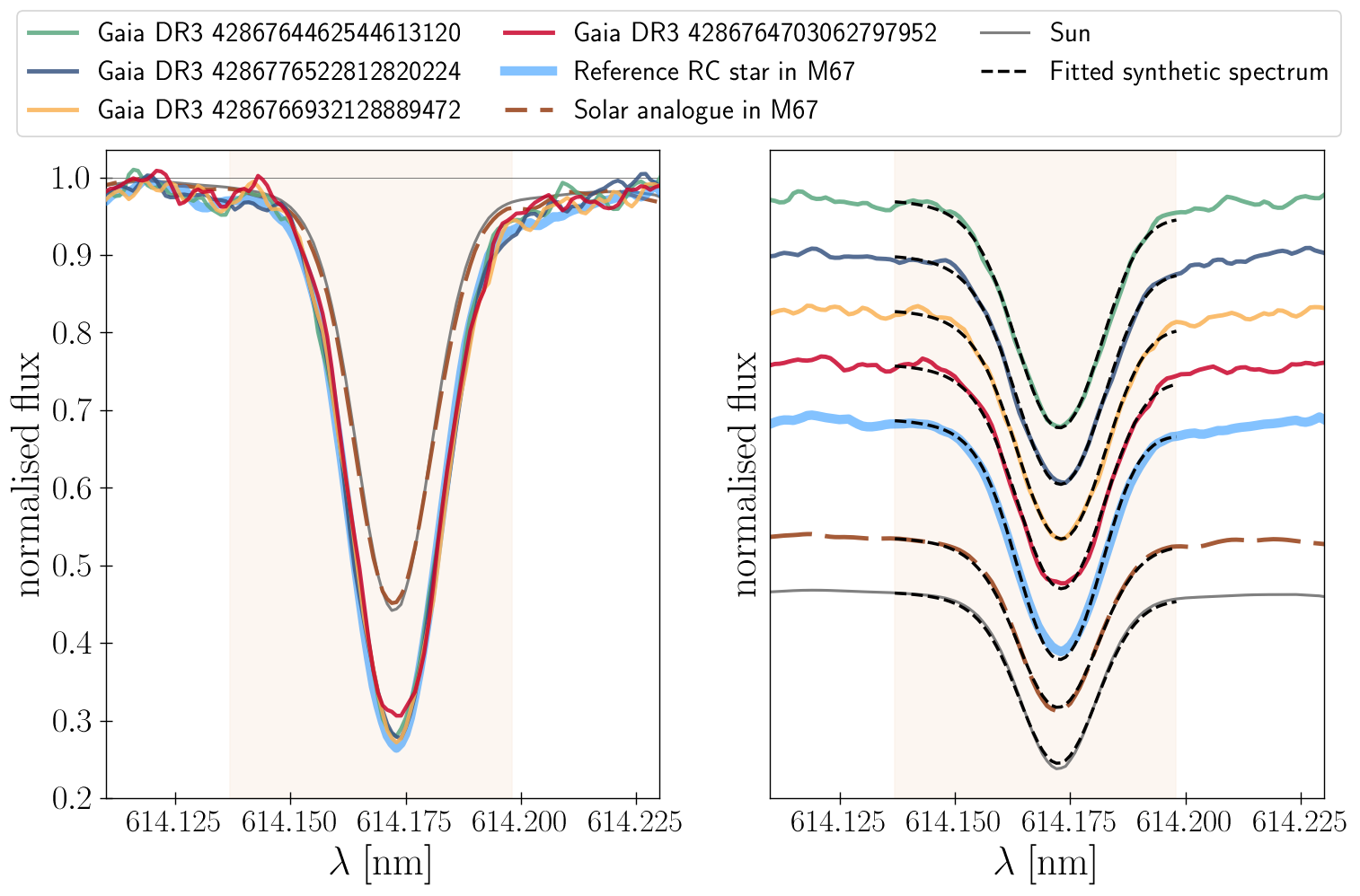}}
\caption{Spectra in the region of the \ion{Ba}{II} line at $\lambda = 6141.713$ \text{\AA} for the four high-S/N stars in UBC~1052, the RC star in M~67 used as reference for their differential analysis, the solar analogue in M~67, and its reference (the Sun). \textit{Left}: Superposition of all normalised spectra. \textit{Right}: Same but now offsetting the spectra for better visibility and showing the synthetic spectra fit to each of them within the shaded region as dashed black lines.}
\label{fig:spectra_oneline}
\end{figure}

In a first step, for each spectrum we computed the absolute chemical abundance for each of the absorption lines\footnote{The absolute abundance of a chemical species X derived from one of its absorption lines $i$ is defined as $A_{{\rm X}_i}=\log_{10}( N_{{\rm X}_i}/N_{\rm H}) + 12$, 
where $N_{{\rm X}_i}$ and $N_{\rm H}$ are the number densities of absorbers of the species X (derived from the line $i$) and of hydrogen, respectively.}.
Individual absolute chemical abundances were obtained using the APs fixed to the values obtained previously, and using the same radiative transfer code (SPECTRUM), model atmospheres (MARCS), atomic line list, and spectral synthesis fitting method; varying the abundance of the studied element for the synthetic spectrum until finding the best fit to the line of the observed spectrum. As an example, in the right panel of Fig.~\ref{fig:spectra_oneline} we show the fits to the \ion{Ba}{II} absorption line at 6141.7 \text{\AA} for the four high-S/N RC stars in UBC~1052, the RC and solar-type stars in M~67, and the Sun. Neither non-LTE nor 3D corrections were taken into account to compute the abundances.

In a second step, we computed line-by-line differential abundances. 
As explained above, for the differential abundance analysis of the RC stars in UBC~1052 the reference star is a RC star in M~67; and for the solar analogue in M~67 the reference star used is the Sun. For each of the lines in common between a studied star and its corresponding reference star, the differential abundance was calculated as $A_{{\rm X}_i, ~\rm star} - A_{{\rm X}_i, ~\rm ref}$.

In order to compute for each star its mean differential abundance of a chemical species X, we first carried out a selection of the suitable lines among all the ones for which we were able to measure $A_{{\rm X}_i, ~\rm star} - A_{{\rm X}_i, ~\rm ref}$. The details are explained in Appendix~\ref{app_subsec:method_line_selection}, and in Table \ref{table_selected_lines} we provide the final sample of selected lines for the five RC stars in UBC~1052 and for the solar analogue M67-1194, and their respective estimated absolute abundances.
Then, the mean differential abundance of a chemical species X for each star ([X/H]$_{\textrm {star wrt ref}}$) was calculated as the mean of the absolute abundance difference with respect to the reference star for each selected line:
\begin{equation}
    \textrm{[X/H]}_{\textrm{star wrt ref}} = \frac{1}{N_{\textrm{lines X}}} \sum_{i=1}^{N_{\textrm{lines X}}} (A_{{\rm X}_i, ~\rm star} - A_{{\rm X}_i, ~\rm ref}). 
\label{eq:mean_star_XoverH_diff_analysis}
\end{equation}

Its uncertainty was computed as described in Appendix~\ref{app_subsec:method_uncertainties_abunds}, taking into account not only the line-to-line scatter of the differential abundances, but also the scatter arising from the uncertainties in the APs and the spectrum’s flux.
For the elements for which we derived abundances for two ionisation states, we chose to estimate the abundances for that element using the ion that had more selected lines and, consequently, smaller uncertainties in [X/H]$_{\textrm {star wrt ref}}$. Namely, from now on we use \ion{Fe}{I}, \ion{Sc}{II}, \ion{Ti}{I}, and \ion{V}{I} as representatives of these elements’ abundances.

Finally, for each UBC~1052 RC star, we translated the mean abundance of each ion to the solar scale by adding up the abundances of the two differential analyses: ${\textrm{[X/H]}_{\textrm{star}}}= {\textrm{[X/H]}}_{\textrm {star wrt RC star in M~67}} + {\textrm{[X/H]}}_{\textrm {solar twin in M~67 wrt Sun}}$. Its reported uncertainty corresponds to the square root of the quadratic sum of the uncertainties in each summand (each computed as described in Appendix~\ref{app_subsec:method_uncertainties_abunds}):
$\sigma_{\textrm{[X/H]}_{\textrm{star}}} = \sqrt{ (\delta{\textrm{[X/H]}}_{\textrm {star wrt RC star in M~67}})^2 + (\delta {\textrm{[X/H]}}_{\textrm {solar twin in M~67 wrt Sun}})^2}$

We derived solar-scaled abundances through this differential method for 22 ions (all studied elements except for Na). For Na, for each UBC~1052 RC star we computed its mean absolute abundance (averaging the $A_{{\rm Na}_i}$ of its selected lines) and referred it to the Sun by directly subtracting from it the Sun’s mean absolute Na abundance estimated from our solar spectrum. For each RC star in UBC~1052 and the Sun, the uncertainties in the mean Na absolute abundance were estimated analogously to the differential abundances (as described in Appendix~\ref{app_subsec:method_uncertainties_abunds} but with $A_{{\rm Na}_i, ~\rm star}$ as variable instead of $A_{{\rm Na}_i, ~\rm star} - A_{{\rm Na}_i, ~\rm ref}$), and the uncertainty in the final solar-scaled Na abundance was also computed as the square root of the sum in quadrature of these uncertainties.

\subsection{Final cluster abundances}
\label{subsec:method_final_cluster_abunds}
The mean cluster abundances were computed based on the abundances derived for the four RC stars with high-S/N spectra (combined S/N $\simeq$ 70-80). We excluded {\it Gaia} DR3 4286778137720526080, the RC star with the poorest combined S/N ($\sim$ 35) for which only three out of the nine planned exposures could be acquired (see Sect.~\ref{subsec:data_VLT}), because its derived abundances are much more uncertain and less accurate than for the other four stars (see Sects.~\ref{subsec:atmospheric_params} and \ref{subsec:abundances_chemical_homogeneity}). For each ion, the mean cluster abundance with respect to the Sun was computed as a weighted mean of its mean solar-scaled abundance for each of the stars. The uncertainty was computed as the sample standard deviation, calculated taking into account the correction for small sample sizes as explained in \citealt{roesslein+2007} (their Eq. 5).

\begin{figure}[h!]
\includegraphics[width=.464\textwidth]{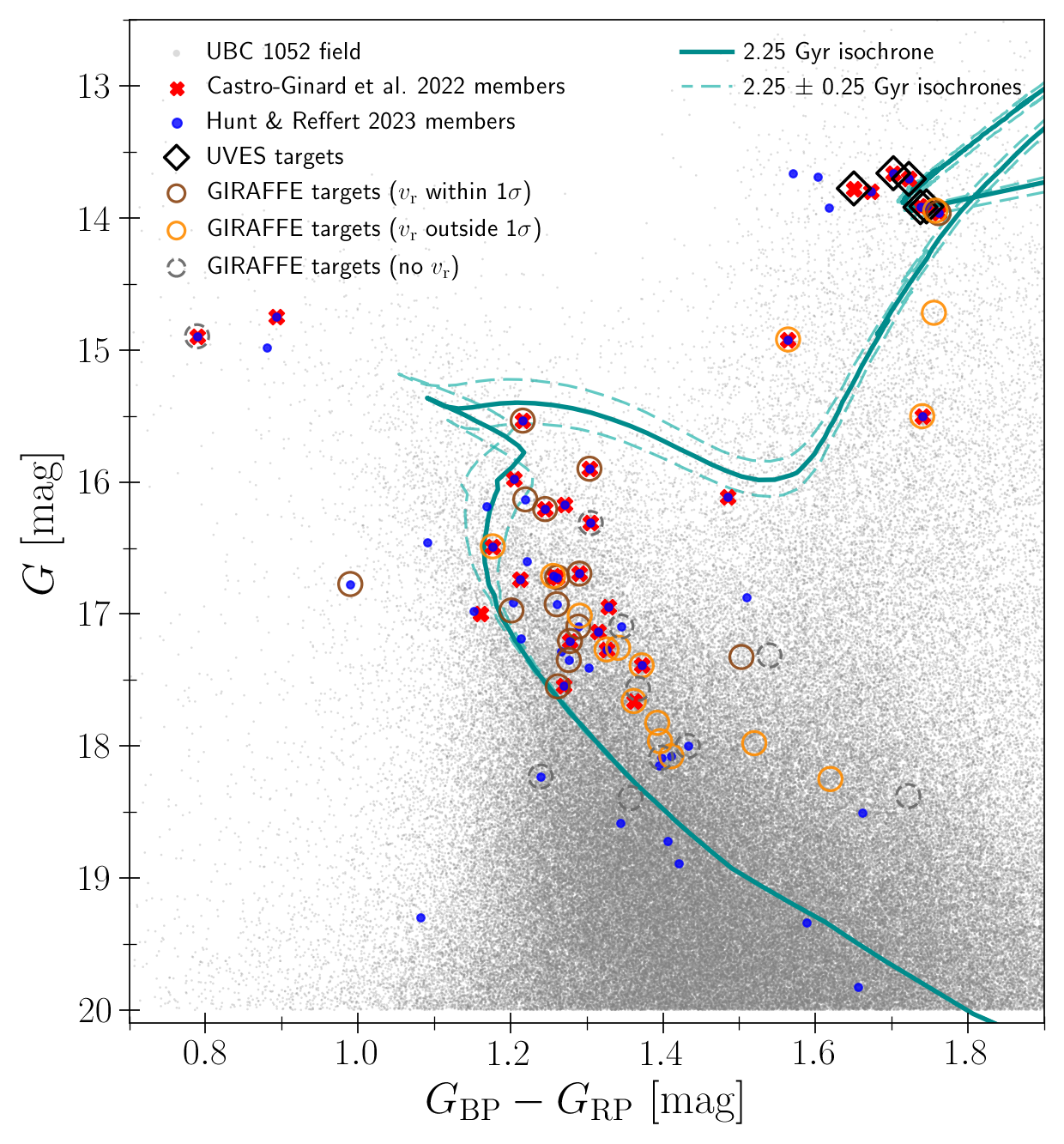} 
\caption{Observed {\it Gaia} DR3 colour-magnitude diagram of UBC~1052. The cluster members and observed stars are colour-coded as in Fig.~\ref{fig:gaiadr3}, and the field stars brighter than $G = 20$ mag are also represented (grey dots). The cyan solid line is a PARSEC isochrone of age 2.25 Gyr and metallicity [M/H] $=+0.05$ dex, shifted by $d=3.11$ kpc and $\AV=1.23$ mag. The dashed lines are the isochrones obtained by changing the age of the solid isochrone to 2 Gyr and 2.5 Gyr.}
\label{fig:cmd}
\end{figure}

\begin{table*}[h!]
\caption{UBC~1052 mean parallax, distance, extinction ($A_V$), age, and mean radial velocity ($\overline{\vr}$) in the literature and in this study. \label{table_params_ubc1052}}
\centering
\begin{tabular}{lcccc}
\hline \hline
 & \citet{castroginard+2022} & \citet{hunt+2023} & \citet{Cavallo+2024} & Donada et al. (this study) \\ 
\hline
Mean parallax  [mas] & 0.29 $\pm$ 0.02 & 0.30 $\pm$ 0.04 & & \\
Distance [kpc] & 3.8 $\pm$ 0.3 & 2.99 [2.95, 3.04] & 2.9 [2.6, 3.3] & 3.11 $\pm$ 0.07 \\
$A_V$ [mag] & 1.30 $\pm$ 0.15 & 1.6 [1.4, 1.9] & 1.4 [1.3, 1.6] & 1.23 $\pm$ 0.03 \\
Age [Gyr] & 1.6 [1.1, 2.2] & 0.9 [0.7, 1.4] & 1.9 [1.5, 2.5] & $2.25 \pm 0.25$ \\
$\overline{\vr}$ [$\kms$] & 35.2 $\pm$ 1.9  \tablefootmark{a} & 36.7 $\pm$ 3.3  \tablefootmark{b} &  & 34.0 $\pm$ 0.6 \\
\hline
\end{tabular}
\tablefoot{We report the parameters in \citetalias{castroginard+2022}, \citetalias{hunt+2023}, and \citealt{Cavallo+2024} (who selected $G \leq 18$ mag members in \citetalias{hunt+2023}), as well as the ones adopted in this study (last column, see Sects.~\ref{subsec:cluster_parameters} and \ref{subsec:vr_orbital_parameters}). The uncertainties in distance, $\AV$, and age for \citetalias{castroginard+2022} are the mean of the typical uncertainty range reported in \citet{Cantat-Gaudin2020}. \tablefoottext{a}{Computed using the {\em Gaia} DR3 $\vr$ values available for seven of the members in \citetalias{castroginard+2022}.} \tablefoottext{b}{Computed using the {\em Gaia} DR3 $\vr$ values of eight members, disregarding an outlier ({\em Gaia} DR3 4286776351014151552, $\vr = -34\pm3$ $\kms$).}}
\end{table*}

\section{Results}\label{sec:results}

\subsection{Cluster distance, extinction, and age}
\label{subsec:cluster_parameters}

We estimated the cluster distance ($d$) using the members in \citetalias{castroginard+2022}. We corrected their parallaxes for the {\it Gaia} EDR3 parallax zero point offsets \citep{Lindegren+2021} and then performed a Bayesian distance estimation assuming negligible parallax correlations and cluster diameter ($r_{\textrm{cluster}}\ll d$). The result is virtually the same using the exponentially decreasing volume density prior in \citealt{BailerJones2015} (${3.12}^{+0.07}_{-0.06}$ kpc) and the uninformative prior in \citealt{Weiler2025} (${3.11}^{+0.07}_{-0.06}$ kpc) using the members from \citetalias{castroginard+2022}. The membership catalogue, on the other hand, has a larger effect on the derived distance, which would be $2.99 \pm 0.05$ kpc with both priors for the members in \citetalias{hunt+2023}. The distance of $d=3.11 \pm 0.07$ kpc corresponds to $\rgc = 6.07 \pm 0.04$ kpc and $Z = 337 \pm 7$ pc (adopting the solar values: distance to the Galactic Centre $R_{\odot, \mathrm{GC}}$ = 8.178 kpc, \citealt{GravityCol2019}; and height above the Galactic plane $Z_{\odot} = 21$ pc, \citealt{BennettBovy2019}). 
We derived the age of the cluster through visual fit of {\tt PARSEC v2.0} isochrones\footnote{\url{https://stev.oapd.inaf.it/cgi-bin/cmd}} \citep{Nguyen+2022} in the observed colour–magnitude diagram (CMD, Fig.~\ref{fig:cmd}). We used isochrones with metallicity $\mh =+0.05$ dex (our derived [Fe/H] abundance, see Table~\ref{table_mean_cluster_abunds}) and initial rotation rate $\omega_\mathrm{i} = 0.3$; and assumed a cluster distance $d = 3.11$ kpc and line-of-sight interstellar extinction $\AV =$ 1.23 mag, computed as the mean extinction in {\tt StarHorse} catalogue \citep{Anders+2022} for the members in \citetalias{castroginard+2022}. We obtain an approximate age of $2.25 \pm 0.25$ Gyr (Fig.~\ref{fig:cmd}, solid curve for the best-fit isochrone and dashed curves for the isochrones defining the uncertainty range). No significant differences are found in the isochrone fit for $\omega_\mathrm{i}$ values in the range $\sim$[0, 0.7].
To test the effect of differential extinction on the isochrone fit, we used the 10 pc-resolution three-dimensional extinction map by \citet{Vergely+2022} to compute individual $\AV$ for the members in \citetalias{castroginard+2022}. They range between [1.25, 1.61] mag, the median is 1.40 mag, and the 0.25 and 0.75 quantiles are 1.36 mag and 1.42 mag, respectively. We represented the intrinsic CMD of the members in \citetalias{castroginard+2022}, correcting the magnitudes of each star for its estimated $\AV$ in \citet{Vergely+2022} and considering a distance $d = 3.11$ kpc to compute the absolute $G$ magnitude. We found that it does not show any relevant difference compared with the observed CMD. If we assume the median $\AV$ in \citealt{Vergely+2022} (1.40 mag), the best-fit isochrone in the observed CMD has an age of $2.0 \pm 0.25$ Gyr. This age is compatible with the one obtained assuming the smaller $\AV$ in {\tt StarHorse}, but the fit in the RC region and MS ridge location is poorer. Therefore, in this study we adopt $\AV = 1.23$ mag and $2.25 \pm 0.25$ Gyr as the approximate age of UBC~1052.

The parameters we derive for UBC~1052 are reported in Table~\ref{table_params_ubc1052}, where we also include the mean cluster $\vr$ (see Sect.~\ref{subsec:vr_orbital_parameters}). For comparison, we also include the parameters in \citetalias{castroginard+2022}, \citetalias{hunt+2023}, and \citet{Cavallo+2024}. These studies use different neural networks to estimate distance, $\AV$, and age for large samples of OCs. Such automatic methods for the estimation of the cluster’s parameters are known to perform poorly in some instances (e.g. \citealt{Carbajo+2025}). The much younger age found in \citetalias{hunt+2023}, for example, may be caused by a poor isochrone fit due to the presence of candidate blue straggler members (see Fig.~\ref{fig:cmd}). Our derived parameters for UBC~1052 are within the uncertainties of the values reported in the three studies except for the distance in \citetalias{castroginard+2022} and the age in \citetalias{hunt+2023}.

\subsection{Radial velocities and orbital parameters}
\label{subsec:vr_orbital_parameters}
The radial velocities of the five UBC~1052 RC stars are reported in Table~\ref{table_vr_APs_mean_star_abunds}. For the four stars with high-S/N spectra, they are $1\sigma$ compatible with those of {\it Gaia} (which has much larger uncertainties than our estimates). For the star with the lowest S/N ({\it Gaia} DR3 4286778137720526080), on the other hand, our derived $\vr = 40.4 \pm 0.3$ $\kms $ is compatible with {\it Gaia}'s $\vr = 35 \pm 3$ $\kms $ within $2\sigma$. In light of its high \texttt{ruwe} = 1.96, this star is a potential unresolved binary. Considering this and the fact that its $\vr$ is much higher than for the other four stars, we do not take it into account to compute the mean radial velocity of the cluster. Instead, this value is computed as the weighted average over the radial velocities of the four RC stars with high-S/N spectra, obtaining $\overline{\vr} = 34.0 \pm 0.6$ $\kms$ (reported in Table~\ref{table_params_ubc1052}), where the associated error is the sample standard deviation. It is compatible with the mean $\vr$ over the available {\em Gaia} DR3 RVS radial velocities of the members in \citetalias{castroginard+2022} and \citetalias{hunt+2023} (disregarding a clear outlier for the latter, see Table~\ref{table_params_ubc1052}). The $\vr$ for the star with the lowest S/N is $>3\sigma$ discrepant with respect to the cluster’s $\overline{\vr}$.

Regarding the cluster's orbit, we analytically computed the current orbital parameters using its mean coordinates and proper motions in \citetalias{castroginard+2022}, and the distance and $\overline{\vr}$ from our study. The calculation was performed using {\tt galpy} \citep{Bovy2015}, assuming the axisymmetric Galactic potential {\tt MWpotential2014} and via the Staeckel approximation. To transform the input parameters into galactocentric coordinates and velocities, we adopted the following solar values: $R_{\odot, \mathrm{GC}}$ = 8.178 kpc \citep{GravityCol2019}, $Z_{\odot} = 21$ pc \citep{BennettBovy2019}, and velocity in the galactocentric frame $(\varv_{\odot,\mathrm{x}}, \varv_{\odot,\mathrm{y}}, \varv_{\odot,\mathrm{z}}) = (11.1, 241.24, 7.25)$ $\kms$ \citep{Schonrich+2010, Eilers+2019}. The observational uncertainties were taken into account by sampling
10\,000 orbit instances from Gaussian distributions centred on UBC~1052’s mean $\alpha$, $\delta$, $\mu_{{\alpha}^*}$, $\mu_{\delta}$ (from \citetalias{castroginard+2022}), distance and $\overline{\vr}$ (from our study), and with standard deviation equal to the corresponding parameter's uncertainty. We obtain the following orbital parameters:
guiding-centre radius $\rguide = 6.18^{+0.06}_{-0.07}$ kpc, pericentre $\rperi = 5.84^{+0.06}_{-0.05}$ kpc, apocentre $\rapo = 6.75^{+0.09}_{-0.08}$ kpc, eccentricity $e= 0.072\pm0.006$, and maximum excursion from the Galactic plane $\zmax = 0.54\pm0.02$ kpc.

To test the dynamical influence of the Galactic bar on the orbit of UBC~1052, we performed several orbit integration tests using {\tt galpy} and GravPot16\footnote{\url{https://gravpot.utinam.cnrs.fr/}} \citep{Fernandez-Trincado2017}. With {\tt galpy} we added the Dehnen’s bar potential in 3D ({\tt DehnenBarPotential}, \citealt{Dehnen2000}; \citealt{Monari+2016}) to {\tt MWPotential2014}, and considered two different bar models: a weak bar with a pattern speed $\omegabar = 40$ $\kmskpc$, bar radius $\rbar = 5.1$ kpc and bar strength $\Af = 621$ $\mathrm{km}^2 \ {\mathrm{s}}^{-2}$ \citep{Wegg+2015}, and a strong Dehnen bar with $\omegabar = 52$ $\kmskpc$, $\rbar = 3.4$ kpc and $\Af = 2101$ $\mathrm{km}^2 \ {\mathrm{s}}^{-2}$  \citep{Monari+2016}. The potential in GravPot16 includes a rotating prolate ‘boxy/peanut’ bar, for which we adopted a total bar mass of $1.2 \cdot 10^{10} \ {\mathrm M}_{\odot}$ (e.g. \citealt{Portail+2017}) and tested different $\omegabar$ between 30 and 50 $\kmskpc$, which cover the range of values found in the literature \citep[e.g.][]{Dehnen2000, Bovy+2019, ClarkeGerhard2022}.  For both {\tt galpy} and GravPot16, we tested values of the bar’s present-day orientation $\phibar$ between 20$^{\circ}$ and 30$^{\circ}$ \citep[e.g.][]{Bland-HawthornGerhard2016, Bovy+2019, GaiaCollaboration2023Drimmel}. Integrating the orbit of UBC~1052 back in time to $t = -2.25$ Gyr in each test, in most of the cases we find $\rperi \in [5.6, 6.0]$ kpc, $\rapo \in [6.3, 7.8]$ kpc, $\zmax \in [0.53, 0.58]$ kpc, and $e \in [0.03, 0.13]$. These orbital parameters are close to the ones obtained for {\tt MWPotential2014} alone, indicating that our results are largely independent of the inclusion of the bar in our orbit modelling. The remaining few cases are bar configurations for which UBC~1052’s orbit in the non-inertial reference frame (where the bar is at rest) has an unusual ‘banana’ shape, and they are very sensitive to $\omegabar$. For instance, with GravPot16 and $\phibar = 20^{\circ}$, for $\omegabar =40$ $\kmskpc$ UBC~1052 displays this kind of orbit (with similar parameters $\rperi =5.1$ kpc, $\rapo= 7.0$ kpc, $\zmax= 0.57$ kpc, and $e = 0.03$), while for $\omegabar =41$ $\kmskpc$ it does not. In conclusion, UBC~1052 has likely suffered only little dynamical heating (‘blurring’ in the terminology of \citealt{Sellwood2002}; see discussion in Sect. \ref{sec:migration}).

\subsection{Atmospheric parameters}\label{subsec:atmospheric_params}

The APs of the five UBC~1052 RC stars are reported in Table~\ref{table_vr_APs_mean_star_abunds}. In Fig.~\ref{fig:kiel_diagram} we represent the Kiel diagram with our inferred $\teff$ and $\logg$. The locations of all five stars are in agreement with the overplotted theoretical isochrone of age 2.25 Gyr and $\mh = +0.05$ dex. However, the APs of the star with the lowest S/N are not compatible with those of the four stars with high-S/N spectra. The mean $\teff$ and $\logg$ over these four members are $4701 \pm 18$ K and $2.49 \pm 0.02$ dex, respectively, where the uncertainties are the standard deviations of the values for the four stars. The RC star in M~67 used as reference for the differential abundance analysis ({\it Gaia} DR3 604918144751101440) has $\teff = 4719 \pm 22$ K and $\logg = 2.48 \pm 0.04$ dex. Both parameters are within the range covered by the four high-S/N RC stars in UBC~1052. In contrast, the difference between the APs of the star with the lowest S/N and those of the RC reference star in M~67 is much larger. Hence, the differential abundances derived for this star are less accurate than those of the other four stars.

\begin{figure*}[h!]
\centering
{\includegraphics[width = 0.975\textwidth]{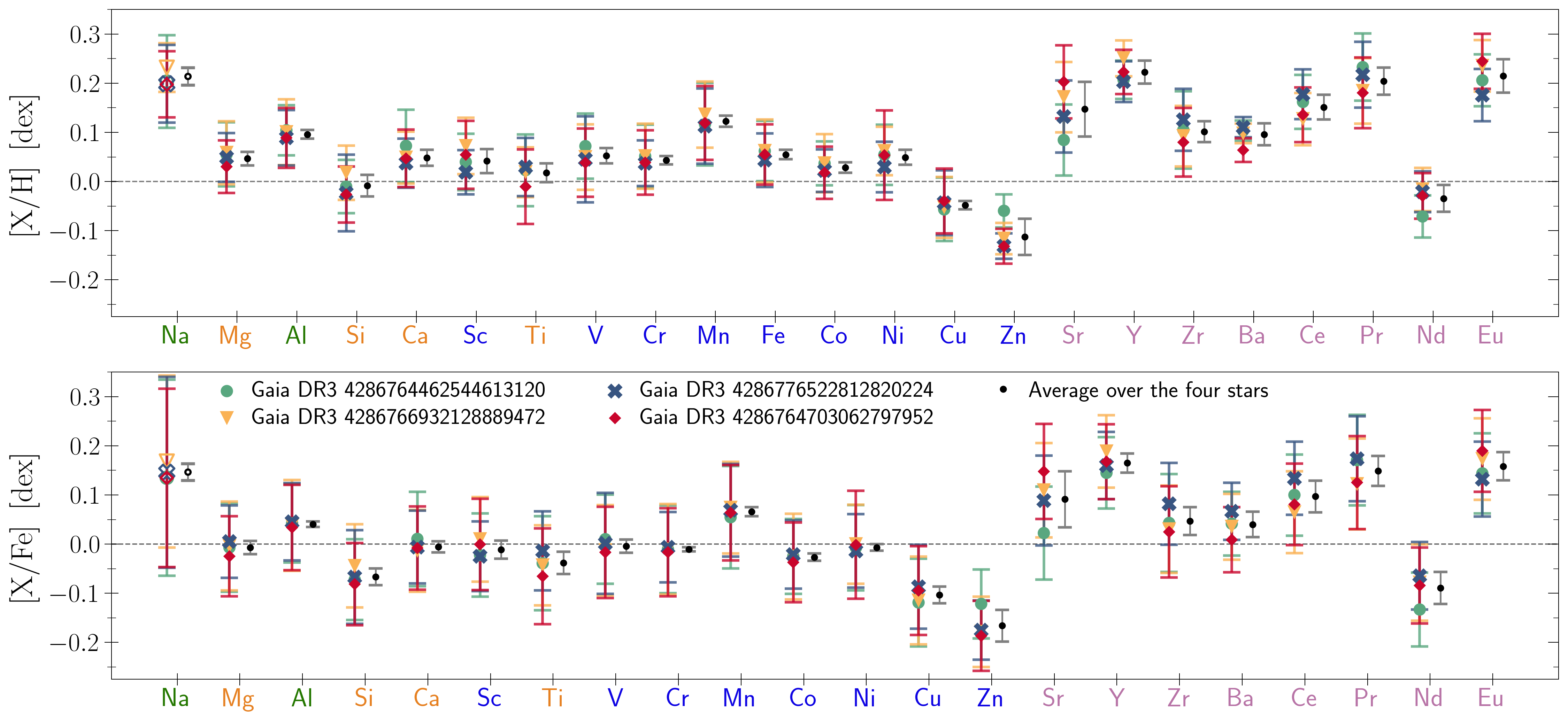}}
\caption{Chemical abundances (\textit{top}: [X/H]; \textit{bottom}: [X/Fe]) of the four RC stars in UBC~1052 with high-S/N UVES spectra (see Table~\ref{table_vr_APs_mean_star_abunds}) and mean cluster abundances (small black circles, the grey error bars are the corrected sample standard deviation over the four stars; see Table~\ref{table_mean_cluster_abunds}). Solid coloured shapes indicate elements for which abundances have been derived through differential analyses. Empty shapes (Na) indicate that the abundance has been derived from the absolute abundances. The colour of each element corresponds to its respective dominant nucleosynthesis channel (blue: iron-peak, yellow: $\alpha$, green: odd-Z, and violet: neutron-capture elements).}
\label{fig:individual_stars_abunds}
\end{figure*}

\subsection{Abundances and chemical homogeneity}
\label{subsec:abundances_chemical_homogeneity}

The solar-scaled abundances of the five RC stars in UBC~1052 observed with UVES, computed as described in Sect.~\ref{subsec:method_diff_abunds}, are reported in Table~\ref{table_vr_APs_mean_star_abunds}.

For the star with the lowest S/N ({\it Gaia} DR3 4286778137720526080), abundances can be derived for less elements (19) than for the four stars with high-S/N spectra (23). And for most of the elements in common, this star has fewer studied lines. Its uncertainties in the mean stellar abundances are much larger than for the four high-S/N stars (the median uncertainty in [X/H] over the 18 elements studied differentially is 0.1 dex).
Nevertheless, its [X/H] abundances are compatible with those of the four high-S/N stars for all common chemical species except for Cu, Zn, and Ba. Given the significantly lower precision in its derived differential chemical abundances and also the lower accuracy (Sect.~\ref{subsec:atmospheric_params}), this star was not considered to compute the mean cluster abundances. Hence, we report its abundances in Table~\ref{table_vr_APs_mean_star_abunds} but exclude it from the discussion for the remainder of the study.

Figure~\ref{fig:individual_stars_abunds} shows the solar-scaled abundances of the 23 studied ions for the four RC stars in UBC~1052 with high-S/N spectra. For each of the four stars, the median uncertainty in [X/H] considering all the 22 ions studied differentially is $\sim$0.06 dex, and the uncertainty values range between 0.02 dex and 0.09 dex. Expressed as [X/Fe], the uncertainties in the differential abundances are notably larger (median value of $\sim$0.09 dex, considering all the ions). \ion{Ba}{II} is the ion with the smallest [X/H] uncertainty for the four stars ($\sigma_{\textrm{[Ba/H]}} = 0.02$ dex). Two species with large [X/H] uncertainties for all four stars are \ion{Sr}{I} ($\sigma_{\textrm{[Sr/H]}} = 0.07$ dex) and \ion{Mn}{I} ($\sigma_{\textrm{[Mn/H]}} \sim 0.08$ dex).

\begin{figure*}[h!]
{\includegraphics[width = 0.994\textwidth]{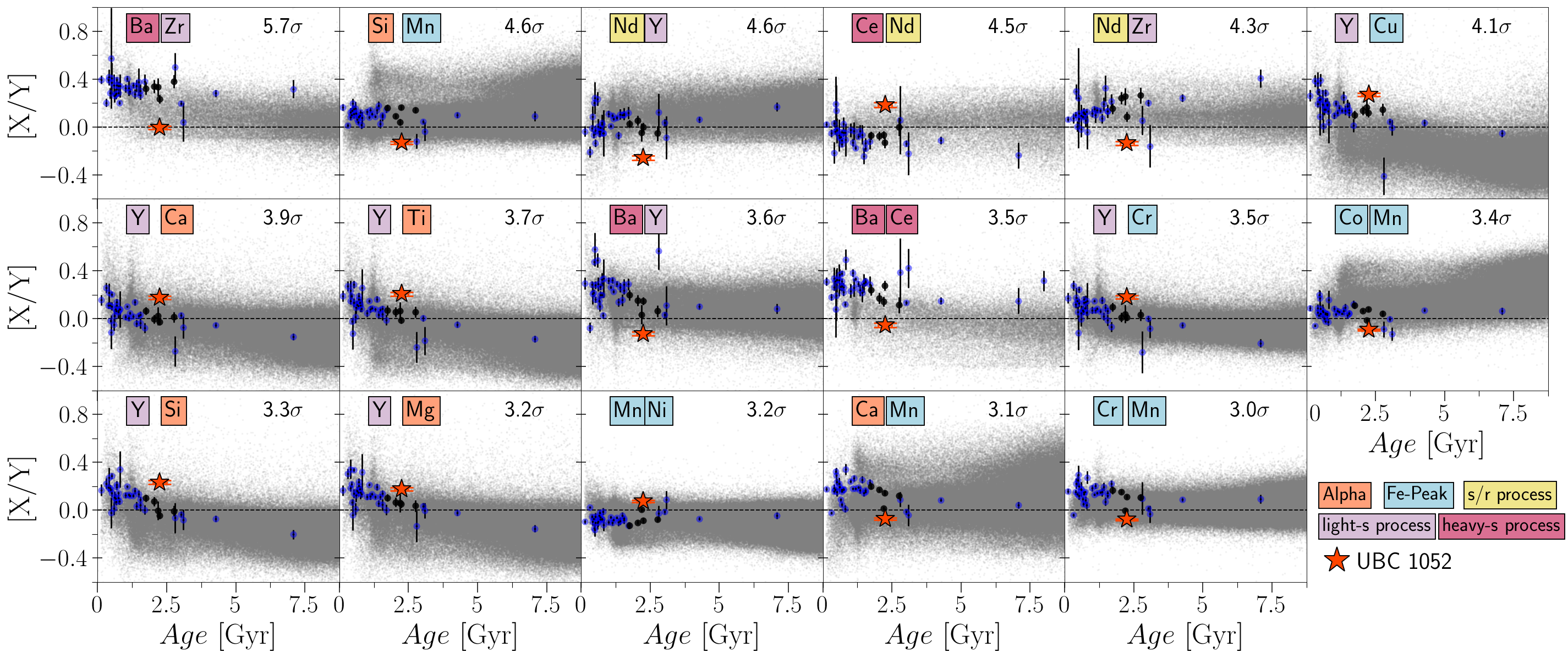}}
\caption{Abundance ratios as a function of age for UBC~1052 (red star) and for the OCs in \citetalias{casamiquela+2021} with more than one studied star for each element involved in [X/Y] (blue circles except for the OCs with ages between $1.75-2.75$ Gyr, which are shown as black circles). The elements of the abundance ratio [X/Y] in each panel are indicated in the boxes (X: left box, Y: right box), colour-coded by nucleosynthetic family, and we only show the abundance ratios for which UBC~1052 is more discrepant with respect to the OCs in \citetalias{casamiquela+2021} with similar ages (black circles). Grey background data points are giant stars from GALAH DR4 detrended abundances in \citealt{Kos+2025} (see text for selection).}
\label{fig:XoverY_vs_age_casamiq21_galahdetrend_sigmadiscrepancy}
\end{figure*}

The mean cluster abundances, computed as described in Sect.~\ref{subsec:method_final_cluster_abunds}, are reported in Table~\ref{table_mean_cluster_abunds}. They are represented as black circles in Fig.~\ref{fig:individual_stars_abunds} and used to compute the abundance ratios shown as the red star in Fig.~\ref{fig:XoverY_vs_age_casamiq21_galahdetrend_sigmadiscrepancy}. We find that UBC~1052 has a slightly super-solar [Fe/H] $=0.05 \pm 0.01$ dex, where the reported uncertainty, notably small, is the standard deviation over the four RC stars. Such nearly-solar [Fe/H] contrasts with its small $\rgc \sim 6$ kpc, as is discussed in Sect.~\ref{subsec:radial_FeH_gradient}.

All stars exhibit remarkably similar abundances in all analysed elements, not only for Fe. As can be seen in Fig.~\ref{fig:individual_stars_abunds}, the standard deviations of the abundances of the four stars are significantly smaller than the abundance uncertainties of each individual star. The average value of the standard deviations of the abundances of the four stars is 0.02 dex for both [X/H] and [X/Fe] (over the ions studied differentially). All elements that belong to the nucleosynthetic groups $\alpha$, odd-Z, and Fe-peak have [X/H] standard deviations over the four stars < 0.025 dex, with the only exception of \ion{Zn}{I} (0.04 dex). These standard deviations are typically larger for the neutron-capture elements (0.03 dex on average), the maximum value being 0.06 dex for \ion{Sr}{I}.

\ion{Zn}{I} and all neutron-capture elements (except \ion{Zr}{I}) are precisely the elements whose derived abundances rely on a single line in the differential analysis with respect to M~67. Although their [X/H] uncertainties for the individual RC stars are not larger than those of the elements belonging to other nucleosynthetic channels, their [X/H] values are potentially less accurate. The reason is that they critically depend on the quality of (i) a particular line’s atomic data, (ii) the continuum placement for the normalised observed spectrum in that line’s narrow spectral region, and (iii) the fit between the synthetic and the observed line. Regarding atomic data, the line used for each of them has ‘\texttt{gf\_flag}’ = Y in \citealt{heiter+2021} (meaning that the transition probability for this line is accurate or among the most accurate available) except for \ion{Nd}{II} (whose line has not been assigned a ‘\texttt{gf\_flag}’ value). 
\ion{Zn}{I} and \ion{Ba}{II} are the only ions for which one star’s [X/H] is not $1\sigma$ compatible with the [X/H] abundances of all the other three stars; but the [X/Fe] abundances of all four stars are $1\sigma$ compatible for all the ions.

The excellent precision of the UVES data allows us to place strict bounds on the chemical homogeneity of UBC~1052, for the first time for such a distant cluster in the inner Galactic disc. For the interested reader, in Appendix~\ref{app_sec:ubc1052_xoverH_dispersion_pryormeylan} we report conservative estimates for the chemical homogeneity of the cluster for each studied element. The intrinsic [X/H] dispersion in the cluster, averaging over all elements, is 0.04 dex. This value is larger than the standard deviation of [X/H] over the four stars averaged over all the elements, which is 0.02 dex. Still, the standard deviation of [X/H] over the four stars is comprised between the 16th percentile and the median of our estimated cluster intrinsic [X/H] dispersion for almost all elements (see Fig.~\ref{fig:XoverH_cluster_dispersion_Pryor93_mcmc}). Given that the [X/H] uncertainties of each individual star are significantly larger than the standard deviation of [X/H] over the four stars, the upper uncertainty we find for the OC’s intrinsic [X/H] dispersion is considerably large: while its 16th percentile is typically 0.01 dex, its 84th percentile is typically 0.1 dex. 

Regarding the auxiliary differential analysis of the solar analogue in M~67, our derived $\vr$, APs, and abundances are found in Table~\ref{table_vr_APs_mean_star_abunds} and compared with the literature in Appendix~\ref{app_sec:solar_scaled_abunds}.

\section{Discussion}\label{sec:discussion}

In this section we put our results for UBC~1052 in the broader context of Galactic chemical evolution and kinematics. 

\subsection{Comparison with other OCs studied through a line-by-line differential analysis}

In this section we compare the chemical pattern of UBC~1052 with that of other OCs studied in \citet[][hereafter C21]{casamiquela+2021}, which is the most similar spectroscopic study to the one we have carried out for UBC~1052 (see Appendix~\ref{app_subsec:comparison_analysis_casamiquela21RC} for a comparison between both studies).

First, we compare UBC~1052’s [X/H] and [X/Fe] with those of two \citetalias{casamiquela+2021} subsamples: the inner-disc OCs, and the similar-age and high-$\zmax$ OCs. The results are discussed in Appendix~\ref{app_subsec:comparison_abunds_casamiquela21}. We find that UBC~1052’s [Fe/H] is compatible with that of the inner-disc, young cluster NGC~6705 (see also the discussion in Sect.~\ref{subsec:radial_FeH_gradient}); as well as with that of Ruprecht~147, a similarly aged cluster in the solar neighbourhood. Overall, UBC~1052’s mean [X/Fe] abundances for most elements fall within the range covered by \citetalias{casamiquela+2021}. Its [X/Fe] abundance pattern appears similar to other OCs in most elements, but is significantly different for some others (e.g. showing a super-solar [Mn/Fe], a quite high [Y/Fe] and [Pr/Fe], a quite low [Ba/Fe], and a very low [Nd/Fe]).

The second part of the comparison focuses on the abundance ratios between elements from different nucleosynthetic families and their dependence on age (chemical clocks). We compare the abundance ratios of UBC~1052 (computed from Table \ref{table_mean_cluster_abunds}) with those of the five OCs in \citetalias{casamiquela+2021} that have ages similar to UBC~1052 (between 1.75 Gyr and 2.75 Gyr), for all possible combinations of the elements that are common in both studies. To quantify the discrepancy, we compare the absolute value of the difference between UBC~1052’s [X/Y] and the mean [X/Y] of the five similarly aged OCs in \citetalias{casamiquela+2021} with the error of this difference (i.e. the square root of the quadratic sum of UBC~1052’s standard deviation of [X/Y] over its four studied stars and the [X/Y] standard deviation over the five OCs). The abundance ratios for which this absolute difference is larger than three times its error are represented as a function of cluster age in Fig.~\ref{fig:XoverY_vs_age_casamiq21_galahdetrend_sigmadiscrepancy}, ordered by decreasing discrepancy (the absolute difference in terms of its error is indicated in each panel). The ages we adopt for the OCs studied in \citetalias{casamiquela+2021} are taken in most of the cases from \citet{Cantat-Gaudin2020}, who do not provide individual age uncertainties for each OC. In \citetalias{casamiquela+2021} they found that the ages, extinctions, and distances in \citet{Cantat-Gaudin2020} were visually consistent with the CMDs for most of their studied OCs, and provided manually-refined values of the extinction and the age for the remaining few OCs. For these cases, we adopt the ages in \citetalias{casamiquela+2021} instead of those in \citet{Cantat-Gaudin2020}. Additionally, for comparison, in Fig.~\ref{fig:XoverY_vs_age_casamiq21_galahdetrend_sigmadiscrepancy} we also include the GALAH DR4 detrended abundances in \citet{Kos+2025} for giant stars (‘\texttt{detrend\_model}’ = 1), selecting only those with ‘\texttt{snr\_px\_ccd3}’ > 30, ‘\texttt{flag\_sp}’ = 0, and ‘\texttt{flag\_X\_fe}’ = 0 for element X (recommended flags), and also restricting it to those with an uncertainty in [X/Y] smaller than 0.1 dex. Most of the abundance ratios in Fig.~\ref{fig:XoverY_vs_age_casamiq21_galahdetrend_sigmadiscrepancy} involve neutron-capture elements, so they comprise at least one element for which the abundance has been derived using only one line in the differential analysis of UBC~1052 with respect to M~67. The abundance ratios for which UBC~1052 deviates most significantly from OCs with similar ages are [Ba/Zr], [Si/Mn], [Nd/Y], and [Ce/Nd]. Considering that UBC~1052’s chemical pattern appears to be peculiar in some abundance ratios, we briefly discuss the possibility of chemical tagging in Sect.~\ref{sec:chemical_tagging}.

\subsection{Radial metallicity gradient}
\label{subsec:radial_FeH_gradient}

To compare UBC~1052 with the radial [Fe/H] gradient traced by OCs (hereafter referred to as the radial metallicity gradient), we use two recent catalogues that compile and homogenise abundances from several high-resolution spectroscopic surveys for a large number of OCs. One study is \cite{Spina+2022}, which provides [Fe/H] abundances for 251 OCs based on five surveys: APOGEE DR16 \citep{Donor+2020}, {\it Gaia}-ESO iDR4, iDR5 and iDR6 \citep{Spina+2017, Baratella+2020, Magrini+2021}, GALAH DR3 \citep{Spina+2021}, OCCASO \citep{Casamiquela+2019}, and SPA \citep{Frasca+2019, DOrazi+2020, Casali+2020, Zhang+2021SPA}. The other study is \citet{Carbajo+2024}, based on the latest OCCASO survey results, in which the authors derive abundances from high-resolution spectra ($R > 60\,000$) for 36 OCs with at least four RC stars. In order to enlarge the spatial coverage, they build the OCCASO+ sample (99 OCs) by adding those OCs in {\it Gaia}-ESO DR5 \citep{Magrini+2023}, APOGEE DR17 \citep{Myers+2022}, and GALAH DR3 \citep{Spina+2021} whose [Fe/H] have also been derived using at least four stars in the RC.
Additionally, for comparison, we also consider the OCs in \citetalias{casamiquela+2021} discussed in the previous section. In contrast to the two previous studies, their [Fe/H] are derived through a line-by-line differential analysis, and they span a more limited range in $\rgc$.

\begin{figure*}
{\includegraphics[width = 0.494\textwidth]{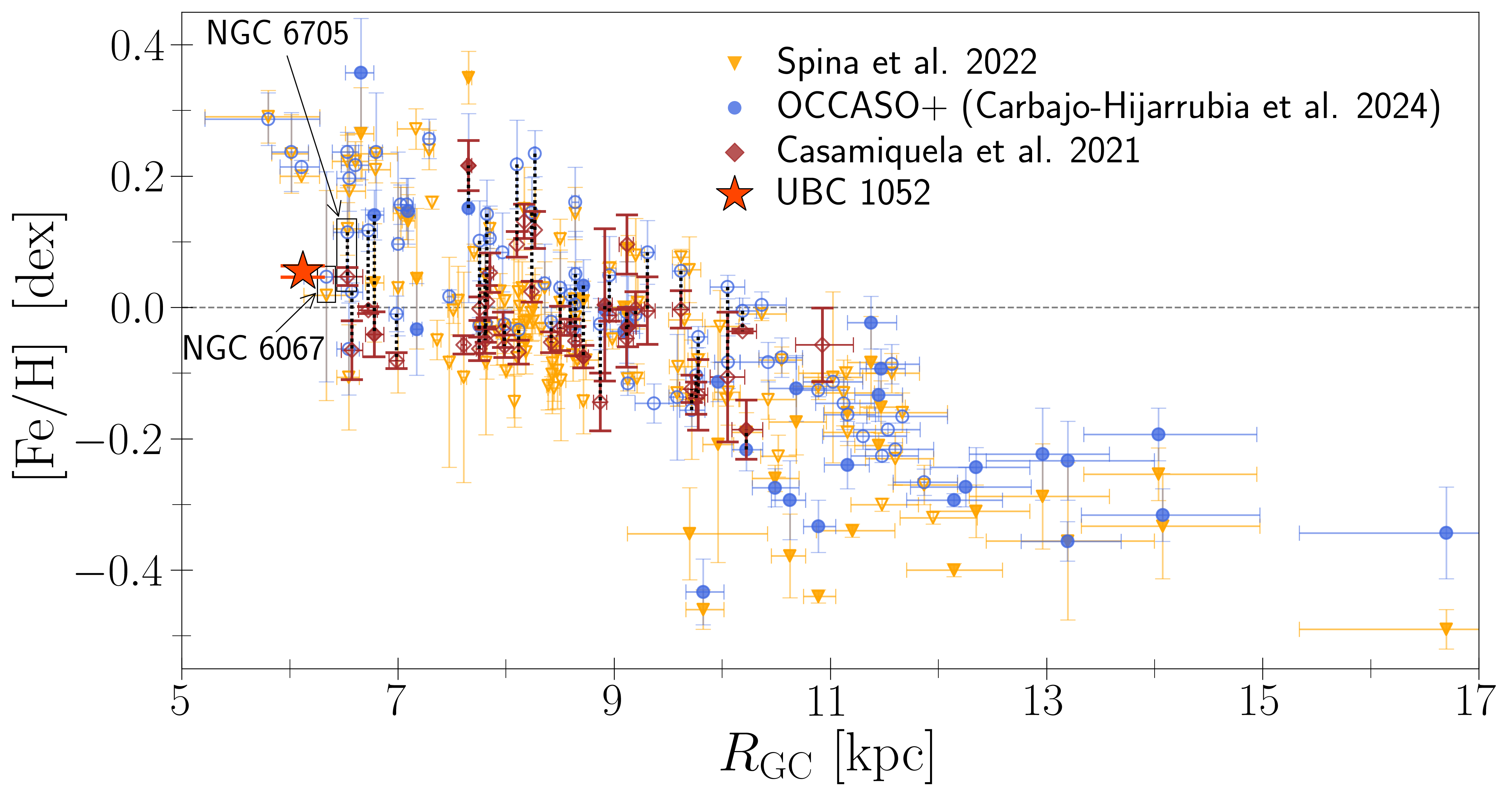}}
{\includegraphics[width = 0.494\textwidth]{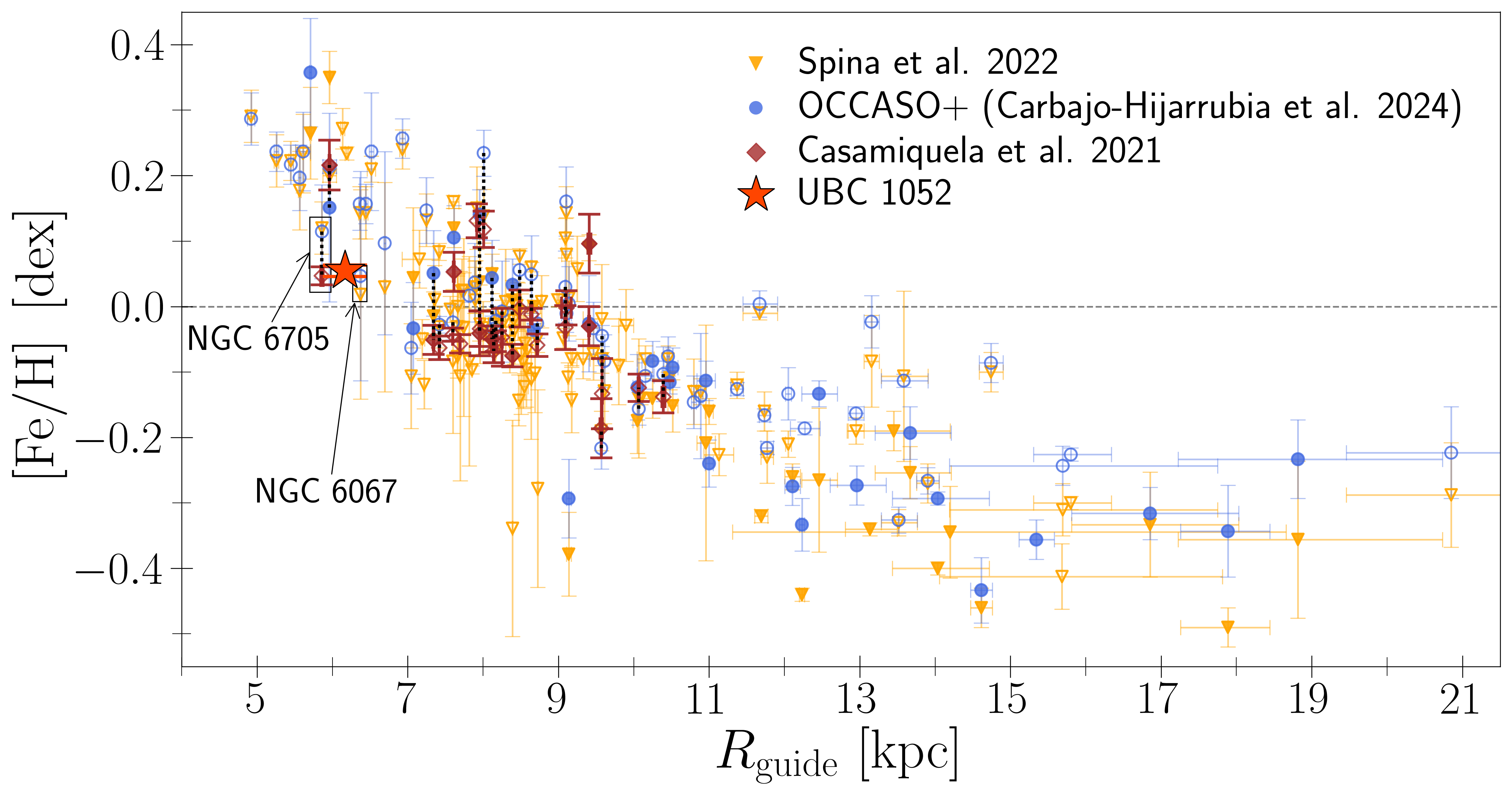}}
\caption{Radial metallicity gradient traced by OCs with [Fe/H] estimates derived from high-resolution spectroscopy of at least two members, adding UBC~1052 (red star). \textit{Left}: Solar-scaled iron abundances for OCs as a function of their galactocentric radius in \citet{Cavallo+2024}. \textit{Right}: Same but as a function of their guiding-centre radius in \citet{Spina+2022}. The uncertainties in [Fe/H] are the standard deviation over the studied members' [Fe/H]. The vertical black dotted lines join the same OCs as studied in OCCASO and in \citetalias{casamiquela+2021}.  Open symbols are OCs younger than $2$ Gyr, filled symbols are OCs that are at least $2$ Gyr old; ages are taken from \citealt{Cavallo+2024} (\textit{left}) and from \citealt{Cantat-Gaudin2020} (\textit{right}).}
\label{fig:FeH_radial_gradient}
\end{figure*}

From these three catalogues, we select only those OCs whose [Fe/H] is based on spectra of at least two members. For the OCs in \citet{Spina+2022} whose [Fe/H] are taken from \citet{Donor+2020}, we additionally require that they belong to the ‘high-quality sample’ (as defined in \citealt{Donor+2020} based on the appearance of their CMDs). We retain 145 OCs in \citet{Spina+2022} and 41 in \citetalias{casamiquela+2021}. The $\rgc$ values are adopted from the recent study of \citet{Cavallo+2024}, and we further select only those OCs with reliable cluster parameters (age, metallicity, extinction, and distance) in this study (their ‘gold’ sample), retaining 132 OCs from \citet{Spina+2022}, 96 from OCCASO+ and 41 from \citetalias{casamiquela+2021}. The radial metallicity gradient traced by these samples is represented in the left panel of Fig.~\ref{fig:FeH_radial_gradient}, adding UBC~1052 and differentiating the OCs that are younger or older than $2$ Gyr (our lower bound on UBC~1052’s age) using open and filled symbols, respectively.
We also represent the metallicity gradient as a function of the clusters’ $\rguide$, adopted from \citealt{Spina+2022} (right panel of Fig.~\ref{fig:FeH_radial_gradient}, the ages are now taken from \citealt{Cantat-Gaudin2020}). Having performed these crossmatches, this right panel includes 143 OCs from \citet{Spina+2022}, 83 OCs from OCCASO+, and 27 OCs from \citetalias{casamiquela+2021}.

To discuss UBC~1052’s [Fe/H] in the context of the Galactic metallicity gradient traced by OCs (Fig.~\ref{fig:FeH_radial_gradient}), we start with some general considerations:

\begin{enumerate}
  \item In Fig.~\ref{fig:FeH_radial_gradient}, there is a considerable abundance scatter at any certain $\rgc$, which is only slightly reduced when the [Fe/H] gradient is represented against $\rguide$ (see e.g. \citealt{Spina+2022}). Most of this scatter is generally attributed to gradual radial mixing inside the Galactic disc, which causes the current $\rgc$ of an OC to be not always representative of its birth location (especially for old OCs). Radial mixing can be of two types: blurring (due to an increase in the amplitude of the star’s epicyclic oscillations around $\rguide$, which does not change because angular momentum is conserved); and churning (migration caused by a gain or loss of angular momentum from resonant interactions with the potential of non-axisymmetric structures, which causes secular changes in $\rguide$; \citealt{Sellwood2002}).
  \item The [Fe/H] obtained at high resolution through different analyses for the same OC can differ by $\sim0.1 - 0.2$ dex, which is beyond the quoted uncertainties in [Fe/H] in many cases. These systematics arise from the different instrumentation and samples used in each survey, as well as from the different assumptions and methodologies used in the analyses and the adopted solar abundances \citep[e.g.][]{jofre+2019, Soubiran+2022}. As an example, in Fig.~\ref{fig:FeH_radial_gradient} we link with dotted black lines the different results for the same cluster using \ispec (through a differential analysis in \citetalias{casamiquela+2021} and a non-differential one in OCCASO).
  \item The inner part of the Galaxy, where UBC~1052 resides, is still poorly sampled.
\end{enumerate}

The [Fe/H] abundance of UBC~1052 is lower than the typical ones of OCs at $\rgc\sim6$ kpc (its uncertainty being also remarkably smaller); and it is also at the lowest end of the [Fe/H] distribution of OCs with similar $\rguide$.
For instance, UBC~1052 falls right outside of the 68\% confidence intervals of the linear models fit to the metallicity gradient (both with respect to $\rgc$ and $\rguide$) in \citet{Spina+2021}. In Fig.~\ref{fig:FeH_radial_gradient} we label the two OCs that are also located at small $\rgc$ and have the largest discrepancies from the metallicity gradient with respect to both $\rgc$ and $\rguide$: NGC~6705 and NGC~6067. In contrast to UBC~1052, these two OCs are much younger (age $< 350$ Myr; e.g. \citealt{Cantat-Gaudin2020}) and have smaller maximum excursions from the Galactic plane ($\zmax < 100$ pc; \citealt{Spina+2022}). For NGC~6067, the uncertainties in [Fe/H] in the two studies ($\sim0.15$ dex) are much larger than what we obtain for UBC~1052. Moreover, the value itself was found to be biased in \citet{Magrini+2023}, who claim that the recommended {\it Gaia}-ESO abundance for this OC adopted in both studies ([Fe/H] = $+0.03 \pm 0.16$ dex) is underestimated due to an overestimation of the microturbulence parameter derived from the spectral analysis. If the microturbulence parameter is fixed at a theoretical value instead, the [Fe/H] obtained for NGC~6067 is higher and consistent with the Galactic metallicity gradient (e.g. \citealt{Alonso-Santiago+2017}). The latter authors find an average [Fe/H] $= +0.19 \pm 0.05$ dex for the giant stars in this OC, thus debunking NGC~6067 as an outlier comparable to UBC~1052. In addition to a systematic increase in the microturbulence parameter for the youngest OCs in \citet{Magrini+2023}, they also find a decrease in [Fe/H] for decreasing $\logg$ for the giants in these OCs, which likely causes an underestimation of [Fe/H] for the young OCs whose studied giants have all low $\logg$ (approximately below $2.5$ dex), as is the case for NGC~6067 and also NGC~6705. Besides, \citet{guerco+2025} find that NGC~6705 likely describes a chaotic orbit, since its disc-like orbit shows oscillations in and out of the bar corotation radius and is moving around the bar's Lagrange points L4 and L5, implying that this OC is trapped in resonance with the bar. Nonetheless, as in our analysis for UBC~1052, they also find that its orbit has a strong dependence on the bar pattern speed.

The degree of discrepancy between UBC~1052’s [Fe/H] and the metallicity gradient traced by OCs depends on the uncertainties in the clusters’ [Fe/H], $\rgc$ and $\rguide$; and on the systematic offsets in [Fe/H] between samples. Some of these aspects are investigated in Appendix~\ref{app_sec:radial_metallicity_gradient_other_dists_lessreliableOCs}. In Fig.~\ref{fig:FeH_radial_gradient_OCs1member} we represent the metallicity gradient with respect to $\rgc$ using two alternative catalogues for the clusters’ distances and also including the OCs whose [Fe/H] have been derived using one single member. Figure~\ref{fig:FeH_radial_gradient_Rguide_OCs1member} shows the result of including these less-reliable OCs in the metallicity gradient with respect to $\rguide$ shown in the right panel of Fig.~\ref{fig:FeH_radial_gradient}. Our conclusion is that, regardless of the distance estimations, UBC~1052 is an outlier of the radial metallicity gradient traced by OCs and the other OCs that are comparable outliers at small $\rgc$ are much younger (NGC~6705 and NGC~6067, at slightly larger $\rgc$ than UBC~1052) or have very uncertain [Fe/H] measurements and/or cluster parameters (Berkeley~43 and Berkeley~44, each with only one member studied spectroscopically at high resolution).

Hence, UBC~1052 has a lower [Fe/H] compared with other OCs at similar galactocentric radii, being persistently the most extreme outlier among the old OCs with reliable parameters studied to date (both in the metallicity gradient with respect to $\rgc$ and $\rguide$). A natural hypothesis for this discrepancy is that UBC~1052 was born farther from the Galactic centre than its current $\rgc$ and that it has significantly migrated inwards (see Sect.~\ref{sec:migration}). 
In addition to its [Fe/H], UBC~1052 also stands out at its location in the metallicity gradient due to its old age and high $\zmax$. In the metallicity gradient with respect to $\rgc$ in Fig.~\ref{fig:FeH_radial_gradient}, UBC~1052 is the innermost OC older than 2 Gyr. This is also the case using other distance and age catalogues (Fig.~\ref{fig:FeH_radial_gradient_OCs1member}). In the metallicity gradient with respect to $\rguide$ in Fig.~\ref{fig:FeH_radial_gradient}, there are two OCs that are older than UBC~1052 and located at a smaller $\rguide$: NGC~6791 and NGC~6253. In contrast to UBC~1052, they are metal-rich and are not candidates for inward-migrated OCs. Regarding $\zmax$, according to \citet{Spina+2022} UBC~1052 has the highest value among all OCs at $\rgc<6.5$ kpc in the bottom panel of Fig.~\ref{fig:FeH_radial_gradient_OCs1member}, and the second highest $\zmax$ – after the peculiar OC NGC~6791 – at $\rguide < 7$ kpc in Fig.~\ref{fig:FeH_radial_gradient_Rguide_OCs1member}. The peculiarity of UBC~1052’s age and orbit at its inner $\rgc$ also becomes evident in Sect.~\ref{subsec:kinematic_properties}, where they are compared with other OCs and field stars.

In terms of the radial [X/Fe] profile, using the OCCASO+ sample \citep{Carbajo+2024}, UBC~1052’s [X/Fe] abundances fall within the scatter of the OCs in OCCASO+ at similar $\rgc$, except for [Si/Fe] (underabundant) and [Mn/Fe] (overabundant).

\subsection{Kinematic properties}
\label{subsec:kinematic_properties}

\begin{figure*}[h!]
{\includegraphics[width = 0.994\textwidth]{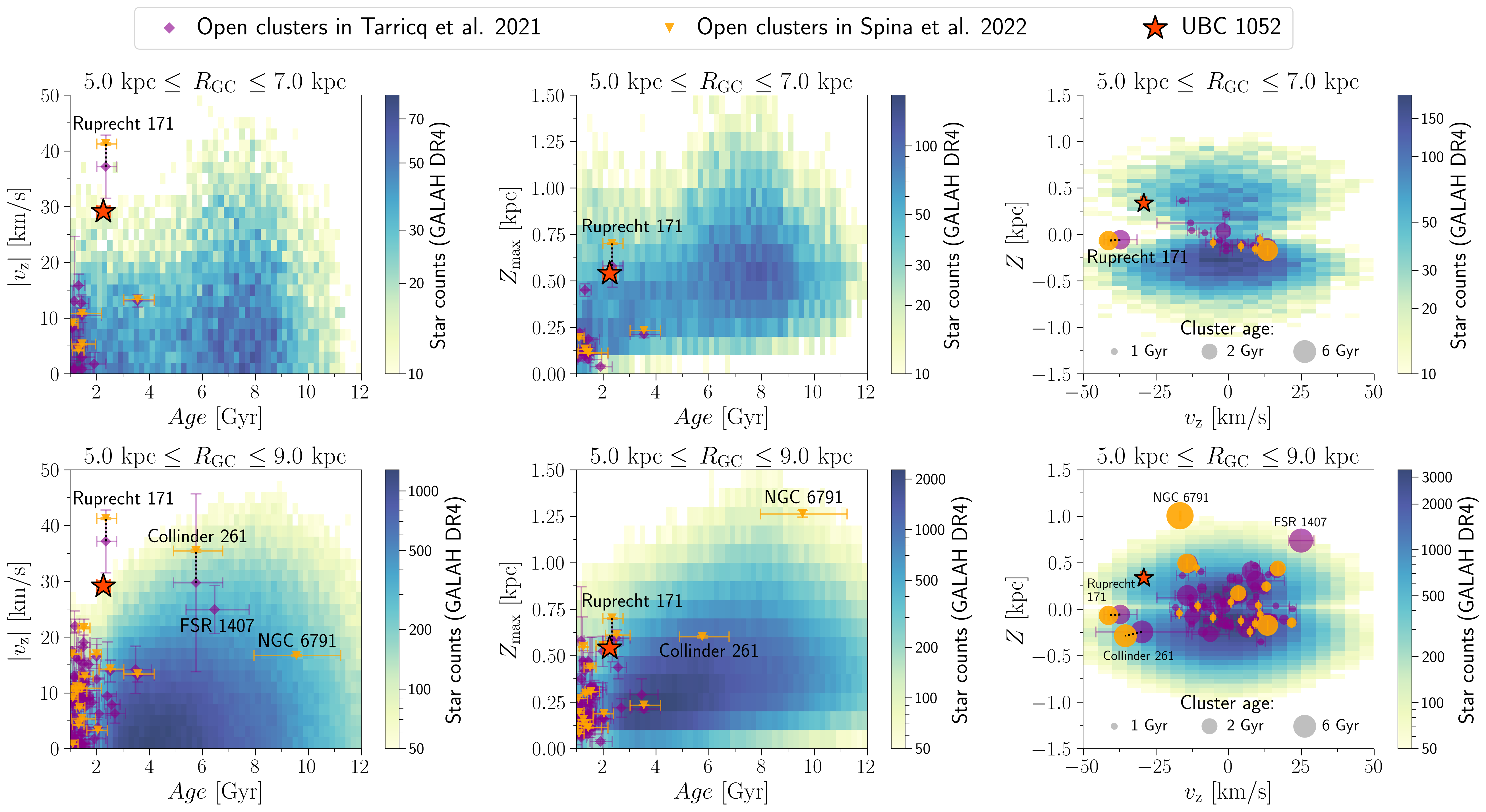}}
\caption{Comparison of the orbital properties and age between UBC~1052 (red star), other OCs (errorbars), and GALAH DR4 thin disc stars (background density) in two different galactocentric radius ranges: [5, 7] kpc (\textit{top row}) and [5, 9] kpc (\textit{bottom row}). \textit{Left column}: Absolute value of the vertical velocity as a function of age. \textit{Central column}: Orbit’s maximum excursion from the Galactic plane as a function of age. \textit{Right column}: Current height above the Galactic plane as a function of the vertical velocity (the clusters’ sizes are proportional to their age). The ages and $\rgc$ for the OCs are from \citet{Cavallo+2024}, and we represent only those OCs with ‘gold’ parameters in this study. The rest of the parameters are taken from the studies indicated in the legend (\citealt{Tarricq+2021} and \citealt{Spina+2022}). For the OCs in \citet{Tarricq+2021}, we further select only those with a maximum error of 20 $\kms$ in $\vz$ (left and right columns) and of 1 kpc in $\zmax$ (central column). The thin disc stars come from GALAH DR4 \citep{Buder+2025} with [$\alpha$/Fe] < 0.1 dex (computed as a weighted mean over the $\alpha$-elements X with ‘\texttt{flag\_X\_fe}’ = 0), ‘\texttt{snr\_px\_ccd3}’ > 30, and ‘\texttt{flag\_sp}’ = 0. In all panels the represented GALAH stars and OCs are older than 1 Gyr.}
\label{fig:kinematics}
\end{figure*}

UBC~1052 has a cylindrical galactocentric radial velocity of $\varv_{\mathrm{R}} = 14 \pm 2$ $\kms$, and a high cylindrical galactocentric vertical velocity of $\vz = -29 \pm 2$ $\kms$. Its cylindrical galactocentric rotational velocity is $\varv_{\Phi} = 244 \pm 1$ $\kms$, fully consistent with the kinematics of the thin disc population.
In Fig.~\ref{fig:kinematics} we compare the kinematic properties and age of UBC~1052 with those of other Galactic OCs and GALAH DR4 thin disc stars. Limiting the comparison to the 5 $ \leq \rgc$ [kpc] $\leq$ 7 range (top row), UBC~1052 is a clear outlier among the OCs (alongside Ruprecht~171) and is also located towards the outskirts of the field star distribution in the three panels. Ruprecht~171 has a similar age to UBC~1052, but is located at a larger $\rgc$ ($\sim$ 6.7 kpc) and is not a remarkable outlier in the metallicity gradients with respect to $\rgc$ traced by \citet{Spina+2022} and OCCASO+. It also falls within the [Fe/H] distribution of OCs with similar $\rguide$ ($\sim$ 8 kpc, \citealt{Spina+2022}).
Extending the comparison to the 5 $ \leq \rgc$ [kpc] $\leq$ 9 range (bottom row), UBC~1052 still stands out as one of the OCs with higher $\zmax$ and vertical action. Compared with the other old OCs with large vertical actions (labelled in the bottom right panel of Fig.~\ref{fig:kinematics}), UBC~1052 is not only the OC located at a smallest $\rgc$, but also the only OC whose [Fe/H] points to a galactocentric radius at birth, $\rbirth$, larger than the current one (see Sect.~\ref{sec:migration}). The other more extreme outliers are (i) NGC~6791, which is an outward migrator candidate suspected to be born in the very inner disc/bulge \citep[e.g.][]{Jilkova2012, Villanova2018}, (ii) Ruprecht~171 and Collinder~261, whose abundances fall within the typical scatter expected for the metallicity gradient with respect to $\rgc$ and $\rguide$ in Fig.~\ref{fig:FeH_radial_gradient}, and (iii) FSR~1407, a high-altitude ($Z \sim$ 650 pc) cluster at $\rgc$ $\sim$ 8.9 kpc with no [Fe/H] estimate from high-resolution spectroscopy.

Therefore, UBC~1052 stands out as the oldest and highest-|$Z$| inner-disc OC among the clusters with reliable high-resolution [Fe/H] estimates, being located in the poorly sampled inner Galactic region where old OCs and OCs with high $\zmax$ are scarce.

\subsection{UBC~1052 in the context of radial migration}\label{sec:migration}
A few clear outliers of the radial metallicity gradient have been pointed out as migrating OCs in the literature. The most extreme and well-studied case is NGC~6791, one of the oldest and most metal-rich OCs that also has high eccentricity and high $\zmax$ \citep{Carraro2006, Jilkova2012}. Other examples of OCs that are believed to have migrated significantly based on their deviation from the [Fe/H] vs. $\rgc$ or [Fe/H] vs. $\rguide$ gradients are NGC~2243 and Berkeley~32 \citep{Spina+2021, Spina+2022}, both inward migrator candidates located at $\rgc$ > 9 kpc.
The identification of such OCs as migrators is based on the assumption that most of the abundance scatter at any certain $\rgc$ or $\rguide$ is caused by radial mixing. However, other factors also contribute to this scatter: the uncertainties in [Fe/H], as well as in $\rgc$ and $\rguide$ (especially for the OCs in the outermost disc; where additionally the Galactic warp is generally ignored), possible azimuthal variations in metallicity \citep[e.g.][]{Poggio2022, Hawkins2023}, and the possible local chemical enrichment in specific regions (e.g. \citealt{Maderak+2015}). Additionally, the metallicity scatter can also vary depending on the spectroscopic survey used to trace the metallicity gradient, its completeness, and the selection of OCs that is made (regarding the minimum number of observed members, maximum uncertainties in [Fe/H], etc.). Thus, except for very extreme outliers, the classification of an OC as likely to have undergone significant radial migration depends on a relatively arbitrary threshold set to the discrepancy with respect to the radial gradient.

Recently, some studies have systematically investigated radial migration for large samples of OCs, increasing the number of candidate migrators. Two of them are \citealt{Chen_Zhao2020} (146 OCs) and \citealt{Zhang+2021Gradient} (225 OCs). However, the inner disc region is poorly sampled in both, and the stricter limits imposed on the uncertainties of the clusters' [Fe/H] and radial velocities in \citet{Zhang+2021Gradient} cause the rejection of all OCs at $\rgc$ < 7 kpc which were studied in \citet{Chen_Zhao2020}. Both studies identify OCs that could have significantly migrated through churning as those with |$\rguide - \rbirth$| > 1 kpc, where $\rbirth$ is computed using the variation in interstellar medium (ISM) metallicity with time at the solar radius and the variation in ISM metallicity gradient with time from \citet{Minchev+2018}. The inferred $\rbirth$, however, is quite uncertain since it relies on the assumptions made in \citet{Minchev+2018} to compute the ISM metallicity gradient, such as that the ISM is well mixed azimuthally, and the profile was obtained by \citet{Minchev+2018} using field stars (so it could systematically differ for OCs). In both studies, inward migrators ($\rguide - \rbirth < -1$ kpc) are younger than about 1.5 Gyr (except Berkeley 32 in \citealt{Chen_Zhao2020}, which is at $\rgc \sim$ 9 kpc).
A similar approach was adopted by \citet{Netopil+2022}, who also used the model by \citet{Minchev+2018} and classified 51\% of the clusters as probable migrators (|$\rguide - \rbirth$| > 1 kpc). Only a small fraction of them are inward-migrating OCs, which are mostly younger than 1 Gyr. They estimated a mean migration rate of 1 kpc ${\textrm{Gyr}}^{-1}$ for objects younger than $\sim$ 2 Gyr, for which the change in $\rguide$ is the main contributor to the radial mixing.

We estimated UBC~1052's $\rbirth$ following \citet{Lu+2024}, \citet{Ratcliffe+2025}, and \citet{Ratcliffe+2026}, which successively improved the method proposed by \citet{Minchev+2018}. All methods provide the birth radius of a star based only on its age and [Fe/H], for a given evolution with cosmic time $\tau$ of the birth radial [Fe/H] gradient ($\nabla$[Fe/H]($\tau$)) and of the central disc's [Fe/H] ([Fe/H]($\rbirth = 0$, $\tau$)). Using our derived values and uncertainties for UBC~1052’s [Fe/H] and age, we find a $\rbirth = 6.0 \pm 0.2$ kpc following \citet{Lu+2024}. Using {\tt Rbirth}\footnote{\url{https://github.com/BridgetRatcliffe/Rbirth}} package to apply the method in \citet{Ratcliffe+2025}, we find $\rbirth = 6.4 \pm 0.3$ kpc if \citet{Anders+2023} ages are used to derive $\nabla$[Fe/H]($\tau$) and [Fe/H]($\rbirth = 0$, $\tau$), and $\rbirth = 7.0^{+0.3}_{-0.1}$ kpc when \citet{Stone-Martinez+2024} ages are used. Following \citet{Ratcliffe+2026}, the result is $\rbirth = 6.5 \pm 0.2$ kpc. The values we find for $\rbirth$ imply that UBC~1052 was born at a slightly larger galactocentric radius than its current $\rgc$, or at least at a roughly compatible one. However, the uncertainties in these $\rbirth$ estimates do not account for uncertainties in the modelling or its assumptions; they only consider the uncertainties in the model parameters (age and [Fe/H]).
If UBC~1052 is now located at $\rgc < \rbirth$, given its low orbital eccentricity ($e \lesssim  0.13$, see Sect. \ref{subsec:vr_orbital_parameters}) the predominant radial mixing process has likely been churning (i.e. it has undergone a neat inward radial migration). The quantification of churning would require the comparison of accurate estimates of $\rbirth$ and $\rguide$, which still remain quite uncertain.
In such a scenario where UBC~1052 has significantly migrated inwards since its birth, this cluster would be a rare and interesting case because since the inner part of the Galaxy has strong gravitational potentials (e.g. central bar, spirals, and dense giant molecular clouds), it has been suggested that inward-migrating clusters are quickly disrupted, and at higher rates than outward-migrating clusters \citep[e.g.][]{Anders+2017, Spina+2022}. UBC~1052 would then stand out as one of the oldest and innermost inward migrator candidates compared with the candidates identified in \citet{Chen_Zhao2020}, \citet{Zhang+2021Gradient}, and \citet{Netopil+2022}.

We also estimated $\rbirth$ for the OCs in OCCASO+ and in \citealt{Spina+2022} (see Appendix~\ref{app_sec:Rbirth}). Regardless of the method used to compute $\rbirth$ and the adopted ages and $\rgc$, we find that most OCs located at $\rgc$ smaller than UBC~1052 have $\rbirth < \rgc$, and for the remaining OCs, $\rbirth$ and $\rgc$ are compatible within the uncertainties (see Fig.~\ref{fig:Rbirth_vs_Rgc} for an example). In contrast to the scenario where UBC~1052 has migrated inwards, these results point to an alternative explanation for its lower [Fe/H] compared with OCs located at similar galactocentric radius: most of the OCs that are now located in the inner disc were born at even smaller $\rgc$, thus having a higher [Fe/H] than expected at their current $\rgc$. This supports the survival bias scenario (e.g. \citealt{Anders+2017}), which would cause a steepening of the measured [Fe/H] gradient in the inner disc, since very few studied OCs are located near their $\rbirth$.

Overall, due to the complexity added by the survival bias, the quantification of past radial migration in OC samples still remains quite uncertain. The relative importance of the blurring and churning radial mixing processes is unclear (although the situation is expected to be similar to field stars for which churning is much more effective; see \citealt{Frankel2020}), and so it is their dependency on the clusters’ positions and ages. Furthermore, the relationship between radial migration and orbital properties that allow OCs to survive enough time to undergo significant migration has not been fully established. The understanding of the mixing processes operating across the Galactic disc and the potential differences in the redistribution of field stars and OCs has important implications for the models of the evolutionary history of the Milky Way (e.g. \citealt{Gallart2024}). To this end, accurate metallicities are crucial. The lack of precise measurements and the significant discrepancies between the values found in different spectroscopic surveys for some OCs still limit the capability to draw conclusions, especially in the inner disc, where there is a scarcity of OCs studied at high resolution (e.g. the cases of Berkeley 43, Berkeley 44, and NGC~6583, discussed in Appendix~\ref{app_sec:radial_metallicity_gradient_other_dists_lessreliableOCs}).

\subsection{UBC~1052 in the context of chemical tagging}
\label{sec:chemical_tagging}

In a concept study using high-resolution, high-S/N data of OC members and field stars from the OCCASO and APOGEE surveys, \citet{casamiquela+2021tagging} found substantial evidence against the possibility of strong chemical tagging - defined as the search for stars that were born in the same cluster based solely on their chemical abundance patterns. In particular, the authors found that different stellar birth sites can have overlapping chemical signatures, even when the most up-to-date methods and high-resolution abundances of many different nucleosynthesis channels are used. Only for some particular birth clusters with truly peculiar abundance patterns was it possible to recover their members in a blind search. Later studies achieved similar results \citep[e.g.][]{Bhattarai+2024, Manea+2025}. Slightly more optimistic conclusions were reached by \citet{Spina+2022tagging, Spina+2025}.

In this context, we suggest that UBC~1052 might be an interesting candidate to attempt a search for past cluster members based on their chemical abundances. In Fig.~\ref{fig:XoverY_vs_age_casamiq21_galahdetrend_sigmadiscrepancy} we show the dependence on age of the abundance ratios [X/Y] for which UBC~1052 deviates most significantly with respect to OCs with similar ages in \citetalias{casamiquela+2021} (in the range [1.75, 2.75] Gyr), including also the GALAH DR4 detrended abundances from \citet{Kos+2025} for field giant stars. Remarkably, UBC~1052’s [Y/$\alpha$] is overabundant for all four studied $\alpha$-elements. These abundance ratios show decreasing trends with age for the OCs in \citetalias{casamiquela+2021}, and UBC~1052’s [Y/$\alpha$] is compatible with the values expected for younger OCs (e.g. for the chemical clock [Y/Mg], its discrepancy with respect to similarly aged OCs is $\sim3 \sigma$). Compared with GALAH, UBC~1052 also deviates to some extent from the [Y/$\alpha$] bulk distribution of field giants. Regarding [Y/Al], another common chemical clock, UBC~1052’s discrepancy with respect to similarly aged OCs is $2.7 \sigma$ (below the minimum discrepancy threshold represented in Fig.~\ref{fig:XoverY_vs_age_casamiq21_galahdetrend_sigmadiscrepancy}). On the other hand, compared with the age-chemical clock relations in \citet{Viscasillas+2022}, which is based on the \gaiaeso survey and includes more OCs than \citetalias{casamiquela+2021}, UBC~1052’s [Y/$\alpha$] and [Ba/$\alpha$] fall within the scatter of all the studied OCs with similar ages (where the $\alpha$-elements studied are Mg, Si, Ca, Ti, and Al - which can be considered an odd-Z element but behaves very similarly to $\alpha$-elements). However, comparing it only with the age-chemical clock relation they derive for the inner-disc clusters (12 OCs at $\rgc < 7$ kpc), UBC~1052 is significantly overabundant in [Y/Si] and [Ba/Si] at its age (lying outside the confidence intervals of the linear regressions). These regressions for inner-disc clusters, however, are especially uncertain for the older OCs since there is only one OC older than 2 Gyr in their samples.

Aside from these typical abundance ratios, UBC~1052 also shows discrepancies in other abundance ratios that are less commonly used. For instance, UBC~1052 shows an extreme discrepancy in abundance ratios of neutron-capture elements, such as [Ba/Zr], [Nd/Y], and [Ce/Nd]. For some of the abundance ratios in Fig.~\ref{fig:XoverY_vs_age_casamiq21_galahdetrend_sigmadiscrepancy}, UBC~1052 not only stands out at the edges of the chemical distribution of all OCs in \citetalias{casamiquela+2021}, but also of the field giants in GALAH.

\section{Conclusions}\label{sec:conclusions}

In this paper, we have characterised the high-altitude, inner-disc open cluster UBC~1052 recently discovered in {\it Gaia} EDR3 data. UVES/VLT spectroscopy of four RC stars allowed us to confirm that UBC~1052 is a real open cluster, and to precisely characterise its chemical abundance pattern through a differential spectroscopic analysis. 

UBC~1052 is located in the inner Galactic disc, which is still poorly sampled at high resolution. The chemical abundances derived in the present study through a strict line-by-line differential analysis are considerably more accurate and precise than the ones derived in large spectroscopic surveys. UBC~1052 has the peculiarity of being the innermost old and high-$\zmax$ OC among the clusters with reliable high-resolution [Fe/H] estimates studied to date. Its lower iron abundance ([Fe/H] $=0.05 \pm 0.01$ dex) compared with other OCs at similar galactocentric radii may suggest that it was born farther from the Galactic centre than its current $\rgc\approx\rguide$ $\sim$ 6.1 kpc. In this scenario, its low orbital eccentricity would suggest it is a strong candidate to have undergone inward radial migration (churning). This comparison, however, is affected by the fact that a large fraction of inner-disc OCs can have undergone strong radial mixing and, thus, now be located far from their $\rbirth$.
UBC~1052 stands out not only as a possible inward-migrating cluster, but also as a chemically peculiar one. Its unique abundance pattern (in particular, its neutron-capture element abundances [Ba/Zr], [Nd/Y], and [Ce/Nd]) makes it a very good candidate for follow-up searches of dispersed member stars that might help reconstruct its past dynamical evolution. Our dynamical analysis shows that UBC~1052’s orbit could be influenced by the Galactic bar.

Our detailed high-precision abundance analysis makes UBC~1052 a benchmark cluster for studying the chemical evolution and radial mixing processes in the inner Galactic disc with the next generation of high-resolution spectroscopic surveys such as WST \citep{Mainieri2024} and HRMOS/VLT \citep{Magrini2023HRMOS}. Many inner-disc OCs still await their discovery \citep{Gupta2024, Hunt2026}, but it will be hard to follow up all new OC discoveries at such a high resolution. Therefore, next-generation survey facilities like WEAVE \citep{Jin2024}, 4MOST \citep{deJong2019, Lucatello2023}, or MOONS \citep{Cirasuolo2020} will conveniently provide homogeneous chemical abundance data for OCs as well as field stars over a wide range of the Galactic disc at intermediate spectral resolution.

\section*{Data availability}
Tables~\ref{table_selected_lines}, \ref{table_vr_APs_mean_star_abunds}, \ref{table_mean_cluster_abunds}, and \ref{table_giraffe_vr_extract} are available in their entirety in electronic format at the CDS via anonymous ftp to \href{cdsarc.u-strasbg.fr}{cdsarc.u-strasbg.fr} (130.79.128.5) or via \href{http://cdsweb.u-strasbg.fr/cgi-bin/qcat?J/A+A/}{http://cdsweb.u-strasbg.fr/cgi-bin/qcat?J/A+A/}.

\begin{acknowledgements}
Based on observations collected at the European Organisation for Astronomical Research in the Southern Hemisphere under ESO programme 111.255H.001.
      
This work was partially supported by the Spanish MICIN/AEI/10.13039/501100011033 and by ‘ERDF A way of making Europe’ by the European Union through grant PID2021-122842OB-C21 and PID2024-157964OB-C21, and the Institute of Cosmos Sciences University of Barcelona (ICCUB, Unidad de Excelencia Mar\'{\i}a de Maeztu) through grant CEX2024-001451-M and the project 2021-SGR-00679 GRC de l'Agència de Gestió d'Ajuts Universitaris i de Recerca (Generalitat de Catalunya). JD acknowledges funding from Universitat de Barcelona’s predoctoral researcher recruitment programme through grant PREDOCS-UB 2022, as well as financial support through the Ambassade de France en Espagne (‘appel à projets scientifiques 2024: Bourses doctorales et post-doctorales’). FA acknowledges funding from MCIN/AEI/10.13039/501100011033 and European Union NextGenerationEU/PRTR through grant RYC2021-031638-I. ACG acknowledges funding by the Horizon Europe HORIZON-CL4-2023-SPACE-01-71 SPACIOUS project funded under Grant Agreement no. 101135205.

This work has made use of data from the European Space Agency (ESA) mission {\it Gaia} (\url{https://www.cosmos.esa.int/gaia}), processed by the {\it Gaia} Data Processing and Analysis Consortium (DPAC, \url{https://www.cosmos.esa.int/web/gaia/dpac/consortium}). Funding for the DPAC has been provided by national institutions, in particular, the institutions participating in the {\it Gaia} Multilateral Agreement.
\end{acknowledgements}

\bibliographystyle{aa}
\bibliography{biblio}

\begin{appendix}

\section{Details of the spectroscopic analysis}\label{app_sec:differential_abunds}

\subsection{Selection of spectral lines}
\label{app_subsec:method_line_selection}

\begin{table*}[h!]
\centering
\small
\setlength{\tabcolsep}{3.4pt}
\begin{tabular}{ccccccccc}
\hline \hline
\small{Element} & \thead{$\lambda$ \\ $\textrm{[nm]}$}& \thead{log $gf$ \\ \,}&  \thead{ \scriptsize{A(X) \textit{Gaia} DR3}\\ {\scriptsize 4286764462544613120}} &  \thead{ \scriptsize{A(X) \textit{Gaia} DR3}\\ {\scriptsize 4286764703062797952}} &  \thead{ \scriptsize{A(X) \textit{Gaia} DR3}\\ {\scriptsize 4286766932128889472}} &  \thead{ \scriptsize{A(X) \textit{Gaia} DR3}\\{\scriptsize 4286776522812820224}} &  \thead{ \scriptsize{A(X) \textit{Gaia} DR3}\\ {\scriptsize 4286778137720526080}} &  \thead{ \scriptsize{A(X) \textit{Gaia} DR3}\\ {\scriptsize 604914949295282816}} \\
\hline
\ion{Fe}{I} & 491.8012 & -1.26 & 7.26 & 7.27 & 7.27 & 7.25 & 7.28 & 7.45 \\
\ion{Fe}{I} & 491.8994 & -0.342 & 7.39 & 7.38 & 7.42 & 7.39 & 7.46 & 7.40 \\
... & ... & ... & ... & ... & ... & ... & ... & ... \\
\hline
\end{tabular}
\caption{Preview of the table containing the spectral lines used in the differential analyses of the RC stars in UBC~1052 and of the solar analogue in M~67. For each line we indicate the element, the wavelength of the transition in air, the logarithm (base 10) of the product of the oscillator strength of the transition and the statistical weight of the lower level (taken from \citealt{heiter+2021}), and the estimated absolute abundance for the five RC stars in UBC~1052 (fourth to eighth columns) and for the solar analogue in M~67 (last column). The full table is available at the CDS.}
\label{table_selected_lines} 
\end{table*}

To compute for each star its mean differential abundance of a chemical species X, we only used some of the lines among all the ones for which we were able to measure $A_{{\rm X}_i, ~\rm star} - A_{{\rm X}_i, ~\rm ref}$. The selection was done in the following way: for the species with at least two available lines, we retained only the lines with an uncertainty in the absolute abundance $\delta A_{{\rm X}_i, ~\rm star} < 0.25$ dex. Then, we discarded the lines with ‘\texttt{gf\_flag}’ = N in \citealt{heiter+2021} (when such flag was available, i.e. for the lines belonging to their ‘preselected line
list’), since their transition probability was found to have low accuracy. But, given that we are performing a line-by-line differential analysis, we did not directly reject all the lines that are not recommended in \citet{heiter+2021} due to blending (‘\texttt{synflag}’ = N). For the elements with at least 10 lines with ‘\texttt{gf\_flag}’ different from N, we additionally performed an automatic rejection of the discrepant lines in most of the studied stars. For the elements with less than 10 lines with ‘\texttt{gf\_flag}’ different from N, on the other hand, the final selection was performed visually, assessing the goodness of the synthetic spectrum’s fit to each observed line (both for the studied star and the reference one). We took into consideration the number of available lines for that species, their ‘\texttt{gf\_flag}’ and ‘\texttt{synflag}’ flags in \citet{heiter+2021} and the discrepancies in their $A_{{\rm X}_i, ~\rm star}$ to make this visual line selection, which is less restrictive for the elements with very few lines. The final selected lines for the five studied RC stars in UBC~1052 and for the solar analogue M67-1194 and their respective estimated absolute abundances are detailed in Table \ref{table_selected_lines}.

\subsection{Uncertainty estimation for the chemical abundances}
\label{app_subsec:method_uncertainties_abunds}

The uncertainty in the mean differential abundance of a chemical species X for a certain star was computed as
\begin{equation}
    \delta\textrm{[X/H]}_{\textrm{star wrt ref}} = \sqrt{ { \left( \sigma_{X,\textrm{star wrt ref}}^{ \{ \textrm{flux}\} } \right)}^2  +  { \left( \sigma_{X,\textrm{star wrt ref}}^{ \{ \textrm{APs} \} } \right)}^2} ,
\label{eq:mean_star_XoverH_diff_analysis_err}
\end{equation}
where each term is the sample standard deviation of the differential abundances derived for the selected lines under different conditions:

\begin{equation}
    \sigma_{X,\textrm{star wrt ref}}^{ \{ \textrm{set} \} } = \sqrt{ \frac{ \sum_{i=1}^{N_{\textrm{lines X}}^{ \{ \textrm{set} \}}} \left[ (A_{{\rm X}_i,\rm star} - A_{{\rm X}_i,\rm ref}) - ( \overline{A_{{\rm X}_i,\rm star} - A_{{\rm X}_i,\rm ref}} )  \right]^2}{N_{\textrm{lines X}}^{ \{ \textrm{set} \} }-1}}. 
\label{eq:standard_dev_diff_abunds_set_lines}
\end{equation}

\{set\} = \{flux\}: This term takes into account how the flux errors influence the derived differential abundances. Absolute abundances were derived as explained in Sect.~\ref{subsec:method_diff_abunds} for 10 random realisations of each spectrum (both the studied star’s and the reference’s), which were created sampling randomly from the flux errors assuming they follow a Gaussian distribution. The set of lines includes the selected lines for each of these 10 random realisations (i.e. each selected line appears 10 times, and each time a differential abundance has been derived for a slightly different spectrum).

\{set\} = \{APs\}: This term quantifies the errors in the differential abundances due to uncertainties in the stellar APs of the studied star. In addition to determining abundances for the studied star assuming the nominal APs derived in Sect.~\ref{subsec:method_APs}, abundances were also derived for the same observed spectrum assuming four different combinations of the APs: ($\teff + \delta \teff$, $\logg + \delta \logg$), ($\teff - \delta \teff$, $\logg + \delta \logg$), ($\teff + \delta \teff$, $\logg - \delta \logg$), and ($\teff - \delta \teff$, $\logg - \delta \logg$). The set of lines considered to compute this dispersion are the selected lines for these four analyses as well as the selected lines for the analysis with the nominal APs (i.e. each line appears five times for the same observed spectrum but different APs are assumed to derive its abundance).

Therefore, $\delta\textrm{[X/H]}_{\textrm{star wrt ref}}$ is a standard deviation of the differential abundances for this species that includes the abundance scatter over the available lines for this species (if it has more than one selected line), as well as the abundance scatter for each line resulting from the uncertainties in the stellar APs and the spectrum flux. In order to provide conservative uncertainties on the mean stellar abundances, we choose to measure them as this standard deviation instead of a standard error.

\subsection{Final solar-scaled abundances}
\label{app_sec:solar_scaled_abunds}

\begin{figure}
\includegraphics[width=.49\textwidth]{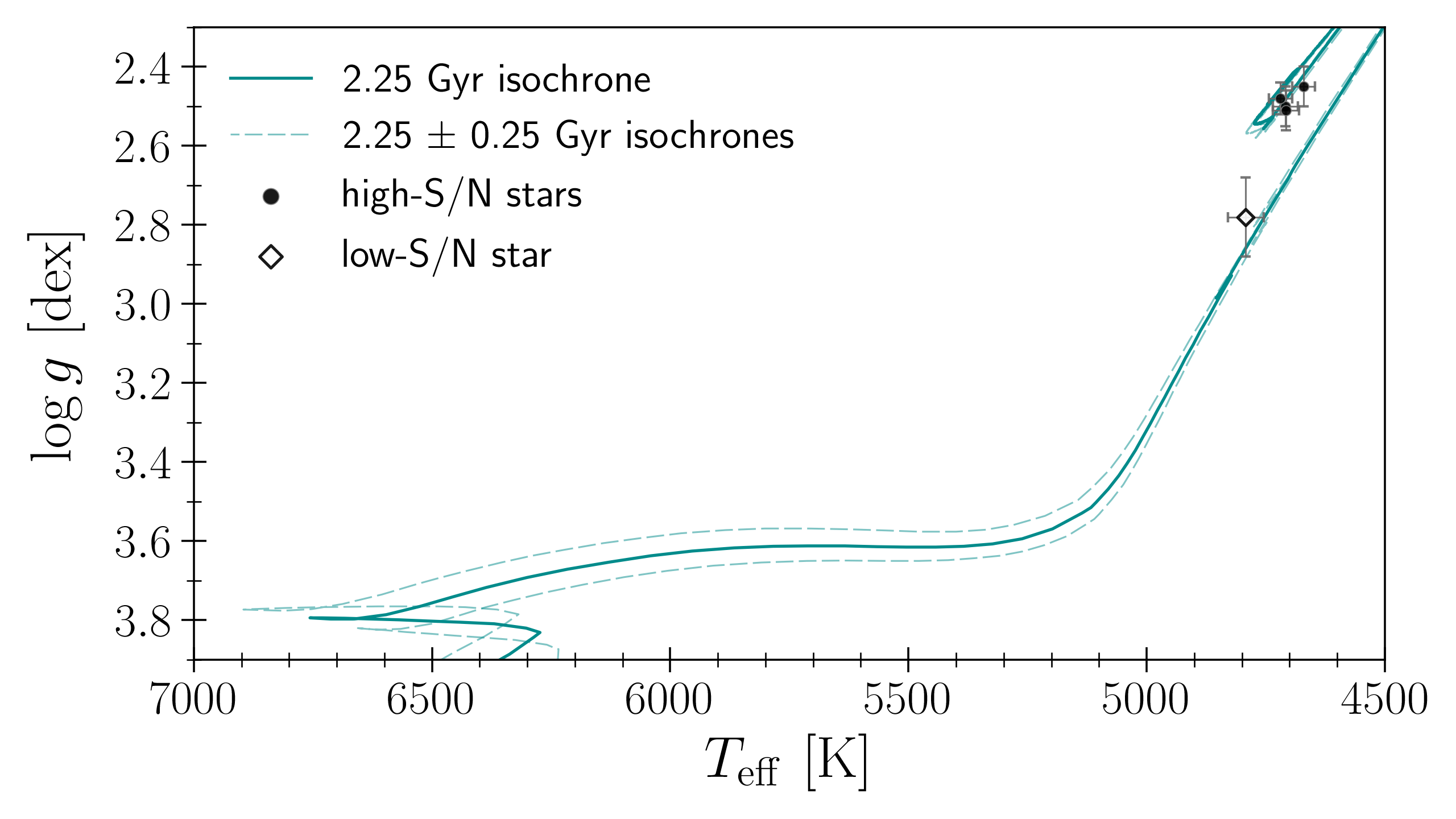} 
\caption{Kiel diagram of the five RC stars in UBC~1052 observed with UVES. [M/H] $=+0.05$ isochrones of ages 2.25 Gyr (solid curve) and 2 and 2.5 Gyr (dashed curves) are overplotted.}
\label{fig:kiel_diagram}
\end{figure}

The radial velocities, APs, and solar-scaled abundances for the five RC stars in UBC~1052 studied with UVES and for the solar analogue in M~67 are found in Table \ref{table_vr_APs_mean_star_abunds}. The Kiel diagram of the five RC stars in UBC~1052 is shown in Fig.~\ref{fig:kiel_diagram}, with an isochrone of the derived cluster age and metallicity equal to our derived iron abundance overplotted. The solar-scaled mean cluster abundances for UBC~1052 are found in Table \ref{table_mean_cluster_abunds}. The abundances of the RC stars and the mean cluster abundances have been discussed in the main text (Sects.~\ref{subsec:abundances_chemical_homogeneity} and \ref{sec:discussion}). Here, we discuss our results for the spectroscopic analysis of the solar analogue in M~67.

\begin{table*}[h!] 
\centering
\small
\begin{tabular}{ccccccccccc} 
\hline \hline
\textit{Gaia} DR3 source ID & \thead{$\vr$ \\ $[\kms]$ } & \thead{$\teff$ \\ $\textrm{[K]}$} & \thead{$\logg$ \\ $\textrm{[dex]}$} & \thead{$\vmic$ \\ $[\kms]$} & \thead{N\_Fe \\ \,} & \thead{[Fe/H] \\ $\textrm{[dex]}$} & \thead{N\_Ca \\ \,} &  \thead{[Ca/H] \\ $\textrm{[dex]}$} &  \thead{[Ca/Fe] \\ $\textrm{[dex]}$} & ...\\
\hline
4286764462544613120 & 33.46$\pm$0.25 & 4709$\pm$26 & 2.50$\pm$0.05 & 1.31$\pm$0.04 & 77 & 0.06$\pm$0.06 & 19 & 0.07$\pm$0.07 & 0.01$\pm$0.10 &...\\
4286764703062797952 & 34.03$\pm$0.24 & 4707$\pm$28 & 2.51$\pm$0.05 & 1.31$\pm$0.05 & 75 & 0.05$\pm$0.06 & 19 & 0.05$\pm$0.06 & -0.01$\pm$0.08 & ...\\
4286766932128889472 & 33.69$\pm$0.24 & 4671$\pm$23 & 2.45$\pm$0.05 & 1.37$\pm$0.05 & 72 & 0.06$\pm$0.06 & 20 & 0.05$\pm$0.05 & -0.01$\pm$0.08 &...\\
4286776522812820224 & 34.75$\pm$0.24 & 4719$\pm$24 & 2.48$\pm$0.04 & 1.38$\pm$0.05 & 76 & 0.04$\pm$0.05 & 20 & 0.04$\pm$0.05 & -0.01$\pm$0.07 &...\\
4286778137720526080 & 40.41$\pm$0.32 & 4792$\pm$38 & 2.78$\pm$0.10 & 1.12$\pm$0.08 & 56 & 0.10$\pm$0.12 & 19 & 0.09$\pm$0.10 & -0.01$\pm$0.16 &...\\
604914949295282816 & 34.56$\pm$0.30 & 5752$\pm$28 & 4.47$\pm$0.03 & 1.11$\pm$0.04 & 106 & -0.02$\pm$0.02 & 16 & -0.03$\pm$0.02 & -0.01$\pm$0.03 &...\\
\hline
\end{tabular} 
\caption{Preview of the table containing the radial velocities, APs, number of spectral lines used for each element X (N\_X), and mean elemental abundances with respect to the Sun for the five RC stars in UBC~1052 (first five rows) and for the solar analogue in M~67 (last row). The abundances for {\it Gaia} DR3 4286778137720526080 are less precise and accurate than for the other four stars in UBC~1052 and have not been considered to compute the mean cluster abundances in Table~\ref{table_mean_cluster_abunds}. The full table containing all the 23 studied elements is available at the CDS.}
\label{table_vr_APs_mean_star_abunds}
\end{table*}

\begin{table}[h!] 
\centering
\begin{tabular}{lrrc} 
\hline \hline
Element & [X/H] [dex] & [X/Fe] [dex] & Number of lines \\ 
\hline
\ion{Fe}{I} & $0.05 \pm 0.01$ &  & $78$ \\
\ion{Na}{I} & $0.21 \pm 0.02$ & $0.15 \pm 0.02$ & $4$ \\
\ion{Mg}{I} & $0.05 \pm 0.01$ & $-0.01 \pm 0.01$ & $6$ \\
\ion{Al}{I} & $0.10 \pm 0.01$ & $0.04 \pm 0.01$ & 3 \\
\ion{Si}{I} & $-0.01 \pm 0.02$ & $-0.07 \pm 0.02$ & $7$ \\
\ion{Ca}{I} & $0.05 \pm 0.02$ & $-0.01 \pm 0.01$ & $20$ \\
... & \multicolumn{1}{c}{...} & \multicolumn{1}{c}{...} & ...  \\
\hline
\end{tabular} 
\caption{Preview of the table containing the mean cluster abundances with respect to the Sun for UBC~1052 and the total number of different spectral lines used for each element. The full table containing the 23 elements (all studied through a differential abundance analysis except for Na) is available at the CDS.}
\label{table_mean_cluster_abunds} 
\end{table}

The stellar parameters we derive for the solar-type star in M~67 (M67-1194) are $\teff = 5752 \pm 28$ K and $\logg = 4.47 \pm 0.03$ dex, in excellent agreement with the solar values and with its estimated parameters in \citealt{onehag+2011} ($\teff = 5780 \pm 27$ K and $\logg = 4.44 \pm 0.035$ dex) and in \citealt{Liu+2016} ($\teff = 5786 \pm 13$ K and $\logg = 4.46 \pm 0.02 $ dex). These two studies carried out a strict line-by-line differential chemical abundance analysis of M67-1194 with respect to the Sun and found that, in addition to its APs, its chemical pattern is also very similar to that of the Sun. Performing the same type of analysis, we find that M67-1194 has [Fe/H] = $-0.02 \pm 0.02$ dex, compatible within 1.4$\sigma$ with the value derived in \citealt{onehag+2011} ($0.023 \pm  0.015$ dex) and within 0.5$\sigma$ with the value derived in \citealt{Liu+2016} ($-0.005 \pm 0.010$ dex), and in agreement with M~67's average cluster [Fe/H] which is found to be very similar to the solar one (e.g. [Fe/H] = $0.00 \pm 0.02$ dex in \citealt{Randich+2022}). Our derived [X/Fe] abundances are 1$\sigma$ compatible with those in \citet{Liu+2016} for all the studied elements except for Ba, Co, Eu, and Pr, which are compatible within 2$\sigma$.

\section{Bounds on UBC~1052's chemical homogeneity}
\label{app_sec:ubc1052_xoverH_dispersion_pryormeylan}

The observed abundance dispersion in UBC~1052 is a superposition of the true abundance dispersion in the cluster and the measurement uncertainties. To measure the underlying abundance distribution in the cluster we follow \citet{PryorMeylan1993}. We assume that each ${\textrm{[X/H]}_{i}}$ measurement for a certain star $i$ in UBC~1052 and element X, with an estimated uncertainty $\sigma_{{\textrm{[X/H]}}_i}$ (as defined in Sect.~\ref{subsec:method_diff_abunds}), is drawn from the normal distribution:
\begin{multline}
    f({\textrm{[X/H]}}_i) = \Biggl[ 2\pi \biggl[ {\left( \sigma_{\textrm{[X/H]}}^{\textrm{UBC~1052}} \right)}^2 + {\left( \sigma_{{\textrm{[X/H]}}_i} \right)}^2 \biggr] \Biggr] ^{-1/2} \cdot \\ \exp \left( - \frac{ \Bigl( {\textrm{[X/H]}}_i - {\textrm{[X/H]}}^{\textrm{UBC~1052}} \Bigr) ^2}{2 \Bigl[ {\left( \sigma_{\textrm{[X/H]}}^{\textrm{UBC~1052}} \right)}^2 + {\left( \sigma_{{\textrm{[X/H]}}_i} \right)}^2 \Bigr]} \right),
\end{multline}
where ${\textrm{[X/H]}}^{\textrm{UBC~1052}}$ and $\sigma_{\textrm{[X/H]}}^{\textrm{UBC~1052}}$ are the parameters to be estimated: mean [X/H] abundance with respect to the Sun and intrinsic [X/H] abundance dispersion in the cluster (respectively). The log-likelihood function is thus:
\begin{multline}
    \textrm{log} \mathcal{L} = -\frac{1}{2} \sum_{i=1}^{N_{\textrm{stars}}} \biggggl\{ \frac{ \left( \textrm{[X/H]}_{i} -  \textrm{[X/H]}^{\textrm{UBC~1052}} \right)^2}{ {\left( \sigma_{\textrm{[X/H]}}^{\textrm{UBC~1052}} \right)}^2 + {\left( \sigma_{{\textrm{[X/H]}}_i} \right)}^2} + \\ \log \left( 2\pi \left[{\left( \sigma_{\textrm{[X/H]}}^{\textrm{UBC~1052}} \right)}^2 + {\left( \sigma_{{\textrm{[X/H]}}_i} \right)}^2 \right] \right) \biggggr\},
\end{multline}
\label{eq:loglikelihood_ubc1052_XoverH_dispersion}
with $N_{\textrm{stars}}=4$ RC stars in UBC~1052 with UVES spectra.

\begin{figure*}
\begin{center}
{\includegraphics[width = 0.99\textwidth]{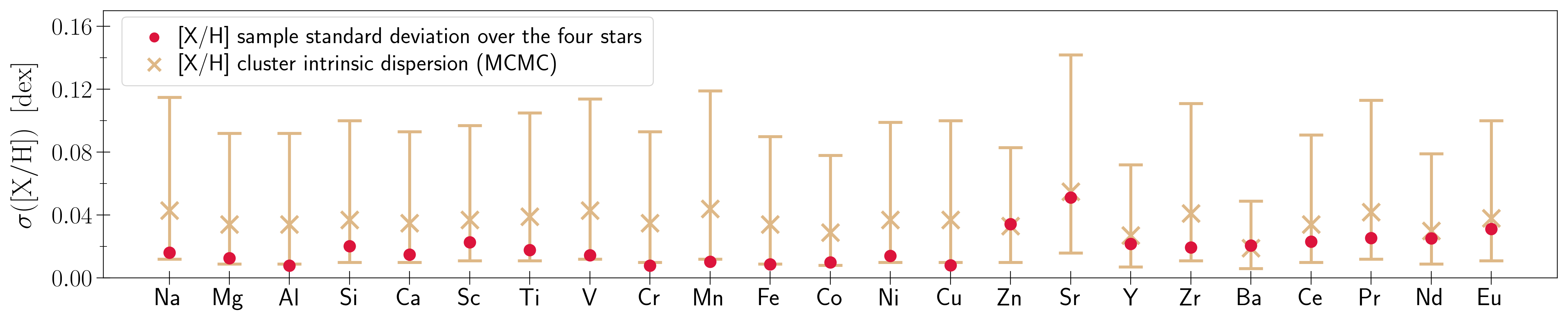}}
\caption{Estimated intrinsic [X/H] abundance dispersions in UBC~1052 (light brown series, crosses are the median of the posterior and the errorbars are the 16th and 84th percentiles) based on the abundances for 23 elements derived for the four high-S/N RC stars studied with UVES and their uncertainties. The standard deviations of the [X/H] values for these four stars are shown as red circles.}
\label{fig:XoverH_cluster_dispersion_Pryor93_mcmc}
\end{center}
\end{figure*}

For each studied element, we sampled this likelihood function using the Markov Chain Monte Carlo code \texttt{emcee} \citep{Foreman-Mackey+2013}, using the uninformative flat priors |${\textrm{[X/H]}}^{\textrm{UBC~1052}}$| < 1 dex and $\sigma_{\textrm{[X/H]}}^{\textrm{UBC~1052}}$ > 0 dex, and using 50 walkers with 10\,000 steps each and \texttt{burnin} = 500. As the best estimate for UBC~1052’s true [X/H] dispersion we use the median of the posterior distribution, and its uncertainty is measured with the 16th and 84th percentiles. The results are represented in Fig.~\ref{fig:XoverH_cluster_dispersion_Pryor93_mcmc}, where they are compared with the abundance dispersion computed as the standard deviation of the [X/H] values over the four stars with high-S/N UVES spectra.

\section{Comparison with \citet{casamiquela+2021}}
\label{app_sec:comparison_casamiquela21_47OCs}

In Appendix~\ref{app_subsec:comparison_analysis_casamiquela21RC} we compare our analysis with that in \citetalias{casamiquela+2021}, and in Appendix~\ref{app_subsec:comparison_abunds_casamiquela21} we compare the [X/H] and [X/Fe] abundances we derive for UBC~1052 with those of the OCs studied in \citetalias{casamiquela+2021}.

\subsection{Comparison of our spectroscopic analysis and the one in \citetalias{casamiquela+2021}}
\label{app_subsec:comparison_analysis_casamiquela21RC}

\citetalias{casamiquela+2021} analysed high-resolution spectra of RC stars belonging to 47 OCs and derived 1D, LTE abundances of 25 chemical species. The ages of the studied OCs range between $\sim$120 Myr and $\sim$7 Gyr, with 80\% of the OCs being younger than 2 Gyr; and their $\rgc$ range between $\sim$6.5 kpc and $\sim$11.5 kpc, with only 5 OCs at  $\rgc<7.5$ kpc (considering the ages and  $\rgc$ in \citealt{Cantat-Gaudin2020}).
Their study and the present one use the same pipeline that runs \ispec to carry out a strict line-by-line differential analysis through spectral synthesis fitting, and the radiative transfer code and atmospheric models used are also the same. The atomic line list used is from \gaiaeso in both cases, but in \citetalias{casamiquela+2021} they used an older version \citep{Heiter+2015}. Despite the similarities between both spectroscopic analyses, there are two remarkable differences.

First, in \citetalias{casamiquela+2021} the Hyades was used as the auxiliary cluster to obtain solar-scaled abundances. The mean absolute differences in $\teff$ and $\logg$ between their full sample of RC stars (considering all OCs) and the Hyades reference star are 116 K and 0.21 dex (respectively); and the maximum differences are 460 K and 0.9 dex. In the present study, the $\teff$ and $\logg$ of the reference star in M~67’s RC are much more similar to those of the RC stars in UBC~1052: mean absolute difference of 17 K and 0.02 dex, and maximum absolute difference of 48 K and 0.03 dex (considering only the four high-S/N stars in UBC~1052).

Second, in the present study the spectra involved in the differential analysis have been acquired with the same instrument (UVES) for both the studied stars and the reference star, thus enabling a minimisation of instrumental effects. This is not the case for \citetalias{casamiquela+2021}: the spectra of the studied RC stars were acquired with nine different spectrographs, and to compute the differential abundance for each of their lines, the considered line abundance of the Hyades reference star was not derived from a single spectrum (instead, it was computed as the average of the abundances for that line obtained for two spectra of this star acquired with two different spectrographs).

These differences, alongside discrepancies in the final sample of selected lines, may cause some offsets between the abundances we derive for UBC~1052 and those derived for the OCs in \citetalias{casamiquela+2021}. The elements for which the abundances have been derived using exactly the same lines in \citetalias{casamiquela+2021} and in our differential analysis of the RC stars in UBC~1052 with respect to the RC star in M~67 are: Al, Ce, Cu, and Zn. Regarding the uncertainties in the individual stars’ abundances, we expect them to be smaller for our study, tailored to UBC~1052, than the ones in \citetalias{casamiquela+2021}. As explained in Appendix~\ref{app_subsec:method_uncertainties_abunds}, the uncertainties we report take into account the effect of the uncertainties in the measured flux and in the stars’ APs, but these contributions were not considered in \citetalias{casamiquela+2021}. Therefore, the uncertainties for the RC stars in UBC~1052 are more conservative. Still, we find them to be comparable to the mean abundance uncertainties for the stars in \citetalias{casamiquela+2021} for all ions except for most of the neutron-capture elements, which have significantly smaller uncertainties for UBC~1052.

\subsection{Comparison of [X/H] and [X/Fe] abundances between UBC~1052 and the clusters studied in \citetalias{casamiquela+2021}}
\label{app_subsec:comparison_abunds_casamiquela21}

\begin{figure*}[h!]
{\includegraphics[width = 0.99\textwidth]{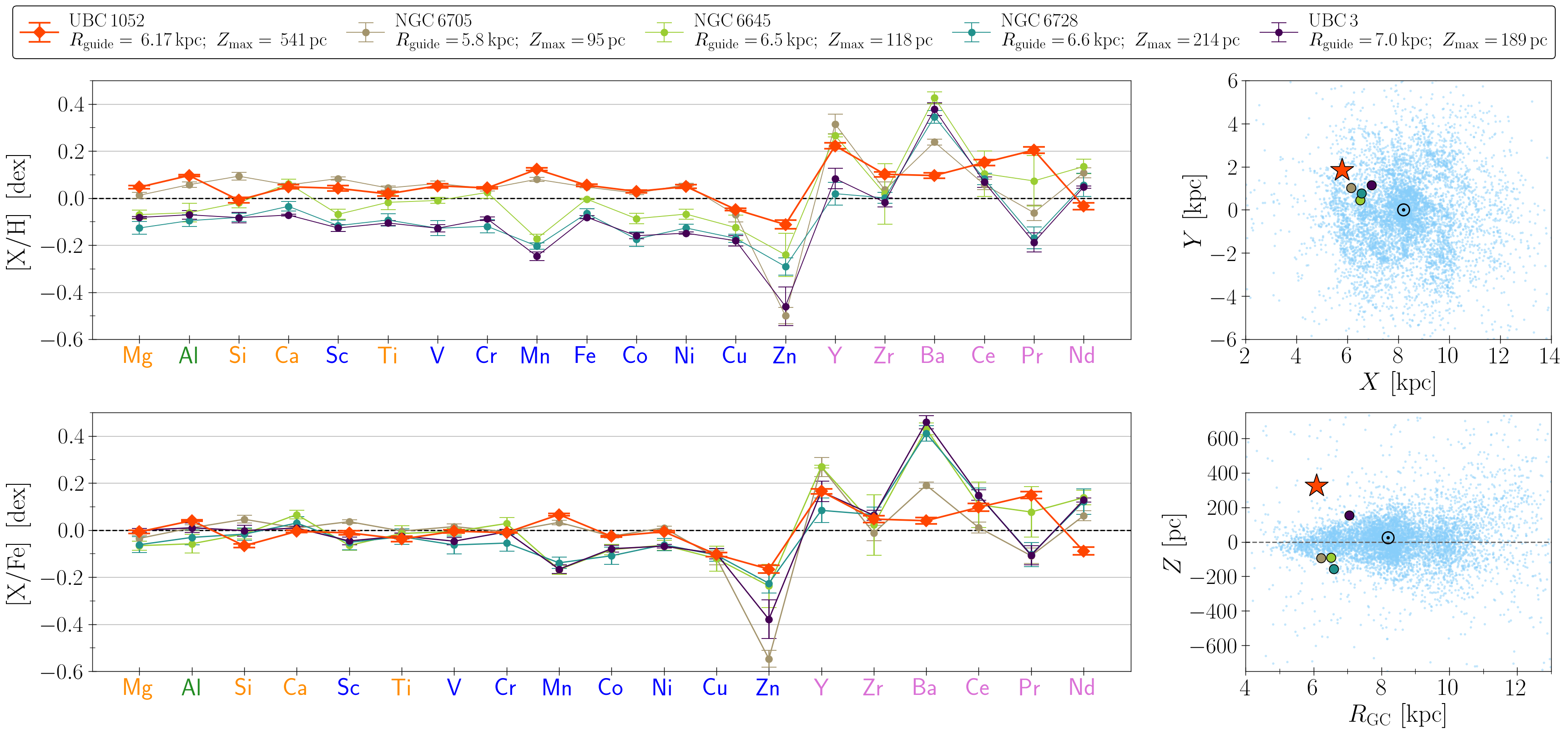}}
\caption{\textit{Left}: Mean cluster abundances (\textit{top}: [X/H]; \textit{bottom}: [X/Fe]) for UBC~1052 (red series), and for the OCs studied in \citetalias{casamiquela+2021} with similar guiding-centre radii (indicated in the legend). Only the elements in common in both studies are represented, and they are colour-coded according to the dominant nucleosynthesis channel as in Fig.~\ref{fig:individual_stars_abunds}. The abundance uncertainties correspond to the standard error.
\textit{Right}: Spatial distribution (\textit{top}: projection onto the $X-Y$ galactocentric coordinates plane; \textit{bottom}: height above the Galactic plane as a function of galactocentric radius) of all the clusters detected in \citetalias{hunt+2023} (light blue dots) and of the OCs represented in the left panel (with UBC~1052 marked as a red star).}
\label{fig:abunds_ocs_similar_Rguide}
\end{figure*}

\begin{figure*}[h!]
{\includegraphics[width = 0.99\textwidth]{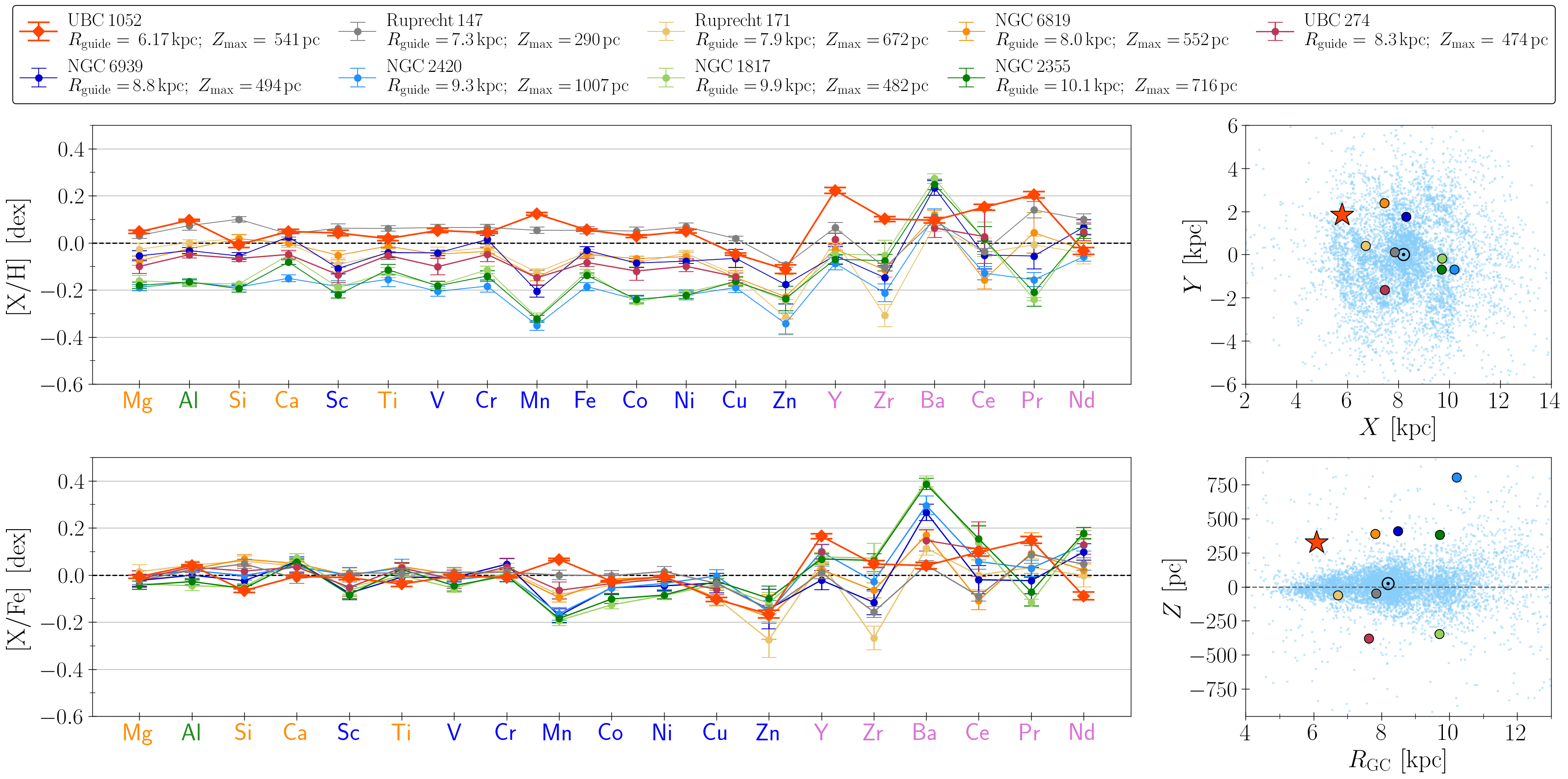}}
\caption{Same as Fig.~\ref{fig:abunds_ocs_similar_Rguide} but now comparing UBC~1052 (red series) with the OCs in \citetalias{casamiquela+2021} with similar age (in the range $\sim[1, 3]$ Gyr) and which have large maximum excursions from the Galactic plane too (indicated in the legend). UBC 274, studied separately in \citet{casamiquela+2022}, verifies this selection criteria and is also included; while Ruprecht 147 is kept for reference (having a similar age to UBC~1052 but small $\zmax$).}
\label{fig:abunds_ocs_similar_age_and_largeZmax}
\end{figure*}

We compare UBC~1052’s [X/H] and [X/Fe] with those of two different subsamples of OCs studied in \citetalias{casamiquela+2021} (Figs.~\ref{fig:abunds_ocs_similar_Rguide} and \ref{fig:abunds_ocs_similar_age_and_largeZmax}). To define these samples, we first estimate $\rguide$ and $\zmax$ for the OCs in \citetalias{casamiquela+2021} using {\tt galpy} and the potential {\tt MWpotential2014}, in the same fashion as done for UBC~1052 (see Sect.~\ref{subsec:vr_orbital_parameters}), adopting the clusters’ $\alpha$, $\delta$, $\mu_{{\alpha}^*}$, $\mu_{\delta}$, and distance from \citet{Cantat-Gaudin2020} and mean $\vr$ from \citet{Tarricq+2021}.

In Fig.~\ref{fig:abunds_ocs_similar_Rguide}, we show the OCs in \citetalias{casamiquela+2021} with the smallest guiding-centre radii ($\rguide \leq 7$ kpc), having discarded the old ($\sim8$ Gyr), super-metal-rich OC NGC~6791 for clarity. This sample, referred to as the inner-disc OCs, comprises four OCs located at $\rgc < 7$ kpc in \citet{Cavallo+2024} which are all much younger than UBC~1052.

In Fig.~\ref{fig:abunds_ocs_similar_age_and_largeZmax}, we compare UBC~1052’s abundances with those of the OCs in \citetalias{casamiquela+2021} that have similar ages (approximately in the range $\sim[1, 3]$ Gyr) and also large maximum excursions from the Galactic plane ($\zmax > 470$ pc). We also represent UBC~274, an OC that meets these selection criteria and whose abundances were derived in a separate study \citep{casamiquela+2022} exactly as done in \citetalias{casamiquela+2021} (the only difference being the version of the \gaiaeso line list used). Additionally, as a reference, we also plot Ruprecht~147 (NGC~6774, in grey), which has a similar age to UBC~1052 ($2.5 \ - \ 3$ Gyr; \citealt{Curtis+2013}, \citealt{Bragaglia+2018}, \citealt{Cantat-Gaudin2020}) but is located in the solar neighbourhood (at a distance $d\sim300$ pc, \citealt{Curtis+2013}, \citealt{GaiaCollaboration2018Babusiaux}, \citealt{Cantat-Gaudin2020}) and has a smaller $\zmax$. We refer to this sample as the similarly aged OCs.

The mean cluster abundances of all elements up to Cu in both samples span a small range of $\sim[-0.2, 0.1]$ dex both in [X/H] (except for Mn) and in [X/Fe]. The [Fe/H] spread, in particular, is small ($\sim0.15$ dex) for the inner-disc OCs; and slightly larger ($\sim0.25$ dex) for the sample of similarly aged OCs, which includes metal-poorer OCs in the outer part of the Galaxy up to $\rgc \sim 10$ kpc. For the neutron-capture elements and Zn, the spread between the clusters’ mean abundances is larger. For UBC~1052, the abundances of these elements (except Zr) rely on a single line in the differential analysis with respect to M~67, and most of them have also been derived using only one line in the auxiliary differential analysis of the solar analogue in M~67. Hence, they are potentially less accurate. Several of these elements have the largest uncertainties in the individual stars’ abundances for the OCs studied in \citetalias{casamiquela+2021}.

The error bars represented in Figs.~\ref{fig:abunds_ocs_similar_Rguide} and \ref{fig:abunds_ocs_similar_age_and_largeZmax} correspond to the standard error over the cluster members' abundances. Despite these small observed abundance spreads in each cluster and the span in the clusters' $\rgc$, $Z$, and age, the overlap in the chemical signatures in both figures is remarkable, especially for elements up to Cu (see bottom left panels). The exception is [Mn/Fe], which tends to be higher for OCs with higher [Fe/H]. \citet{BattistiniBensby2015} also found this trend in our [Fe/H] range in the LTE abundances of F and G dwarfs in the solar neighbourhood, but with a considerable [Mn/Fe] scatter of $\sim$ 0.2 dex. However, when they applied non-LTE corrections the trend became almost flat (for instance, non-LTE corrections for Mn for RC stars in the solar-metallicity OC M~67 are $\sim$ 0.3 dex, \citealt{souto+2019}). In our case, however, we still observe the trend despite having performed a differential analysis that mitigates the effects of the departure from LTE. Therefore, even though the OCs in the two samples cover a relatively wide age range, and also in spite of the wide $\rgc$ and $\rguide$ ranges of the OCs in Fig.~\ref{fig:abunds_ocs_similar_age_and_largeZmax}, it is difficult to distinguish one OC from the other, especially when no neutron-capture elements are considered. \citet{casamiquela+2021tagging} noted that different stellar birth sites can have overlapping chemical signatures, even when high-resolution abundances of many different nucleosynthesis channels are used.

UBC~1052 is the innermost OC in both samples and has the highest [Fe/H], compatible with that of NGC~6705 and Ruprecht~147. These two OCs, however, show substantial differences with respect to UBC~1052. NGC~6705 (also known as M~11) is located at a similar $\rgc$ to UBC~1052 but is much younger ($\sim300$ Myr old; \citealt{Cantat-Gaudin+2014}; \citealt{Dias+2021}) and reaches a smaller $\zmax$ ($\sim 100$ pc). Ruprecht~147, on the other hand, is slightly older than UBC~1052, but it is located further out in the Galaxy, in the solar neighbourhood, and has a larger $\rguide$ ($\sim 7.3$ kpc). Regarding [X/Fe], UBC~1052’s abundances are compatible with those of NGC~6705 and Ruprecht~147 only for some elements (including most of the Fe-peak ones), and the largest differences are found on average for the neutron-capture elements. For several neutron-capture elements, similar variations in [X/Fe] ($\sim 0.2$ dex or larger) have been found among clusters of the same [Fe/H] \citep[e.g.][]{ReddyLambert2019, Viscasillas+2022, Molero+2023}. UBC~1052 is not enhanced in $\alpha$-elements, with even slightly subsolar [Si/Fe] and [Ti/Fe]. All neutron-capture elements except for Nd have super-solar abundances in terms of both [X/H] and [X/Fe]. UBC~1052’s [Pr/Fe] is the highest compared with the inner-disc and similarly aged OC samples, and its [Nd/Fe] is the lowest compared with both samples. Its [Ba/Fe] is compatible with that of Ruprecht~147, and they have the lowest value considering both samples. UBC~1052’s [Y/Fe] is the highest compared with the similarly aged OCs.

\section{Further discussion of UBC~1052 in the context of the radial metallicity gradient in the inner disc}
\label{app_sec:radial_metallicity_gradient_other_dists_lessreliableOCs}

In this section we expand the discussion in Sect.~\ref{subsec:radial_FeH_gradient}. We now adopt different OC distances and also include the OCs with less reliable [Fe/H] estimates, and investigate how this affects the comparison between UBC~1052 and the inner-disc OCs used to trace the radial metallicity gradient.

\begin{figure*}
\begin{center}
{\includegraphics[width = 0.952\textwidth]{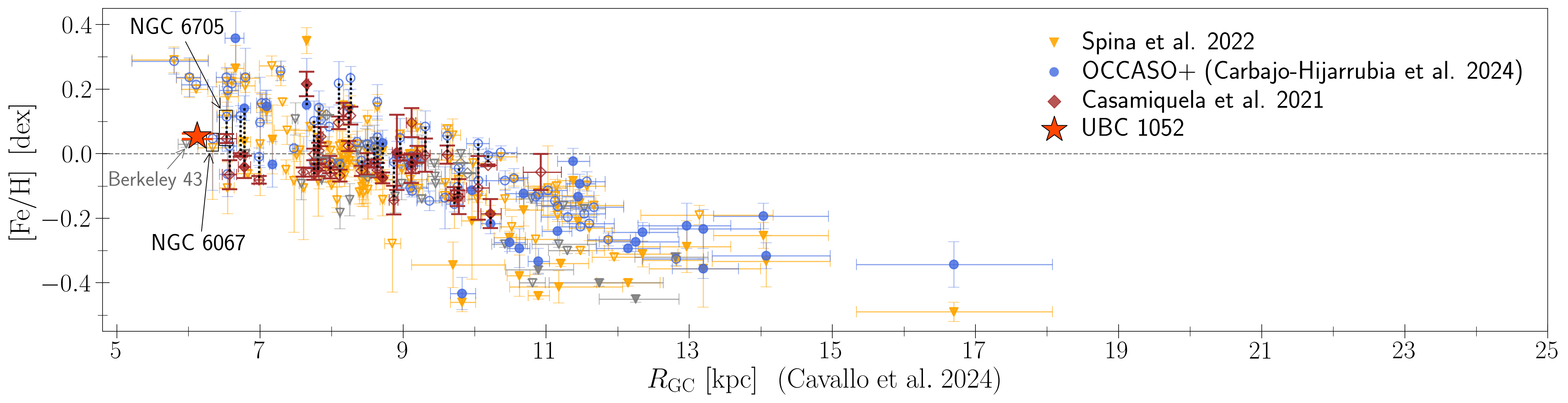}}\\
{\includegraphics[width = 0.952\textwidth]{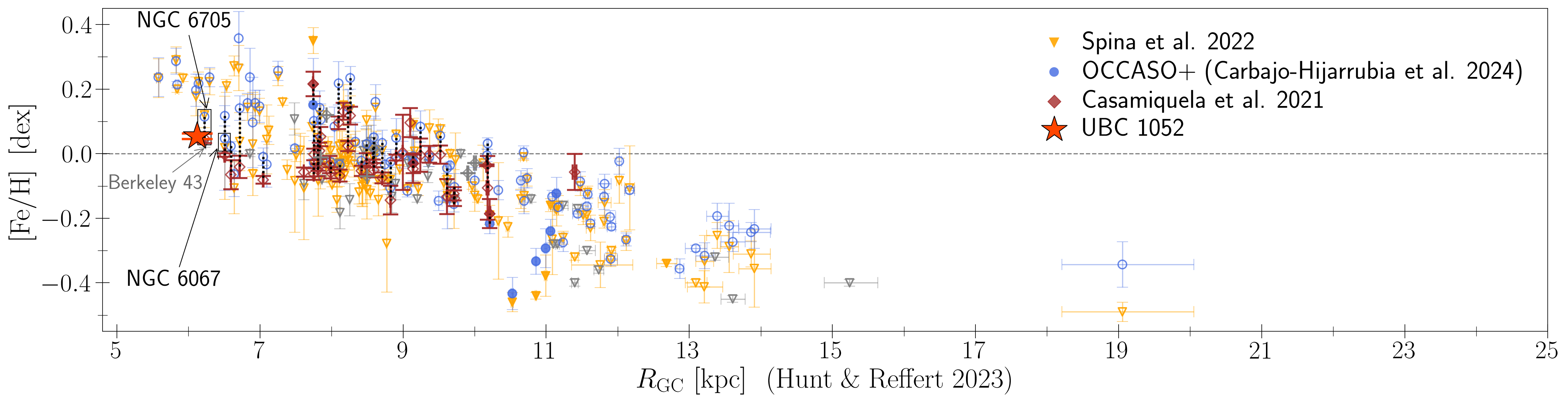}}\\
{\includegraphics[width = 0.952\textwidth]{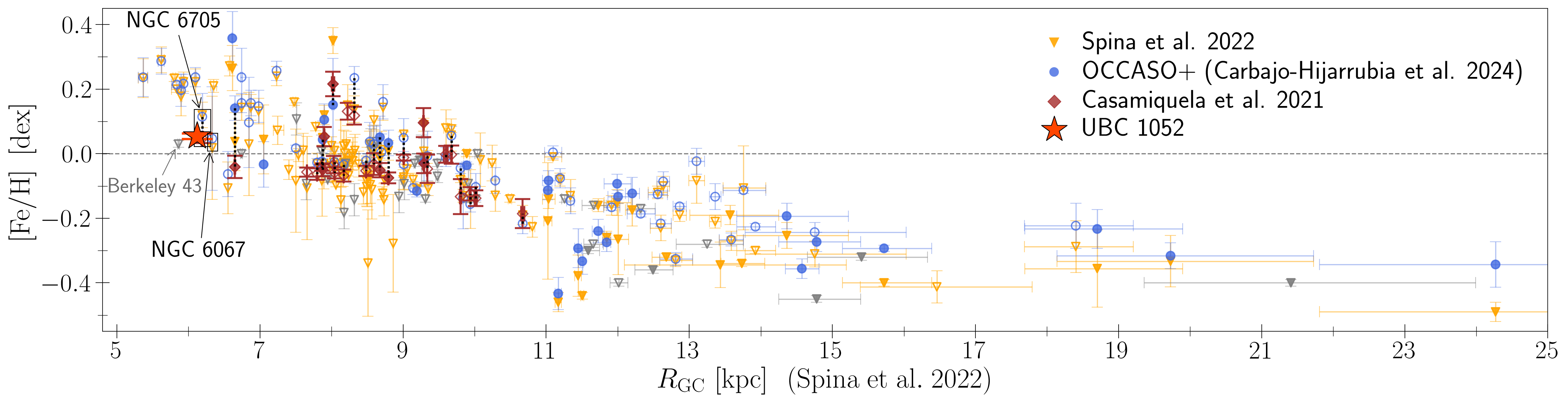}}
\caption{Radial metallicity gradient traced by the OCs in three high-resolution spectroscopic studies (see text for the sample selection) with respect to the galactocentric radius in cylindrical coordinates ($\rgc$) from three different catalogues (\textit{top}: $\rgc$ from \citealt{Cavallo+2024}; \textit{middle}: $\rgc$ from \citetalias{hunt+2023}; \textit{bottom}: $\rgc$ from \citealt{Spina+2022}). UBC~1052 has been added as the red star. Open symbols are OCs younger than 2 Gyr and filled symbols are OCs that are at least 2 Gyr old, the ages are taken from \citealt{Cavallo+2024} (\textit{top}); \citetalias{hunt+2023} (\textit{middle}); and \citealt{Cantat-Gaudin2020} (\textit{bottom}). Grey data points correspond to OCs with less reliable [Fe/H] estimates, derived through only one observed member in \citetalias{casamiquela+2021} or in \citet{Spina+2022}. The represented abundance uncertainties correspond to the standard deviation of [Fe/H] over the studied stars in each cluster.}
\label{fig:FeH_radial_gradient_OCs1member}
\end{center}
\end{figure*}

In Fig.~\ref{fig:FeH_radial_gradient_OCs1member} we show the radial metallicity gradient with respect to $\rgc$ traced by OCs in three high-resolution spectroscopic studies, using three different distance catalogues and distinguishing the OCs that are younger and older than $2$ Gyr according to the ages in three different catalogues. In all the panels, grey data points correspond to OCs with less reliable [Fe/H] estimates, since they have been derived using only one observed member in \citetalias{casamiquela+2021} or in \citealt{Spina+2022} (for the latter, we only represent those OCs with one observed member that were included in their ‘gold’ sample of 180 OCs).

In the top panel, we reproduce the figure in the left panel of Fig.~\ref{fig:FeH_radial_gradient}, now including the less-reliable OCs in grey and without imposing any restriction on the quality of the OCs’ parameters in \citealt{Cavallo+2024} (in contrast to Fig.~\ref{fig:FeH_radial_gradient}, where we only represented those in \citealt{Cavallo+2024} ‘gold’ sample).

The middle panel shows the metallicity gradient with respect to $\rgc$ in \citetalias{hunt+2023}, and the OCs are divided into younger/older than 2 Gyr using the ages in \citetalias{hunt+2023}. In the bottom panel, $\rgc$ is taken from \cite{Spina+2022} and the ages used are those in \cite{Cantat-Gaudin2020}. In the middle and bottom panels, the data points not coloured in grey correspond to the OCs in our selected samples with at least two members with measured [Fe/H] (see the beginning of the second paragraph in Sect.~\ref{subsec:radial_FeH_gradient}) that have also been studied in the catalogues from which $\rgc$ and age are adopted. The uncertainties in $\rgc$ in these two panels for inner-disc OCs ($\rgc \lesssim 7$ kpc) are significantly smaller than for \cite{Cavallo+2024}. In comparison to the two other panels, in the middle panel of Fig.~\ref{fig:FeH_radial_gradient_OCs1member} the known effect that \citetalias{hunt+2023} ages are systematically too young for old OCs can be noticed. In the bottom panel, if we included all the OCs in \citet{Spina+2022} and not only those in their ‘gold’ sample, there is only one OC located in our inner region of interest where UBC~1052 resides: Berkeley~79 ($\rgc = 6.2$ kpc), at the upper end of the [Fe/H] distribution at its $\rgc$ and much younger than UBC~1052 ($\sim 200$ Myr, \citealt{Cantat-Gaudin2020}).

\begin{figure*}[h!]
\begin{center}
{\includegraphics[width = 0.952\textwidth]{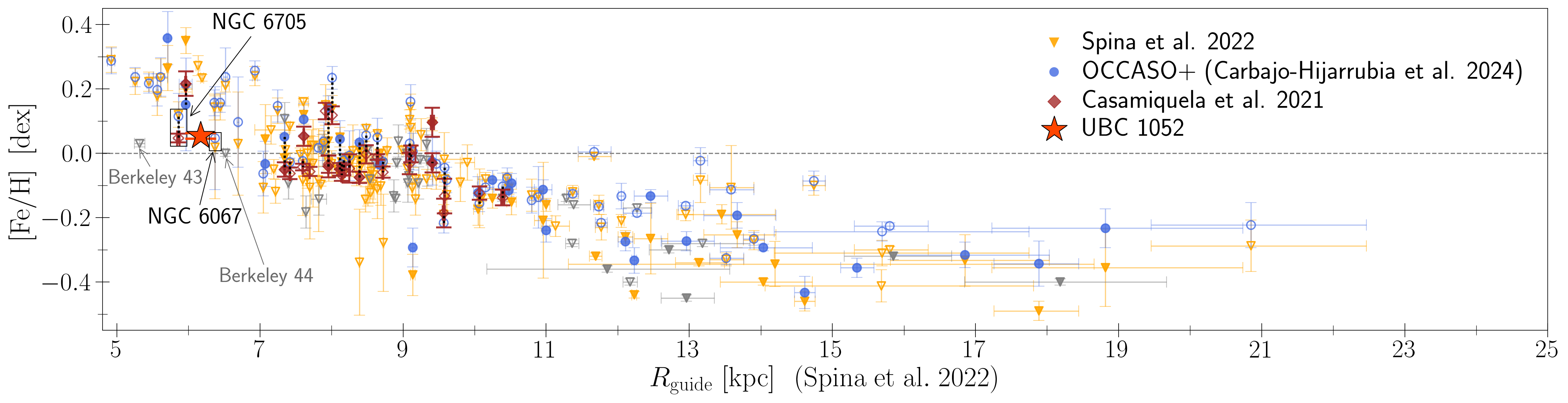}}
\caption{Radial metallicity gradient traced by the OCs in three high-resolution spectroscopic studies (see text for the sample selection) with respect to the guiding-centre radius in \cite{Spina+2022}. UBC~1052 has been added as the red star. Open symbols are OCs younger than 2 Gyr and filled symbols are OCs that are at least 2 Gyr old, using ages from \citet{Cantat-Gaudin2020}. Grey data points correspond to OCs whose [Fe/H] has been derived through only one observed member in \citet{Spina+2022}. The abundance uncertainties correspond to the standard deviation of [Fe/H] over the studied stars in each cluster.}
\label{fig:FeH_radial_gradient_Rguide_OCs1member}
\end{center}
\end{figure*}

Figure \ref{fig:FeH_radial_gradient_Rguide_OCs1member} corresponds to the [Fe/H] gradient with respect to $\rguide$ shown in the right panel of Fig.~\ref{fig:FeH_radial_gradient}, now including in grey the less-reliable OCs whose [Fe/H] has been derived using one member (for \citealt{Spina+2022} we only represent the OCs with one observed member that belong to their ‘gold’ sample).

There are two OCs whose [Fe/H] has been derived from high-resolution spectroscopy of a single member that are at $\rgc$ and $\rguide$ smaller than 7 kpc and located at the lowest part of the [Fe/H] scatter as a function of $\rgc$ (Fig.~\ref{fig:FeH_radial_gradient_OCs1member}) and/or $\rguide$ (Fig.~\ref{fig:FeH_radial_gradient_Rguide_OCs1member}): Berkeley~43 and Berkeley~44, studied in \citet{Spina+2022}.

Berkeley~43 stands out as a potential candidate for being a comparable outlier to UBC~1052 in Fig.~\ref{fig:FeH_radial_gradient_OCs1member} and even a more extreme outlier in Fig.~\ref{fig:FeH_radial_gradient_Rguide_OCs1member} considering its [Fe/H] in \citealt{Spina+2022} ([Fe/H] = $0.03 \pm 0.01$ dex from OCCAM SDSS/APOGEE DR16, \citealt{Donor+2020}). Its [Fe/H] in OCCAM DR17, derived for the same star, is similar ([Fe/H] = $-0.01 \pm 0.01$ dex, \citealt{Myers+2022}); but higher values were obtained for this star using previous APOGEE data releases (e.g. [Fe/H] = 0.18 dex in \citealt{SitNess2020}, which would not make it an outlier). In addition to [Fe/H], the parameters of Berkeley~43 are likewise largely uncertain. This cluster is embedded in a rich Galactic field and there is large reddening towards it \citep{Subramaniam+2010}, being located very close to the Galactic plane ($Z = 6$ pc) at a Galactic longitude $l = 46^{\circ}$ (\citetalias{hunt+2023}). Its CMD in \cite{Cantat-Gaudin2020} and \citetalias{hunt+2023}, though considerably populated (197 and 472 members, respectively), is significantly blurred and does not present any distinguishable RC. Consequently, its parameters could not be estimated by \cite{Cantat-Gaudin2020} who flagged this OC as ‘too red’. In \cite{Cavallo+2024}, Berkeley~43 belongs to the ‘wood’ sample, for which the parameters retrieved by the artificial neural network are the least reliable (deriving an age between $10-20$ Myr). In \citetalias{hunt+2023}, its parameters are estimated (finding an age between $35-175$ Myr), but it does not belong to the highly reliable sample of clusters. However, other studies point to an older age: $281-316$ Myr in \citet{Subramaniam+2010}, 617 Myr in \citet{Kharchenko2013}, and a visually inferred intermediate age in \citet{Netopil+2022}. The latter excluded Berkeley~43 from the sample of OCs used to trace the radial metallicity gradient and investigate radial migration due to the impossibility to perform accurate isochrone fits to its largely scattered CMD. According to the currently available uncertain age estimates, Berkeley~43 would be younger than UBC~1052; and it would reach a smaller $\zmax$ too ($\sim$ 110 pc in \citealt{Spina+2022}). Berkeley~43 deserves further investigation to better constrain its parameters and to obtain reliable abundances, which will enable to explore if it shows evidence of inward radial migration.

Berkeley~44, on the other hand, is an old OC whose RC can be distinguished in its blurry CMD both in \citealt{Cantat-Gaudin2020} (192 members) and in \citetalias{hunt+2023} (351 members, it belongs to their highly reliable sample of clusters). It is $1.3 \pm 0.2$ Gyr old according to \citet{Carraro+2006}, 2.9 [2.4, 3.5] Gyr old in \citet{JanesHoq2011}, 1.5 Gyr old in \citet{Cantat-Gaudin2020}, 1 [0.7, 1.7] Gyr old in \citetalias{hunt+2023}, and $\sim$800 Myr old in \citealt{Cavallo+2024} (wherein it has ‘gold’ quality cluster parameters). It is located at Galactic longitude $l = 53 ^{\circ}$ and $Z=173$ pc (\citetalias{hunt+2023}). Its metallicity in \citealt{Spina+2022} ([Fe/H] = 0.00 $\pm$ 0.01 dex from OCCAM SDSS/APOGEE DR16, \citealt{Donor+2020}) renders it not an outlier of the metallicity gradient at its $\rgc \sim 6.8$ kpc (Fig.~\ref{fig:FeH_radial_gradient_OCs1member}), but it places it at the lowest end of the [Fe/H] distribution at its $\rguide$ (see Fig.~\ref{fig:FeH_radial_gradient_Rguide_OCs1member}). However, its [Fe/H] in the latest OCCAM catalogue, which is also based on a single star, is much higher ([Fe/H] = 0.16 $\pm$ 0.01 dex from OCCAM SDSS-V/MWM DR19, \citealt{Otto+2025}).  This value is compatible with the one in the \gaiaeso survey: [Fe/H] = $0.22 \pm 0.09$ dex, based on UVES spectra for seven stars \citep{Randich+2022}, and does not place Berkeley~44 as an outlier of the metallicity gradient with respect to $\rgc$ or $\rguide$.

From Figs.~\ref{fig:FeH_radial_gradient_OCs1member} and \ref{fig:FeH_radial_gradient_Rguide_OCs1member} we derive the following conclusions:
Regarding the influence of the uncertainties in [Fe/H] and in the OCs’ parameters, if we loosened our selection criteria to include the less reliable OCs with only one observed member in \citet{Spina+2022} or in \citetalias{casamiquela+2021}, and if we also included the OCs having any quality for the parameters in \citet{Cavallo+2024} in the gradient with respect to $\rgc$, there would be only two new OCs at $\rgc$ and $\rguide$ smaller than 7 kpc in Fig.~\ref{fig:FeH_radial_gradient}. These are Berkeley~43 and Berkeley~44 (see top panel of Fig.~\ref{fig:FeH_radial_gradient_OCs1member} and Fig.~\ref{fig:FeH_radial_gradient_Rguide_OCs1member}), whose [Fe/H] adopted in \cite{Spina+2022} were derived from only one member observed in APOGEE and were part of their ‘gold’ sample used to trace the metallicity gradient. As has been discussed, given the current large uncertainties in Berkeley~43’s parameters and [Fe/H], no conclusions can be drawn as to whether it is a comparable outlier to UBC~1052 in the metallicity gradient; and Berkeley~44 would no longer be an outlier of the metallicity gradient according to more recent measurements of [Fe/H].

We find that the catalogue from which the OC distances are taken does not significantly alter the comparison between UBC~1052 and the other inner-disc OCs. For all distance catalogues, if only the reliable OCs are considered, UBC~1052 is the innermost OC having [Fe/H] < 0.1 dex, and NGC~6705 and NGC~6067 are also its closest clusters in the [Fe/H]-$\rgc$ plane. If we also took into account the less-reliable OCs (grey data points in Fig.~\ref{fig:FeH_radial_gradient_OCs1member}), for all three distance estimates Berkeley~43 is the only OC located at a smaller or comparable $\rgc$ as UBC~1052.

Finally, the case of NGC~6583 is worthy of notice as another example of how the lack of accurate and precise [Fe/H] estimations limits our ability to constrain the chemical evolution of the Galactic disc and the radial mixing processes that take place in it. NGC~6583 is a $\sim 1.5$ Gyr old OC, located at $\rgc = 6.8$ kpc and $Z=-22$ pc according to \citet{Cavallo+2024}. Its $\rgc$ in \citetalias{hunt+2023} and \cite{Spina+2022} is smaller ($\rgc \sim 5.9$ kpc), placing it amongst the innermost OCs used to trace the metallicity gradient. Its estimated iron abundance in the \gaiaeso survey is [Fe/H] = $0.22 \pm 0.01$ dex, based on four stars \citep{Randich+2022}. According to its [Fe/H] in \cite{Spina+2022}, adopted from \textit{Gaia}-ESO, its location is fully consistent with the [Fe/H] gradients in Figs.~\ref{fig:FeH_radial_gradient_OCs1member} and \ref{fig:FeH_radial_gradient_Rguide_OCs1member}. However, two previous studies found NGC~6583 to be more metal rich, so much as to render it a potential remarkable outlier of the [Fe/H] gradient. \citet{Zwitter+2018} estimated a [Fe/H] = $0.32 \pm 0.08$ dex based on five stars. \citet{Magrini+2010} derived a compatible but higher super-solar abundance of [Fe/H] = $0.37 \pm 0.03$ dex from a high-resolution spectroscopic study of two giants, placing it amongst the most super-metal-rich known OCs (comparable to NGC~6791). In contrast to NGC~6791, NGC~6583 is much younger, so it should be closer to its $\rbirth$ since the radial migration process has had less time to operate. In addition, it has a more circular orbit and smaller vertical excursions from the Galactic plane ($e= 0.07$ and $\zmax = 0.11$ kpc in \citealt{Spina+2022}). Therefore, its origin was considered perplexing \citep{Maderak+2015}, and so was its survival for several hundreds of Myr in the dense environment of the inner disc \citep{Cantat-Gaudin2020}. \citet{Quillen+2018} found that its high [Fe/H] in \citet{Magrini+2010} is enough to place its estimated $\rbirth$ within the Galactic bar, concluding that it was born in the bar, and the bar helped eject it from the inner Galaxy, or else that it was born near the bar end, and an outward radial migration of 2-3 kpc ${\textrm{Gyr}}^{-1}$ is required to explain its current position. Using $N$-body simulations, they constrained the surface density contrast in the Galaxy’s spiral structure at the Sun’s $\rgc$ required to induce the estimated migration distances and rates for the young, super-solar metallicity OCs NGC~6583, NGC~2632, and the Hyades. Based also on the [Fe/H] in \citet{Magrini+2010}, \citet{Netopil+2022} estimated a larger migration rate for NGC~6583, about 4 kpc ${\textrm{Gyr}}^{-1}$, much larger than the mean radial migration found for OCs of comparable age.

\section{Birth radii of the clusters used to trace the radial metallicity gradient}\label{app_sec:Rbirth}

\begin{figure*}[h!]
    \centering
    {\includegraphics[width = 0.485\textwidth]{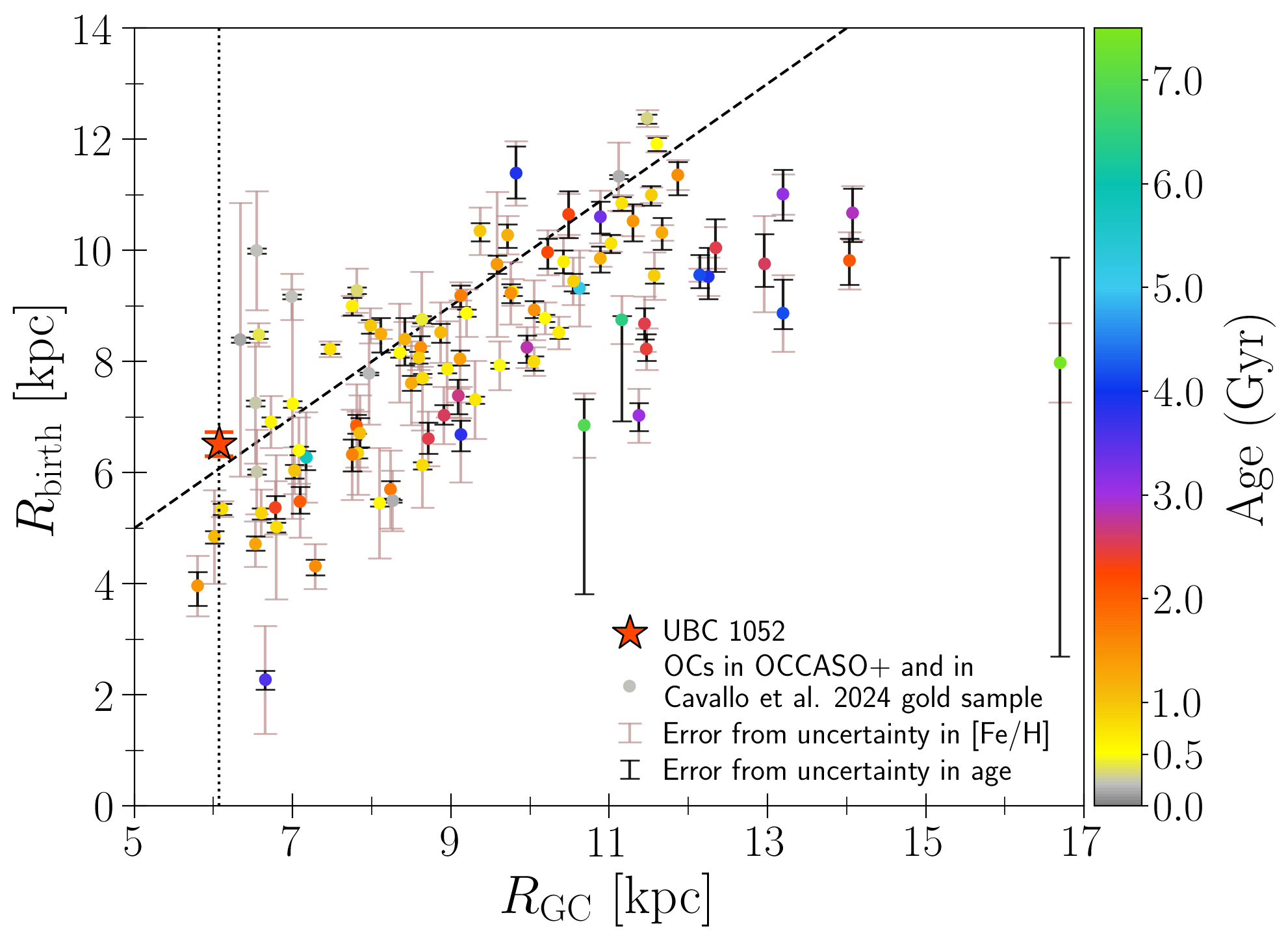}}
    {\includegraphics[width = 0.485\textwidth]{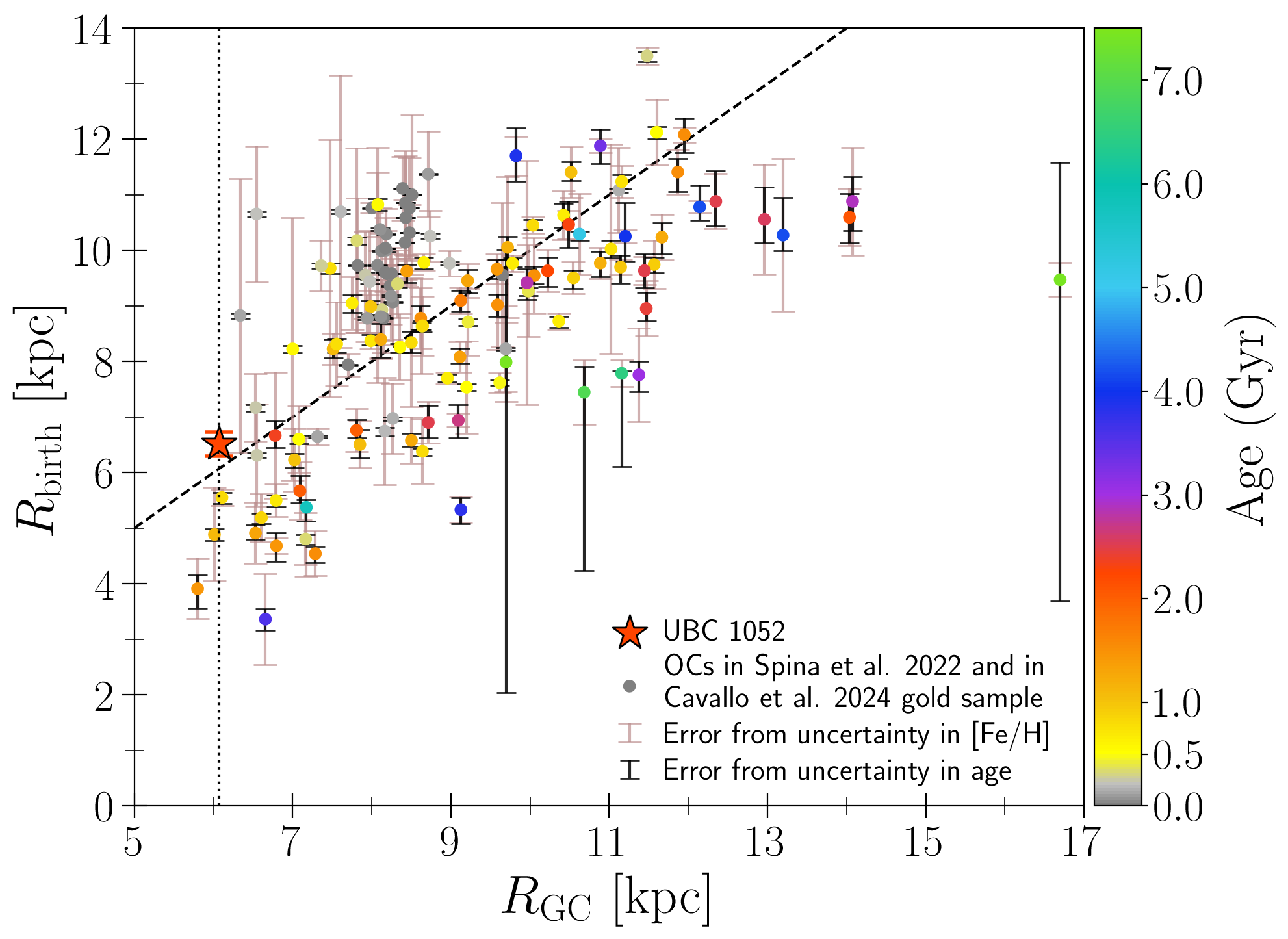}}
    \caption{Birth radius, computed as described in \citet{Ratcliffe+2026} using {\tt Rbirth} package, as a function of the current galactocentric radius in \citet{Cavallo+2024} for UBC~1052 (red star) and for the OCs in OCCASO+ (\citealt{Carbajo+2024}, \textit{left}) and in our selected sample from \citealt{Spina+2022} (\textit{right}) which belong to \citet{Cavallo+2024} ‘gold’ sample. The OCs are coloured according to their age in \citet{Cavallo+2024}. The black dashed line is the identity line, and the vertical dotted line indicates UBC~1052’s $\rgc$. NGC~6791 is not represented because its estimated $\rbirth$ is negative in both panels.}
\label{fig:Rbirth_vs_Rgc}
\end{figure*}

As done for UBC~1052 in Sect.~\ref{sec:migration}, we also estimate $\rbirth$ for the OCs in OCCASO+ and for the 145 OCs in the high-quality sample in \citet{Spina+2022} whose [Fe/H] was derived using at least two members (see Sect.~\ref{subsec:radial_FeH_gradient}). We compute several $\rbirth$ estimates following the methods in \citet{Lu+2024}, \citet{Ratcliffe+2025} and \citet{Ratcliffe+2026}; and using the [Fe/H] in each catalogue and two different age estimates (\citealt{Cantat-Gaudin2020} and \citealt{Cavallo+2024}). Then, for each OC the estimated $\rbirth$ is compared with the current $\rgc$ (adopted from \citealt{Cavallo+2024} and \citealt{Spina+2022}). We find, for all the tested methods, ages, and $\rgc$ estimates, that most OCs located at $\rgc$ smaller than UBC~1052 have $\rbirth < \rgc$. And for the remaining OCs, $\rbirth$ and $\rgc$ are compatible within the uncertainties. An example is shown in Fig.~\ref{fig:Rbirth_vs_Rgc}, where we represent $\rbirth$ (computed following \citealt{Ratcliffe+2026} and using the ages in \citealt{Cavallo+2024}) as a function of the current $\rgc$ from \citet{Cavallo+2024} for the OCs in OCCASO+ and in our selected sample from \citet{Spina+2022} that belong to \citet{Cavallo+2024} ‘gold’ sample.

\section{Analysis of GIRAFFE spectra}
\label{app_sec:giraffe}

\begin{table*}[h!]
\centering
\begin{tabular}{cccccccc}
\hline \hline
{\it Gaia} DR3 source ID & In \citetalias{castroginard+2022} & In \citetalias{hunt+2023} & $G$ [mag] & $G_{\mathrm{BP}}-G_{\mathrm{RP}}$ [mag] & S/N & $\vr \,[\kms]$ & \thead{Discrepancy w.r.t.\\ UBC~1052's mean $\vr$} \\
\hline
4286766936446235264 & Y & Y & 15.54 & 1.22 & 40 & 34.1 $\pm$ 2.8 & 0.03 \\
4286752230476578560 & N & Y & 16.13 & 1.22 & 29 & 33.6 $\pm$ 3.0 & 0.1 \\
4286760339375937920 & Y & Y & 15.90 & 1.30 & 40 & 34.5 $\pm$ 3.8 & 0.1 \\
4286752947709356416 & N & Y & 17.35 & 1.28 & 12 & 33.4 $\pm$ 4.6 & 0.1 \\
4286740616884860544 & Y & Y & 16.21 & 1.25 & 30 & 34.5 $\pm$ 2.7 & 0.2 \\
4286785078388152960 & N & N & 17.55 & 1.26 & 14 & 35.0 $\pm$ 3.8 & 0.3 \\
4286765184099511936 & Y & Y & 16.69 & 1.29 & 24 & 35.0 $\pm$ 3.3 & 0.3 \\
... & ... & ... & ... & ... & ... & ... & ... \\
\hline
\end{tabular}
\caption{Preview of the table containing the information of the targets observed with GIRAFFE. The second and third columns indicate whether the star is (‘Y’) or is not (‘N’) a member of UBC~1052 according to the membership lists by \citet{castroginard+2022} and \citet{hunt+2023}, respectively. The fourth and fifth columns correspond to {\it Gaia} $G$ magnitude and $G_{\mathrm{BP}}-G_{\mathrm{RP}}$ colour, respectively. The S/N in the sixth column corresponds to the median of the ratio between the flux and the flux error over all three GIRAFFE gratings. Our inferred radial velocity and its associated uncertainty is given in the seventh column (for 31 stars), and its discrepancy with respect to UBC~1052’s mean $\vr$, computed as described in the text, is given in the last column. The full table containing the 41 stars is available at the CDS.}
\label{table_giraffe_vr_extract}
\end{table*}

In this section, we describe our radial velocity measurements for the stars observed with GIRAFFE. For each GIRAFFE target we merged the spectra in the three gratings into a single spectrum, retaining the highest-S/N spectrum in the overlap region between gratings HR13 and HR14A. The radial velocities were determined using {\tt iSpec}, following the same method as for the UVES targets; that is, cross-correlating the barycentric-corrected merged spectrum with the high-S/N solar spectrum from NARVAL available in the {\it Gaia} FGK benchmark stars library. This calculation was restricted to the targets with S/N $\geq 10$ (computed as the median of the ratio between the flux and its corresponding error across the merged spectrum), which were all the observed stars except for the two faintest ones (with $G = 18.4$ mag). Thus, we obtained $\vr$ estimates for 39 of the 41 targets. They were further filtered by retaining only those which differed at most 10 $\kms$ (in absolute value) with respect to the $\vr$ value estimated as the weighted average of the radial velocities determined performing the cross-correlation independently for each grating with S/N $\geq 10$. This selection yielded 31 stars, enabling the rejection of seven stars with large dispersions ($> 10$ $\kms$) in the radial velocities derived for the different gratings, and of a star with radial velocities that are compatible among them for the different gratings but not with the $\vr$ value determined using all the gratings at once (merged spectrum).

The results are reported in Table \ref{table_giraffe_vr_extract}. For each star observed with GIRAFFE, we list the membership in \citetalias{castroginard+2022} and \citetalias{hunt+2023}, its {\it Gaia} DR3 photometry, the median S/N across the three gratings used (HR11, HR13, and HR14A), our determined $\vr$ and its uncertainty, and its discrepancy with respect to the cluster’s mean $\vr$ obtained from the four high-S/N RC stars in UBC~1052 observed with UVES ($\overline{\vr} = 34.0 \pm 0.6$ $\kms$). The latter is calculated as the ratio of the absolute value of the difference between the $\vr$ of each star and the cluster’s $\overline{\vr}$ over the error of this difference (i.e. the square root of the quadratic sum of the individual errors).

The 31 stars for which we provide radial velocities have $G$ magnitudes in the range [13.9, 18.3] mag and median S/N in the range [99, 11]. The three brightest ones also have {\em Gaia} DR3 RVS $\vr$ estimates: {\it Gaia} DR3 4286764462544611968 has $\vr = 30.1 \pm 2.6$ $\kms$ (compatible with our $\vr = 38.1 \pm 1.7$ $\kms$ within $3\sigma$), {\it Gaia} DR3 4286764015867996928 has $\vr = 38.4 \pm 2.9$ $\kms$ (compatible with our $\vr = 35.1 \pm 1.6$ $\kms$ within $1\sigma$), and {\it Gaia} DR3 4286766214891693440 has $\vr = 51.6 \pm 5.1$ $\kms$ (compatible with our $\vr = 52.5 \pm 1.6$ $\kms$ within $1\sigma$). The {\tt ruwe} values of the 31 stars range between [0.91, 1.26], and the average uncertainty in their radial velocity is $3.4$ $\kms$.

Out of the 31 stars observed with GIRAFFE for which we provide radial velocities, 20 (65\%) are UBC~1052 members according to \citetalias{castroginard+2022} and/or \citetalias{hunt+2023}. We find that 15 stars (48\%) have a $\vr$ compatible with UBC~1052’s mean $\vr$ within $1\sigma$, 20 (65\%) within $2\sigma$, 21 (68\%) within $3\sigma$, and 10 stars (32\%) have a $\vr$ that differs more than $3\sigma$ from UBC~1052’s mean $\vr$. 80\% of stars with a $\vr$ compatible with UBC~1052’s mean $\vr$ within $1\sigma$ and $2\sigma$ are UBC~1052 members according to \citetalias{castroginard+2022} and/or \citetalias{hunt+2023}, while 60\% of stars that have a $\vr$ that differs more than $3\sigma$ with respect to UBC~1052’s mean $\vr$ are not members of UBC~1052 in any of the two catalogues.

\end{appendix}

\end{document}